\documentclass{aa}
\usepackage[varg]{txfonts}
\usepackage{natbib}
\usepackage{caption}
\usepackage{subcaption}
\usepackage{xcolor}
\usepackage{afterpage}
\usepackage{booktabs}
\usepackage{chemformula}
\usepackage{multirow}
\usepackage{graphicx}
\usepackage{subcaption}
\usepackage{placeins}      
\usepackage{stfloats}

\usepackage[separate-uncertainty = true,multi-part-units=single,group-separator={,}, group-digits = integer, group-four-digits = true,per-mode=reciprocal]{siunitx}
\usepackage{xspace}

\usepackage{afterpage}
\usepackage[markup=bfit]{changes}
\usepackage{xcolor}

\begin{document}
\title{Magnetic field strengths of hot giant exoplanets consistent with Solar System values}

\author{Julia~V.~Seidel\inst{1,2}\thanks{ESO Fellow, Poincaré Fellow}
\and
Vivien~Parmentier\inst{1}
\and
Bibiana~Prinoth\inst{3, 24\thanks{ESO Fellow}}
\and
Thea~Hood\inst{1}
\and
Nishil~Mehta\inst{1}
\and
Valentin~De~Lia\inst{1}
\and
Konstantin~Batygin\inst{15}
\and
Tristan~Guillot\inst{1}
\and
Ragnar~Van~den~Broeck\inst{1}
\and
Hayley~Beltz\inst{11,27} 
\and
Brian~Thorsbro\inst{1,3} 
\and
Florian~Debras\inst{4}
\and
Daniel~D.~B.~Koll\inst{19} 
\and
Thaddeus~D.~Komacek\inst{10} 
\and
Emily~Rauscher\inst{18} 
\and
Lorenzo~Pino\inst{7}
\and
Matteo~Brogi\inst{13,16}
\and
Joost~P.~Wardenier\inst{8,28} 
\and
Jacob~L.~Bean\inst{6} 
\and
Björn~Benneke\inst{8,14} 
\and
Jean-Michel L. B. D\'esert\inst{23,25,26}
\and
Pablo~Drake\inst{1}
\and
Siddharth Gandhi\inst{20,21}
\and
Mark~Hammond\inst{10} 
\and
David~Kasper\inst{6} 
\and
Michael~R.~Line\inst{9} 
\and
Elspeth K.H. Lee\inst{12}
\and
Stefan~Pelletier\inst{5} 
\and
Andreas~Seifahrt\inst{17} 
\and
Adrien~Simonnin\inst{1,3}
\and
Peter~C.~B.~Smith\inst{9} 
\and
Kevin~B.~Stevenson\inst{22} 
}
\offprints{J.V. Seidel, \email{jseidel@oca.eu, the accepted manuscript can be found at https://doi.org/10.1038/s41550-026-02870-1}}

\institute{
Laboratoire Lagrange, Observatoire de la Côte d’Azur, CNRS, Université Côte d’Azur, Nice, France 
\and
European Southern Observatory, Alonso de C\'ordova 3107, Vitacura, Regi\'on Metropolitana, Chile 
\and
Lund Observatory, Division of Astrophysics, Department of Physics, Lund University, Box 118, 221 00 Lund, Sweden 
\and
 IRAP, Université de Toulouse, CNRS UMR 5277, 31400 Toulouse, France 
\and
Observatoire astronomique de l'Universit\'{e} de Gen\`{e}ve, 51 chemin Pegasi 1290 Versoix, Switzerland 
\and
Department of Astronomy \& Astrophysics, University of Chicago, Chicago, IL 60637, USA 
\and
INAF - Osservatorio Astrofisico di Arcetri, Largo Enrico Fermi 5, 50125 Firenze, Italy 
\and
Institut Trottier de recherche sur les exoplan\`etes, D\'epartement de Physique, Universit\'e de Montr\'eal, Montr\'eal, Qu\'ebec, Canada  
\and
School of Earth and Space Exploration, Arizona State University, Tempe, AZ 85281, USA 
\and
Department of Physics (Atmospheric, Oceanic and Planetary Physics), University of Oxford, Oxford, OX1 3PU, UK
\and
Department of Astronomy, University of Maryland, College Park, MD 20742 
\and
Center for Space and Habitability, University of Bern, Gesellschaftsstrasse 6, CH-3012 Bern, Switzerland 
\and
Dipartimento di Fisica, Universit\`a degli Studi di Torino, Torino, Italy 
\and
Department of Earth, Planetary, and Space Sciences, University of California, Los Angeles, CA 90095, USA 
\and
Division of Geological and Planetary Sciences, California Institute of Technology, Pasadena, CA 91125, USA 
\and
INAF -- Osservatorio Astrofisico di Torino, 10025, Pino Torinese, Italy 
\and
International Gemini Observatory/NSF NOIRLab, 950 N. Cherry Ave., Tucson, AZ 85719, USA 
\and
Department of Astronomy and Astrophysics, University of Michigan, Ann Arbor, MI, 48109, USA 
\and
School of Physics, Peking University, Beijing, People's Republic of China 
\and
Department of Physics, University of Warwick, Coventry CV4 7AL, UK 
\and
Centre for Exoplanets and Habitability, University of Warwick, Gibbet Hill Road, Coventry CV4 7AL, UK 
\and
JHU Applied Physics Laboratory, 11100 Johns Hopkins Rd, Laurel, MD 20723, USA 
\and
Leibniz-Institut für Astrophysik Potsdam (AIP), An der Sternwarte 16, 14482 Potsdam, Germany 
\and
European Southern Observatory, Karl-Schwarzschild-Str. 2, 85748 Garching bei München, Germany 
\and
Anton Pannekoek Institute for Astronomy, University of Amsterdam, Science Park 904, 1098 XH Amsterdam, The Netherlands 
\and
DESY, Platanenallee 6, D-15738 Zeuthen, Germany 
\and
Department of Physics and Astronomy, University of Kansas, Lawrence, KS, USA; 
\and
Weltraumforschung und Planetologie, Physikalisches Institut, University of Bern, Sidlerstrasse 5, Bern, 3012, Switzerland 
}

\date{Received September 30, 2025, Accepted April 16th, 2026}

\abstract{
Magnetic fields are ubiquitous in the universe. They play a key role in shaping the activity of stars, the habitability of rocky planets, and the long-term retention of planetary atmospheres. Theoretical scaling laws are largely constrained by the limited set of stars and Solar System planets, leading to a wide range of possible values for hot giant planets outside of the Solar System from fractions of the Jovian field to orders of magnitude larger. Ultra-hot Jupiters, with their highly ionised atmospheres, provide a new avenue to probe magnetic effects, as their atmospheric circulation could be directly sensitive to atmospheric magnetic field strength.
Using high-spectral resolution observations targeting the iron lines of ultra-hot Jupiters we measure the Doppler shift and thus the wind speed of seven transiting ultra-hot Jupiters. We find a clear decrease of wind speed with increasing planetary temperature, a trend inconsistent with purely hydrodynamic mechanisms but naturally reproduced by magnetic drag. From this relation we estimate the possible strength of magnetic fields of hot giant planets to at most a few gauss - comparable to the Jovian equatorial field.

Our results support the idea that magnetic fields affect the atmospheric circulation of ultra-hot Jupiters and could provide a crucial benchmark for scaling laws used to predict magnetic fields in exoplanets, from hot Jupiters to rocky Earths with additional implications for future direct observations. 
}

\maketitle

\section{Introduction}
 For gas giant planets, such as Jupiter, interactions between the atmospheric flow and the atmospheric magnetic field (for Jupiter varying between $\sim 4~G$ at the equator and a maximum of $\sim 21~G$, \cite{connerney_new_2018}) lead to a dissipation of the atmospheric jets at pressures where hydrogen becomes conductive \citep{liu_mechanisms_2010, kaspi_jupiters_2018}. Such dissipation, when applied to hot exoplanets, could be strong enough to explain the long-standing issue of their inflated radii \citep{batygin_inflating_2010, perna_Ohmic_2010}. 

Numerous theories have been proposed to provide a unified description of magnetic field formation in stars and planets via a scaling law that depends on the heat convective flux \citep{christensen_energy_2009}. For hot Jupiters, the predictions based on this scaling law differ widely. On the one hand, models predict magnetic fields of hundreds of gauss for hot giant exoplanets \citep{yadav_estimating_2017, kilmetis_magnetic_2024} based on the implicit assumption of additional heat deposited in the deep, dynamo region. On the other hand, recent works accounting for the feedback between induced currents in hot atmospheres and the convection-supported dynamo field \citep{zaghoo_size_2018, elias-lopez_rossby_2025, vigano_inflated_2025} predict a low magnetic field with at most a few gauss for the same class of hot exoplanets.

Thus far, the wide diversity of magnetic fields among Solar System planets, together with the absence of direct measurements for exoplanets, limits the benchmarking of these two opposed predictions and, in consequence, our understanding of magnetic fields in planets. Another important complication is the high ionisation of these planets, leading to the possibility of large induced fields \citep{dietrich_magnetic_2022}.

\subsection{Current observational evidence}

Observationally, promising results were shown for star-planet-interactions (SPI) measurements based on activity indicators for planets in extremely short orbits \citep{cuntz_stellar_2000, shkolnik_evidence_2003, shkolnik_hot_2005, shkolnik_star-planet_2009} which can e.g. be observed via the modulation of the Ca II doublet lines as a function of orbital phase \citep{cauley_magnetic_2019, cauley_effects_2022} or light curve asymmetries \citep{vidotto_early_2010, vidotto_prospects_2011, llama_shocking_2011}. However, both techniques are highly susceptible to false positives as stated in the respective works themselves, particularly those arising from stellar activity or the underlying dependence on the specific SPI modeling. While their tentative results of hundreds of gauss for the atmospheric magnetic field strength support the high magnetic field regime hypothesis, these results have to be interpreted with caution.

If the order of magnitude result from SPI works holds true, the associated radio frequencies (100~G corresponding to 280~MHz) should be observable routinely with current facilities. Surprisingly, decades of search for radio emission signals \citep{cowley_solar-wind-magnetosphere-ionosphere_2003,griesmeier_predicting_2007, zarka_plasma_2007, zarka_ground-based_2008} have led to a single tentative detection by \cite{turner_search_2021} that could not be confirmed through follow-up observations \citep{turner_follow-up_2024}. Possible reasons for this lack of detection are the naturally transient and directional nature of radio emissions \citep{Collet_new_2024} and potentially the shielding mechanisms of the highly ionised atmospheres of hot planets \citep{narang_ugmrt_2024}. In fact, it is the inherent directional nature requiring beaming in the line of sight that does not allow to place a firm upper limit on magnetic fields from such non-detections as they could simply reflect an unfavourable geometry. Moreover, current ground-based radio observations of exoplanetary magnetic fields are inherently restricted due to strong attenuation of high-frequency radio waves in Earth’s ionospheric D layer, most severe below $10$~MHz. This translates into an observational cut-off corresponding to a minimal observable magnetic field strength of $\sim 3.6~$G \citep{zarka_auroral_1998}. In light of these limitations and the long-standing difficulty of securing a detection, it is worth considering---without excluding the possibility of future breakthroughs---that the absence of firm radio emission observations from hot exoplanets may reflect an intrinsic upper limit to the strength of their magnetic field.

\subsection{Atmospheric dynamics as an alternative probe}

In summary, currently no unambiguous detection of high magnetic field strengths in planets exists and the question on the order of magnitude from scaling laws remains unresolved. As an alternative, the atmospheric magnetic field strength can be deduced indirectly for planets outside of the Solar System. In this context, ultra-hot Jupiters with temperatures above $2000$~K provide an exceptional opportunity. Due to tidal synchronisation, the rotation period of ultra-hot Jupiters is expected to be significantly longer than the Solar System gas giants at the order of days instead of $\sim 10$~hours for Jupiter or Saturn. This implies a much diminished role of the Coriolis force on the winds of ultra-hot Jupiters. Secondly, the energetic input into ultra-hot Jupiter atmospheres exceeds Earth's insolation by three orders of magnitude on average. In consequence, the stellar irradiation dwarfs the intrinsic planetary heat flux. Combining both leads to a strong day-to-night-side temperature gradient which drives sub-stellar to anti-stellar flow - from here on called day-to-night-side wind \citep{showman_doppler_2013}. Without taking other processes into account, in a purely hydrodynamical scenario, wind speeds of this day-to-night-side wind should increase with increasing equilibrium temperature \citep{showman_atmospheric_2002}.

However, in ultra-hot Jupiter atmospheres, thermal ionisation of alkali metals leads to a strong coupling between the atmosphere and their atmospheric magnetic field inducing mainly Ohmic drag \citep[for planetary equilibrium temperatures beyond $1600~$K,][]{batygin_inflating_2010, perna_Ohmic_2010, thorngren_bayesian_2018}. Particularly, similar to the processes in the deep layers of Jupiter, Ohmic drag is expected to slow the wind speed. Given that ionisation increases rapidly with temperature, magnetic drag would lead to decreasing wind speeds with increasing temperature \citep{perna_Ohmic_2010, menou_magnetic_2012, rogers_magnetic_2014}.  

As proposed early on by \cite{batygin_magnetically_2013} and equally shown in \cite{vigano_inflated_2025} directly measuring the speed of the planetary winds for a population of exoplanets beyond the $1600~$K temperature limit, such as ultra-hot Jupiters, allows us to establish if Ohmic drag is the likely main driver of kinetic energy dissipation in hot atmospheres. If this trend can be observed, it provides us with the opportunity to estimate the order of magnitude of atmospheric magnetic fields for planets outside of the Solar System by measuring the speed of the planetary winds and discriminate between the two postulated field strength regimes. Any such method, however, would be sensitive to the total atmospheric field, that is composed of the sum of the deep-seated dynamo field and the field induced by the atmospheric motions and therefore can only provide order of magnitude estimates. 

Wind speed measurements on exoplanets have become routine with the arrival of ultra-stable, ultra-precise high-spectral resolution spectrographs on large telescopes such as ESPRESSO \citep{pepe_espresso_2021} and MAROON-X \citep{seifahrt_-sky_2020}. These instruments are capable of resolving the wind-induced Doppler shifts in the planetary spectral lines, which has allowed the community to characterize the wind profiles in exoplanets in unprecedented detail \citep[e.g.][]{ehrenreich_nightside_2020, seidel_into_2021, kesseli_up_2024, nortmann_crires_2024, seidel_vertical_2025}. \cite{snellen_exoplanet_2025} presents a summary of the reported day-to-night side winds and describes a heterogeneous picture without a clear overarching trend as a function of temperature \citep[see similarly][for implications on Ohmic drag]{knierim_shallowness_2022}. However, these conclusions are drawn from datasets reduced by different groups and are further shaped by the underlying assumption that the various atmospheric tracers used to infer these winds are equivalent in origin and diagnostic power. Yet, different tracers are potentially sensitive to different regions of the atmosphere \citep{borsa_atmospheric_2021, seidel_vertical_2025}, taken with different instruments and wavelength ranges, as well as applied to different classes of planets that might be affected by different physics. This highlights the need for a clearly defined study sample, such as ultra-hot Jupiters in our case, analysed homogeneously with the same atmospheric tracer and the same observational setup to mitigate biases. \\

In this work, we combine the technological and methodological advances in extracting exoplanetary wind speeds from high-resolution spectra for the ultra-hot planet population to show that Ohmic drag likely dominates energy dissipation for ultra-hot atmospheres. This allows us to constrain the maximal atmospheric magnetic field strength of ultra-hot exoplanets as a class for the first time.

\section{Transit spectroscopy observations}
\label{sec:observations}

For this homogeneous population study, we select iron as our atmospheric tracer because its rich forest of spectral lines in the optical makes the observations more robust against local noise than single-line measurements. Compared to other refractory species, iron is found to be a robust tracer of planet metallicity, making our study less sensible to temporary changes \citep{gandhi_retrieval_2023}.  Moreover, after sodium, iron is the metal most frequently detected in exoplanet atmospheres with high-resolution spectroscopy, providing a sufficiently large sample for a population survey. Our analysis focuses exclusively on the residual Doppler shift in the planet’s rest frame. When integrated over the full transit, this residual shift offers a direct measure of the day-to-night wind velocity along the line of sight as a first order displacement of the central wavelength of the signal. Other circulation patterns, such as equatorial jets, would instead manifest as second order effects, such as a symmetric broadening or splitting of the lines when integrated over the transit.

We focus on Jupiter-sized targets observed with either MAROON-X on Gemini-North or ESPRESSO at ESO's VLT on Paranal Observatory because of their high spectral resolution, wavelength coverage and most importantly telescope mirror size and resulting signal-to-noise ratio (S/N). The data was either directly acquired from the informed principal investigators (PIs) or downloaded from the ESO archive in the case of public - and prior published - ESPRESSO data.

Of the original $15$ potential suitable targets in Table \ref{tab:observation_log}, three targets did not show a resolved iron signature likely due to lower temperatures leading to e.g. a smaller scale height, more likely cloud cover, and condensation. Additionally, one target was excluded due to low S/N and four targets due to pulsating or chemically variable host stars which make any extraction of the iron signature difficult and unreliable. The overview of the data with the named targets, as well as the details of the analysis and necessary corrections such as the telluric or the Rossiter-McLaughlin effect corrections are described in the Methods (Section \ref{app:data_prep}). Importantly, this also includes our measurement of the largest uncertainty on the velocity, the system velocity - between the barycenter of the host system and the Solar System - which can be poorly constrained in the literature (see Methods Section \ref{sec:vsys}.

 For targets observed with both spectrographs, as is the case for WASP-76~b and WASP-189~b, we also account for differences in template wavelength coverage during stacking, following the method described in \cite{borsato_mantis_2023}. For example, the MAROON-X red arm contributes less due to fewer iron lines present in this wavelength region. We check for a potential impact from the use of two different spectrograph with WASP-76~b which was observed on multiple occasions with both the ESPRESSO and MAROON-X spectrograph. We find no impact from the choice of spectrograph. We then cross-correlate the iron template from \cite{kitzmann_mantis_2023} with our data using {\tt tayph} \citep{hoeijmakers_hot_2020}. Given the uncertainty on the orbital velocity, we marginalise over the orbital velocity as well as over the planetary wind velocity. The resulting detection maps for each target can be found in the Methods, Figure \ref{fig:kpvsys} as well as further information on the cross-correlation and detection peak extraction.\\
 
\begin{figure}
    \centering
    \includegraphics[width=1.0\linewidth]{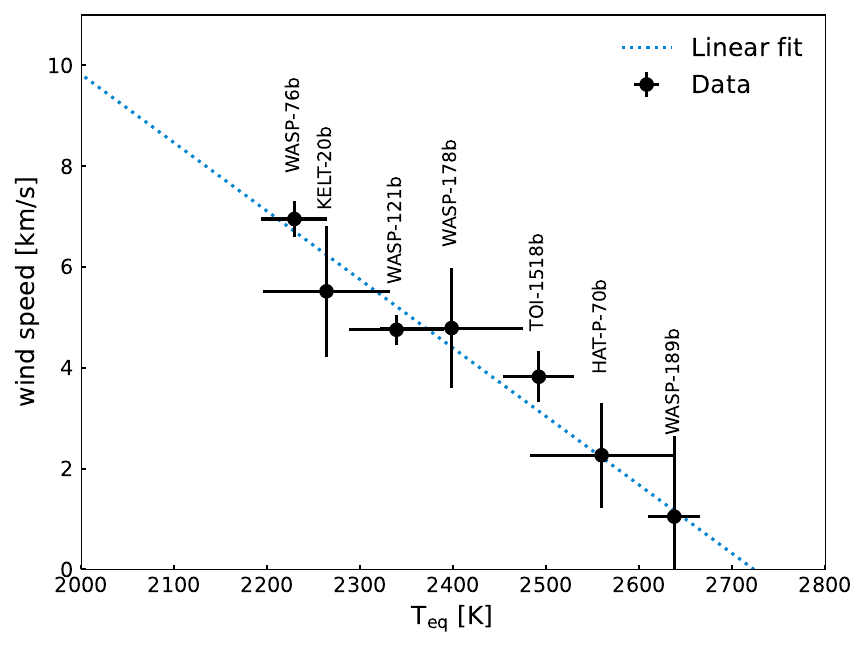}
    \caption{A clear trend of decreasing wind speeds with equilibrium temperature. We show day-to-night line of sight wind speeds for the available population of ultra-hot Jupiters as measured from iron with a clear decrease in wind velocity as a function of equilibrium temperature highlighted by a simple fit to guide the eye (dotted line). The uncertainty of the equilibrium temperature is the error propagation of the effective temperature of the star and the stellar radius to semi-major axis ratio in the calculation of the planet's equilibrium temperature. The uncertainty on the wind speed is the fit uncertainty of the observed cross-correlation peak, taking into account the photon noise of the data, as well as the propagated uncertainty of the system velocity measurement.}
    \label{fig:master_vsys}
\end{figure}

 The resulting line-of-sight atmospheric wind speeds derived from the iron lines' Doppler shift are shown in Fig.~\ref{fig:master_vsys}. We find that the day-to-night wind speeds at the terminator of ultra-hot Jupiters range from approx. 7~km/s (WASP-76~b) to less than 2~km/s (WASP-189~b). Fitting a simple linear model to the sample (see Figure \ref{fig:master_vsys}) shows a decreasing trend of the wind speed as a function of equilibrium temperature for planets with equilibrium temperatures in the studied range above 1600~K, which marks an important result emerging from the study of the population of ultra-hot Jupiters instead of singular objects.

Given the rapidly growing number of line-of-sight Doppler shift velocity measurements reported in recent years, it is crucial to keep two key considerations in mind when placing our results in context with the existing literature. First, absolute velocity values are highly sensitive to the details of data post-processing and analysis—most notably to the choice of spectral template and to how the systemic velocity is derived from the data (see Methods \ref{sec:vsys} and the discussion in \cite{lenhart_pepsi_2025} on differences between their work and the literature on KELT-20~b, caused by the same issue). Second, planetary atmospheres are three-dimensional. The vertical stratification of their flow patterns inherently limits the comparability of wind speeds and directions when different species trace distinct atmospheric layers. Consequently, inter-species comparisons—for example, between Doppler shifts measured for iron, sodium, or helium—should be interpreted with caution, as these species probe markedly different regions of the atmosphere (see e.g. Figure 1 in \cite{seidel_vertical_2025}). This three-dimensionality is equally important in the data selection, as partial transits favour viewing geometries of one hemisphere over the other, arriving at distorted mean velocity values due to the non-symmetric impact of planetary rotation. One such example is the partial transit analysed in \cite{basinger_pepsi_2025} for TOI-1518~b where seemingly their result is in tension with the velocities derived in \cite{simonnin_time_2025} and our work if the viewing geometry of the partial transit is not accounted for. Overall, the strength of our study is the combination of a careful selection of high-quality datasets and the homogeneous data reduction and interpretation process, leading to Doppler-shift values that are comparable between planets.

\section{Interpretation of the observed velocity trend}
\label{sec:modeling}

For tidally locked planets, wind speeds are expected to increase with temperature: higher planetary temperatures amplify the day–night contrast, strengthening the pressure gradient that drives atmospheric flow \citep{showman_doppler_2013}. This trend, however, cannot persist indefinitely, as winds are ultimately limited by dissipative processes. Our observation that wind speeds decrease with increasing temperature indicates that the efficiency of the mechanisms dissipating atmospheric flow must rise sharply with equilibrium temperature across the range probed by our sample. 

\subsection{Evidence for Ohmic dissipation}

Ultra-hot Jupiters are dominated, in their upper atmospheres before the effects of escape are notable, by a global sub-stellar to anti-stellar flow \citep{ehrenreich_nightside_2020, kesseli_confirmation_2021, seidel_vertical_2025, simonnin_time_2025}. The magnitude of this flow is determined by the balance between the input energy source, which is mainly the stellar irradiation, and the dissipation of energy by different mechanisms. These can either be due to hydrodynamical instabilities (e.g. Rayleigh-Taylor instabilities \citep{fromang_shear-driven_2016}, shocks \citep{heng_atmospheric_2011, watkins_gravity_2010}, or, specifically for ionised atmospheres, Ohmic drag \citep{perna_Ohmic_2010, rogers_magnetohydrodynamic_2014}. In such a framework, the scaling of the dissipation mechanisms with the planetary temperature is what determines the scaling of the winds with planetary temperature at the population level. 

One way to quantify this equilibrium is by considering atmospheres as heat engines. This approach has proven fruitful for the understanding of hurricanes on Earth \citep{emanuel_air-sea_1986} and rocky exoplanets \citep{koll_temperature_2016}, and was proposed in \cite{koll_atmospheric_2018} for the application to hot and ultra-hot Jupiters. 

In short, with their approach we assume the atmosphere is a heat engine with work output rate $W = \eta Q$, where $\eta$ is the engine's thermodynamic efficiency and $Q=\sigma T_{\text{eq}}^4$ is the heating rate. With $T_{\text{eq}}$ as the equilibrium temperature of the planet and $\sigma$ as the Stefan-Boltzmann constant, it is simply the absorbed stellar flux. We further assume that, in equilibrium, all work goes into dissipating the atmosphere's large-scale kinetic energy. This leads us to the following relation for the wind velocity in the modelled atmosphere (equation 11 in \cite{koll_atmospheric_2018} where further details on the derivation can be found):

\begin{equation}
\label{eq:rayleigh}
\text{vel} = k_0 \left( \tau_{\text{drag}} \eta \sigma T_{\text{eq}}^4 \frac{g}{p} \right)^{1/2}
\end{equation}

where $T_{\text{eq}}$ is the equilibrium temperature of the planet, $g$ is gravity, and $p$ is pressure. $k_0$ is a scale factor, set to 0.25 based on comparison with a recent grid of global circulation models (see Methods \ref{sec:assumptions}). $\tau_{\text{drag}}$ is the drag timescale that depends on the exact physical mechanism dissipating the energy and finally, $\eta$ is the efficiency of the atmospheric heat engine. For an ideal heat engine, we can place an upper bound on $\eta$. For example, if parcels of air are heated and cooled at constant pressure, and the day-to-night temperature contrast is large, the resulting thermodynamic cycle is called an Ericsson cycle, with efficiency:

\begin{equation}
\eta = \dfrac{
2R/c_p}{1 + 2R/c_p}\approx 0.36 
\end{equation}

where $R$ is the gas constant and $c_p$ is the heat capacity. In practice, $\eta$ is less than the value for an ideal heat engine. This is because atmospheres are not perfectly efficient. However, any difference between the real value of $\eta$ and our calculation would reflect as a constant value across the population and is absorbed into $k_0$. As shown in Methods Section \ref{sec:assumptions}, the heat engine model is an excellent match to complex, non-grey, global circulation models. Deviations between the wind speeds predicted by the heat engine formalism and the GCM outputs are of the order of $\approx 0.2~$km/s at most. Although this points towards a variation of heat engine efficiency with temperature, it is orders of magnitude too small to explain the change in velocity that we observe in our sample.  

The drag timescale used in equation~\ref{eq:rayleigh} encompasses all the dissipation physics. It can be seen roughly as the time for a parcel of gas to lose all its velocity if all other forces were suddenly removed. Large timescales therefore mean weak dissipation, whereas small timescales mean strong dissipation. The timescales necessary to explain the measured wind speed decrease by more than one order of magnitude with temperature across our sample, going from $\approx 7~h$ at 2200~K to less than $15~min$ for our hottest target (see \ref{fig:tau_drag} in the Methods).

Among all the possible mechanisms that can dissipate the winds on hot exoplanets, Ohmic dissipation is the only known one that is expected to have such a strong scaling with temperature due to the strong dependence of electron density with temperature in the Saha equation. In such a case, the drag timescale can be calculated following \cite{perna_Ohmic_2010, rauscher_three-dimensional_2013, beltz_magnetic_2023} and taking into account \cite{christie_geometric_2025}:
\begin{equation}
\label{eq:drag}
    \tau_{\text{drag}}=\frac{4\pi H_e\rho}{B^2}
\end{equation}

where $B$ is the atmospheric magnetic field strength, $H_e$ is the atmospheric magnetic diffusivity, and $\rho$ is the gas density (inversely proportional to temperature at a given pressure level). 
The magnetic diffusivity is described as

\begin{equation}
    H_e = 230\sqrt{T_{\text{eq}}}/x_e [\text{cm}^2/\text{s}]
\end{equation}

in Equation 2 from \cite{beltz_effects_2025}, following \cite{draine_magnetohydrodynamic_1983}, with $x_e$ as the unitless ionisation fraction for which we use chemical equilibrium tables from~\cite{lupu_correlated_2021} with the underlying chemical network from \cite{visscher_atmospheric_2010}. As shown in \ref{fig:alkali_contribution}, the electron density increases by an order of magnitude between $2200$ and $2500$~K. This hinges on the use of the correct temperature, for which we use the equilibrium temperature. This choice and the possible impact of a deviation are discussed in the Methods \ref{sec:assumptions}.

\begin{figure*}
    \centering
    \includegraphics[width=1.0\linewidth]{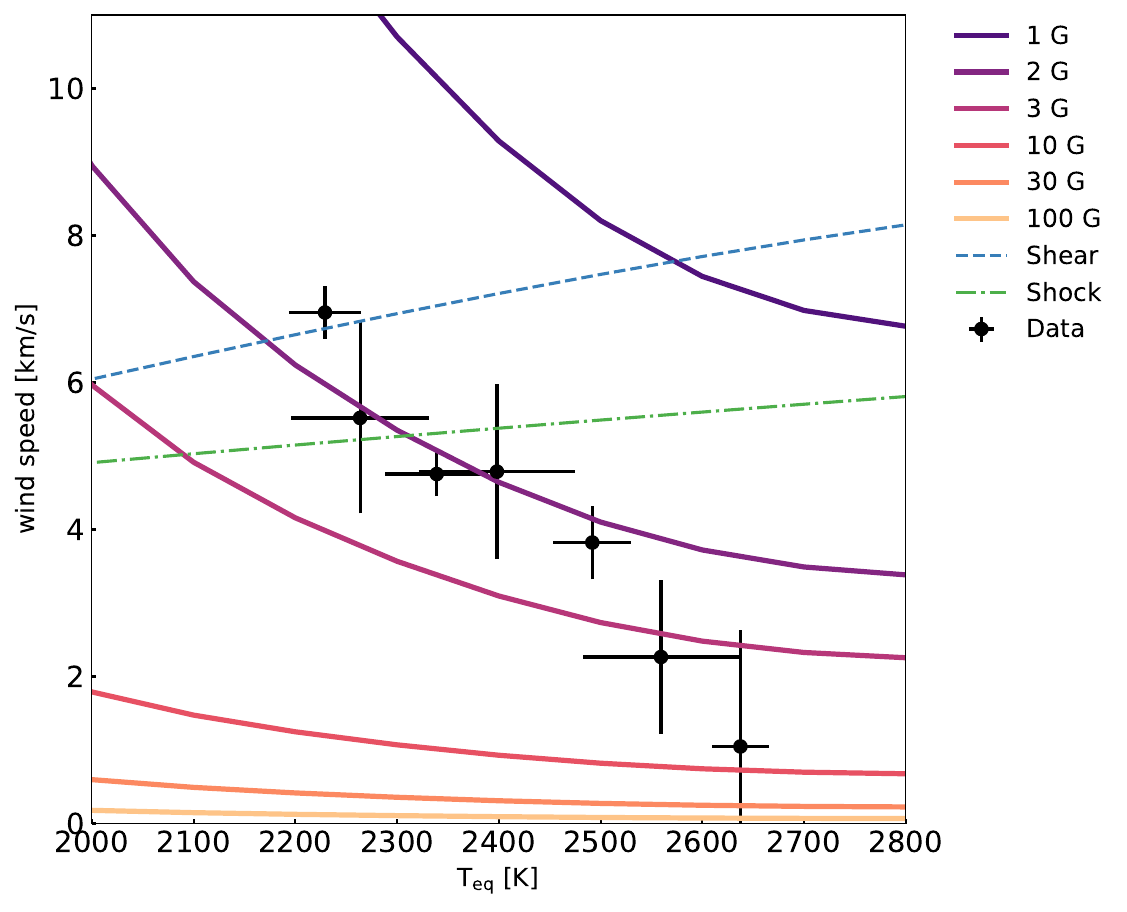}
    \caption{Ohmic drag models reproduce the observed trend of decreasing velocities. We show the measured velocities as data points with the Ohmic drag models for ultra-hot Jupiters colour coded at different atmospheric magnetic field strengths. For comparison the classical models only invoking shear instabilities (dashed) or shocks (sound speed, dashed dotted) to dissipate kinetic energy are shown.}
    \label{fig:master_vsys_model_real}
\end{figure*}

Hydrodynamical processes have been proposed to limit wind speeds in hot Jupiters, with shear instabilities and shocks identified as the most likely candidates. As shown in \cite{koll_atmospheric_2018}, for the case of planetary-scale hydrodynamic instabilities, the wind speed is still expected to increase with temperature as $T_{\rm eq}^{4/3}$ (see their equation 15).  Similarly, if winds were limited by shock dissipation, they would likely scale with the sound speed, as $\sqrt(T_{\rm eq})$. As shown in Figure~\ref{fig:master_vsys_model_real}, the observed decrease in wind speed with temperature is consistent with predictions from the magnetic drag model, while the hydrodynamic shear instability (instability scale as the planetary diameter) and shock-limited models are clearly incompatible with the data. We thus conclude that Ohmic drag is the most likely mechanism that limits the wind speed in ultra-hot Jupiter atmospheres.

\subsection{Estimation of the atmospheric magnetic field strength}

Whereas the slope of the wind–temperature relation provides evidence for Ohmic drag acting to slow the winds, the absolute wind speed can be used to infer the approximate atmospheric magnetic field strength. As shown in equation~\ref{eq:rayleigh}, the wind speed at a given temperature is expected to scale in direct proportion to the atmospheric magnetic field strength. This relation can therefore be inverted to estimate the atmospheric magnetic field of each planet individually.

However, because we are now considering the absolute magnitude of the atmospheric magnetic field rather than its scaling with temperature, careful attention must be paid to the individual terms in equation~\ref{eq:drag}. In particular, the ionisation fraction, used to calculate the magnetic diffusivity, is pressure and metallicity dependent, as well as on the local temperature at the point of main contribution to the dissipated work.  

In order to link the observed wind speed to the atmospheric magnetic field values, we need to determine which pressure levels are probed by the observations. We do this by calculating the contribution with a framework designed specifically for cross-correlation spectroscopy (see  Figure~\ref{fig:contrib_func} in Methods). We find that our observations probe from 1~to~100~mbar with a maximum of the contribution function at 30~mbar. In order to derive our estimate of the atmospheric magnetic field, we averaged the ionisation fraction at all pressure levels using the contribution function as a weighting factor (see \ref{sec:mag_field_assumptions} for more details). Finally, atmospheric retrievals have shown that the refractory abundance in ultra-hot Jupiter atmospheres is similar to the metallicity of their host star~\citep{gandhi_retrieval_2023,pelletier_crires_2024,smith_roasting_2024}. Given that hot Jupiter host stars are slightly enriched in metals, we used a metallicity of 0.15~dex, following the peak of the population ~\citep{osborn_investigating_2020}. 
One important limitation of the applied model is the lack of induced fields which play a significant role on the day-side of ultra-hot atmospheres where the magnetic Reynolds number is large, $\text{Re}_m >> 1$ \citep{rogers_magnetic_2014, dietrich_magnetic_2022, soriano-guerrero_non-ideal_2025}. As a consequence, wind speeds could decrease faster than captured by our model, meaning our work provides an upper limit of atmospheric magnetic field strengths and does not imply a linear relation to the underlying deep-seated dynamo field. We caution against an over-interpretation of the applied model beyond the order of magnitude scale and invite self-consistent MHD follow-up work.

\begin{figure*}
    \centering
    \includegraphics[width=1.0\linewidth]{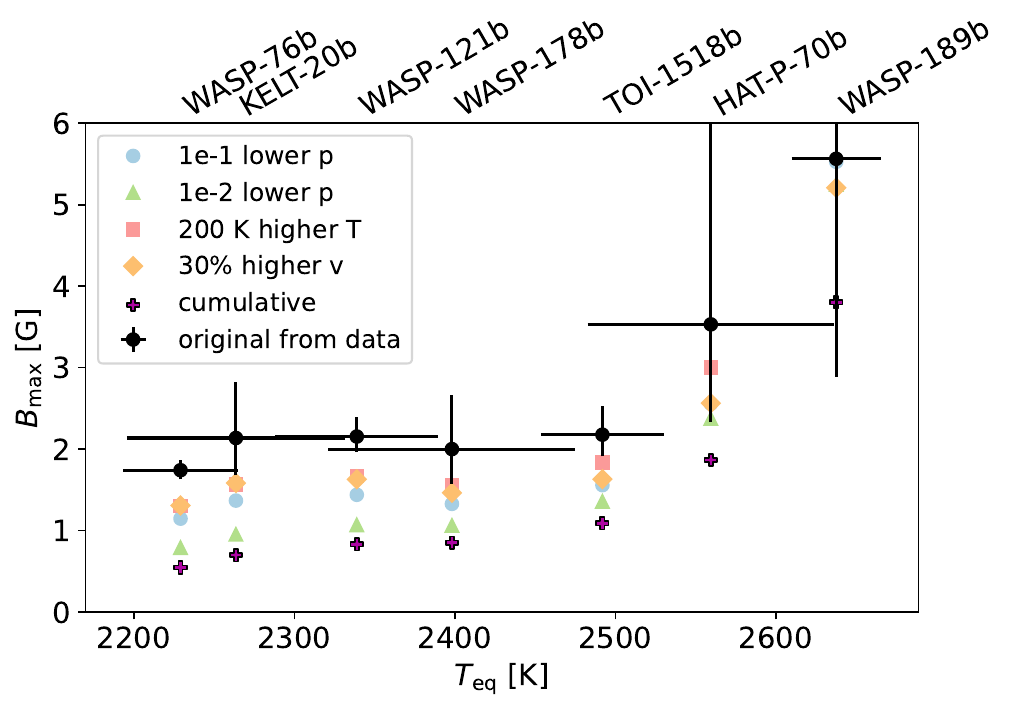}
    \caption{Ultra-hot Jupiters have maximum atmospheric magnetic field strength compatible with Jovian values. We show the derived maximum atmospheric magnetic field strength as a function of equilibrium temperature with the values derived with uncertainties from the data in Figure \ref{fig:master_vsys_model_real}. Additionally, we provide the main impacts that could create deviations on the absolute atmospheric magnetic field strength, notably order of magnitude deviations in pressure, the maximum deviation from the equilibrium temperature, and an underestimation of the absolute wind speed velocity from the line of sight measurement. The cumulative impact of all these effects is marked with a dark cross. These are order-of-magnitude estimates of the average atmospheric field in the equatorial regions, including both the deep-seated dynamo field and the field induced by the atmospheric flow.}
    \label{fig:Bfield_strength}
\end{figure*}

As shown in Figures~\ref{fig:master_vsys_model_real} and~\ref{fig:Bfield_strength}, our wind speed measurement of the five planets between $2200$ and $2500$~K points towards a constant magnetic field of at most 2~G, with little variation between targets. For the hottest two planets we derive larger magnetic field values (up to 6~G), although with much larger errorbars.  


In inverting the wind speed measurements into atmospheric magnetic field strength, we made multiple informed choices about several atmospheric parameters. As discussed in more detail in the Methods \ref{sec:mag_field_assumptions}, the main source of uncertainties in our estimate come from the estimate of the contribution function, the temperature that is relevant to calculate the ionisation fraction, and a possible projection factor between the actual wind speed and the measured winds projected along the line-of-sight. Whereas all these have an effect on the measured wind speed, Figure \ref{fig:Bfield_strength} shows that all caveats would bias us towards lower atmospheric field strength, meaning that we can firmly limit the atmospheric magnetic field of hot gas giants to a few gauss.

Nonetheless, our derivations provide a remarkably narrow range of potential magnetic field strengths and most importantly show that the atmospheric fields are likely homogeneous across the population of ultra-hot gas giants. 

\section{Discussion and Conclusion}

The absolute atmospheric magnetic field strength of exoplanets has been a long standing open question, with direct observations intrinsically limited to large field strengths and indirect methods plagued by false positives. As a consequence, it was unclear up to now whether scaling laws derived from observations of stars and in-situ measurements of solar-system planets could indeed apply to exoplanets and potentially inform habitability as different applications yielded predictions ranging from several hundred to a few gauss. In this work, we observe that wind speeds in ultra-hot Jupiters decrease very strongly with planet temperature. We further show that Ohmic drag is the best explanation for this decrease, due to the strong temperature dependence of the atmospheric conductivity. Our observations rule out the possibility of negligible magnetic fields for ultra-hot planets and demonstrate that atmospheric magnetic fields of ultra-hot Jupiters likely do not exceed field strengths of a few gauss. 

\subsection{From atmospheric to deep-seated dynamo magnetic fields}

Our measurements are sensitive to the averaged dayside atmospheric field, which, for ultra-hot Jupiters, is a combination of the deep-seated dynamo field and the induced field. Indeed, as shown in multiple MHD works, the high magnetic Reynolds number regime of these objects means that the induced field can dominate over the deep-seated dynamo field, and can even self-sustain without a deep field \citep{rogers_magnetic_2014, dietrich_magnetic_2022, soriano-guerrero_non-ideal_2025}. However, our observations can shed light on the order of magnitude of the deep-seated dynamo field. Indeed, if the deep-seated dynamo field were of the order of hundreds of gauss, as proposed by \cite{yadav_estimating_2017, kilmetis_magnetic_2024}, the induced field would have to be fine-tuned to have the opposite sign and very similar magnitude of the background field, in order to have a combined smaller field of a few gauss for \textit{all} planets in the studied population. As this scenario is exceedingly unlikely, we propose that the deep-seated dynamo field of ultra-hot Jupiters cannot be larger than a few gauss. Our estimate is in line with the recent theoretical estimates by~\cite{elias-lopez_rossby_2025}, which show that the Ohmic dissipation can shut down the convective motions in the outer convective zone, reducing the strength of the dynamo field "to or below Jupiter's". 

\subsection{Consequences for future observational programmes}

Decades of large programmes to directly observe the magnetic field strength of exoplanets have thus far only yielded tentative upper limits for potential magnetic fields \citep{turner_search_2021, turner_follow-up_2024}, with the strong caveat that these upper limits assume emission in the line of sight of the observer.

Similarly, modeling the polarimetry signature of helium as proposed in \cite{oklopcic_detecting_2020} results in estimates of the order of $0.1~$G \citep{khodachenko_impact_2021, schreyer_using_2024} to explain the observed atmospheric escape. Considering the difference in altitude between our atmospheric magnetic field strength measurements and their estimates related to the outflow of the exosphere, the weaker field might be a simple result of the attenuation of magnetic fields by the cubed radius distance relation and is likely compatible with our results.
More recently, \cite{savel_new_2024} suggested that the velocity difference between neutral and ionised species could be used to estimate exoplanetary magnetic fields. However, even a relatively strong field of $10~$G - three times the equatorial field of Jupiter - would produce an average velocity difference of only $\sim 1$ kms$^{-1}$, implying a detection threshold comparable to that of direct radio observations. \\

We show, from a homogeneous study of ultra-hot Jupiters, that wind speed decreases strongly with planetary temperature, the best evidence for the presence of Ohmic drag to date. Our measurements are compatible with a near-Jovian atmospheric magnetic field strength for these planets, ruling out the presence of magnetic fields of tens or hundreds of gauss. This is in agreement with recent theoretical studies predicting Jovian-like magnetic fields based on an improved interpretation of the scaling laws applicability to strongly irradiated planets. Our work has strong implications for the observations of exoplanetary coherent radio emission (e.g.,\cite{zarka_plasma_2007, griesmeier_predicting_2007}): since the electron cyclotron frequency is $\nu = 2.8 B$ [G] MHz, potential signals from ultra-hot Jupiters would lie very close to the $\sim 10$ MHz ionospheric cutoff, hampering detectability. Among other factors, this might naturally explain the current lack of confirmed radio observations despite decades-long campaigns.

\begin{acknowledgements}
This manuscript is the preprint version of the paper of the same title published in Nature Astronomy, please go to \url{https://doi.org/10.1038/s41550-026-02870-1} for the accepted version of the manuscript. The authors acknowledge the ESPRESSO project team for its effort and dedication in building the ESPRESSO instrument, as well as the project team and support staff of MAROON-X. This work relied on observations collected at the European Organisation for Astronomical Research in the Southern Hemisphere and at the international Gemini Observatory, a program of NSF NOIRLab, which is managed by the Association of Universities for Research in Astronomy (AURA) under a cooperative agreement with the U.S. National Science Foundation, on behalf of the Gemini partnership of Argentina, Brazil, Canada, Chile, the Republic of Korea, and the United States of America. This work was enabled by observations made from the Gemini North telescope, located within the Maunakea Science Reserve and adjacent to the summit of Maunakea. We are grateful for the privilege of observing the Universe from a place that is unique in both its astronomical quality and its cultural significance. JVS gratefully acknowledges the financial support of the Observatoire de la Côte d'Azur via the Poincaré fellowship, as well as the MERAC foundation. This work was funded by the French National Research Agency (ANR) project EXOWINDS (ANR-23-CE31-0001-01)(VP). BP acknowledges financial support of the Walter Gyllenberg Foundation for computational resources and for funding the research stay that made this project possible. BT acknowledges the financial support from the Wenner-Gren Foundation (WGF2022-0041)(BT). JPW acknowledges support from the Trottier Family Foundation via the Trottier Postdoctoral Fellowship, Swiss National Science Foundation (SNSF) under grant 10002706 (JPW) and from the Canadian Space Agency (CSA) under grant 24JWGO3A-03 (JPW). SP acknowledges support from the Swiss National Science Foundation under grant 51NF40\_205606 (SP) within the framework of the National Centre of Competence in Research PlanetS. 
EKHL is supported by the CSH through the Bernoulli Fellowship.
D.D.B.K. acknowledges support from the National Natural Science Foundation of China (NSFC) under grant 12473064 (DDBK). J.M.D acknowledges support from the research program VIDI New Frontiers in Exoplanetary Climatology with project number 614.001.601 (JMD), which is (partly) financed by the Dutch Research Council (NWO).
\end{acknowledgements}

\textbf{Data Availability.} The MAROON-X data is publicly available from Gemini Observatory with the respective programme number. The ESPRESSO data can be downloaded directly from the ESO data archive by searching for the respective run number given in the methods. The data from the main figures as well as the drag timescales are published at \url{https://doi.org/10.5281/zenodo.19498588} for reproducability.\\

\textbf{Code Availability.} The cross-correlation analysis code \texttt{tayph} which embeds the data preparation, cross-correlation and extraction of the velocity is publicly available on github. molecfit is a standard ESO software embedded in ESOReflex and publicly available.\\

\textbf{Author Contributions.} JVS led the project, including the data acquisition and selection, the data analysis, as well as the modelling work, interpretation, and elaboration of the manuscript. VP has lead the observing proposal for the MAROON-X programme and devised the scientific question. He was instrumental for the overall modelling strategy, as well as contributed to the manuscript writing and general editing. BP led the cross-correlation analysis, and was instrumental in the telluric correction, post-processing, fit of the systemic velocity and in writing the manuscript. TH provided the GCM estimates. NM provided the ionisation tables. VdL benchmarked the contribution function approach and provided additional caveats. BT gave insights into the measurement of the system velocity for fast rotating host stars. KB was instrumental in the initial discussion of magnetic field strength impacts and trends. TG aided the interpretation in relation to the Solar System. RvdB advised on the methods of calculating contribution functions for cross correlation. FD provided data and important feedback on the manuscript. 
HB provided insights into magnetic field theories and MHD GCMs. JLB is the PI of MAROON-X and supervised the data collection and reduction. BB contributed to initial survey design and proposal and provided detailed comments on the manuscript. MB provided key insights in the interpretation and uncertainty assessment. PD benchmarked various GCM assumptions. SG, MH, EL, PS, JW, KBS, JMLBD contributed to the initial survey proposal and provided extensive commentary. DK performed MAROON-X observations and data reduction prior to July 2023. DDBK and TK advised on the theoretical interpretation and limitations of the underlying heat engine model. LP was instrumental in the uncertainty estimate and overall commentary. SP contributed to the target selection, preparation of observing proposals, and planning of observations. ER provided input on the original proposal for the data and comments on the manuscript. AS conducted some of the observations and reduced the data MAROON-X Data (2020-2022). AS provided benchmarking for the data analysis. 

\bibliographystyle{aa}
\bibliography{aanda}

@article{draine_magnetohydrodynamic_1983,
	title = {Magnetohydrodynamic shock waves in molecular clouds},
	volume = {264},
	issn = {0004-637X, 1538-4357},
	url = {http://adsabs.harvard.edu/doi/10.1086/160617},
	doi = {10.1086/160617},
	language = {en},
	urldate = {2026-02-06},
	journal = {The Astrophysical Journal},
	author = {Draine, B. T. and Roberge, W. G. and Dalgarno, A.},
	month = jan,
	year = {1983},
	pages = {485},
}

@article{connerney_new_2018,
	title = {A {New} {Model} of {Jupiter}'s {Magnetic} {Field} {From} {Juno}'s {First} {Nine} {Orbits}},
	volume = {45},
	issn = {0094-8276, 1944-8007},
	url = {https://agupubs.onlinelibrary.wiley.com/doi/10.1002/2018GL077312},
	doi = {10.1002/2018GL077312},
	language = {en},
	number = {6},
	urldate = {2026-02-06},
	journal = {Geophysical Research Letters},
	author = {Connerney, J. E. P. and Kotsiaros, S. and Oliversen, R. J. and Espley, J. R. and Joergensen, J. L. and Joergensen, P. S. and Merayo, J. M. G. and Herceg, M. and Bloxham, J. and Moore, K. M. and Bolton, S. J. and Levin, S. M.},
	month = mar,
	year = {2018},
	pages = {2590--2596},
}

@article{lendl_hot_2020,
	title = {The hot dayside and asymmetric transit of {WASP}-189 b seen by {CHEOPS}},
	volume = {643},
	copyright = {© ESO 2020},
	issn = {0004-6361, 1432-0746},
	url = {https://www.aanda.org/articles/aa/abs/2020/11/aa38677-20/aa38677-20.html},
	doi = {10.1051/0004-6361/202038677},
	language = {en},
	urldate = {2025-12-05},
	journal = {Astronomy \& Astrophysics},
	author = {Lendl, M. and Csizmadia, Sz and Deline, A. and Fossati, L. and Kitzmann, D. and Heng, K. and Hoyer, S. and Salmon, S. and Benz, W. and Broeg, C. and Ehrenreich, D. and Fortier, A. and Queloz, D. and Bonfanti, A. and Brandeker, A. and Cameron, A. Collier and Delrez, L. and Muñoz, A. Garcia and Hooton, M. J. and Maxted, P. F. L. and Morris, B. M. and Grootel, V. Van and Wilson, T. G. and Alibert, Y. and Alonso, R. and Asquier, J. and Bandy, T. and Bárczy, T. and Barrado, D. and Barros, S. C. C. and Baumjohann, W. and Beck, M. and Beck, T. and Bekkelien, A. and Bergomi, M. and Billot, N. and Biondi, F. and Bonfils, X. and Bourrier, V. and Busch, M.-D. and Cabrera, J. and Cessa, V. and Charnoz, S. and Chazelas, B. and Damme, C. Corral Van and Davies, M. B. and Deleuil, M. and Demangeon, O. D. S. and Demory, B.-O. and Erikson, A. and Farinato, J. and Fridlund, M. and Futyan, D. and Gandolfi, D. and Gillon, M. and Guterman, P. and Hasiba, J. and Hernandez, E. and Isaak, K. G. and Kiss, L. and Kuntzer, T. and Etangs, A. Lecavelier des and Lüftinger, T. and Laskar, J. and Lovis, C. and Magrin, D. and Malvasio, L. and Marafatto, L. and Michaelis, H. and Munari, M. and Nascimbeni, V. and Olofsson, G. and Ottacher, H. and Ottensamer, R. and Pagano, I. and Pallé, E. and Peter, G. and Piazza, D. and Piotto, G. and Pollacco, D. and Ratti, F. and Rauer, H. and Ragazzoni, R. and Rando, N. and Ribas, I. and Rieder, M. and Rohlfs, R. and Safa, F. and Santos, N. C. and Scandariato, G. and Ségransan, D. and Simon, A. E. and Singh, V. and Smith, A. M. S. and Sordet, M. and Sousa, S. G. and Steller, M. and Szabó, Gy M. and Thomas, N. and Tschentscher, M. and Udry, S. and Viotto, V. and Walter, I. and Walton, N. A. and Wildi, F. and Wolter, D.},
	month = nov,
	year = {2020},
	pages = {A94},
}

@misc{martinez_kelt-25b_2019,
	title = {{KELT}-25b and {KELT}-26b: {A} {Hot} {Jupiter} and a {Substellar} {Companion} {Transiting} {Young} {A}-stars {Observed} by {TESS}},
	shorttitle = {{KELT}-25b and {KELT}-26b},
	url = {http://arxiv.org/abs/1912.01017},
	doi = {10.48550/arXiv.1912.01017},
	urldate = {2025-12-05},
	publisher = {arXiv},
	author = {Martínez, Romy Rodríguez and Gaudi, B. Scott and Rodriguez, Joseph E. and Zhou, George and Labadie-Bartz, Jonathan and Quinn, Samuel N. and Penev, Kaloyan Minev and Tan, Thiam-Guan and Latham, David W. and Paredes, Leonardo A. and Kielkopf, John and Addison, Brett C. and Wright, Duncan J. and Teske, Johanna K. and Howell, Steve B. and Ciardi, David R. and Ziegler, Carl and Stassun, Keivan G. and Johnson, Marshall C. and Eastman, Jason D. and Siverd, Robert J. and Beatty, Thomas G. and Bouma, Luke G. and Pepper, Joshua and Lund, Michael B. and Villanueva, Steven and Stevens, Daniel J. and Jensen, Eric L. N. and Kilby, Coleman and Cohen, David H. and Bayliss, Daniel and Bieryla, Allyson and Cargile, Phillip A. and Collins, Karen A. and Conti, Dennis M. and Colon, Knicole D. and Curtis, Ivan A. and DePoy, Darren L. and Evans, Phil A. and Feliz, Dax and Gregorio, Joao and Rothenberg, Jason and James, David J. and Penny, Matthew T. and Reed, Phillip A. and Relles, Howard M. and Stephens, Denise C. and Joner, Michael D. and Kuhn, Rudolf B. and Stockdale, Chris and Trueblood, Mark and Trueblood, Patricia and Yao, Xinyu and Zambelli, Roberto and Vanderspek, Roland and Seager, Sara and Winn, Joshua N. and Jenkins, Jon M. and Henry, Todd J. and James, Hodari-Sadiki and Jao, Wei-Chun and Wang, Sharon X. and Butler, R. Paul and Crane, Jeffrey D. and Thompson, Ian B. and Schectman, Stephen and Wittenmyer, Robert A. and Bedding, Timothy R. and Okumura, Jack and Plavchan, Peter and Bowler, Brendan P. and Horner, Jonathan and Kane, Stephen R. and Mengel, Matthew W. and Morton, Timothy D. and Tinney, C. G. and Zhang, Hui and Scott, Nicholas J. and Matson, Rachel A. and Everett, Mark E. and Tokovinin, Andrei and Mann, Andrew W. and Dragomir, Diana and Guenther, Maximilian N. and Ting, Eric B. and Fausnaugh, Michael and Glidden, Ana and Quintana, Elisa V. and Manner, Mark and Marshall, Jennifer L. and McLeod, Kim K. and Khakpash, Somayeh},
	month = dec,
	year = {2019},
	note = {arXiv:1912.01017},
}

@article{west_three_2016,
	title = {Three irradiated and bloated hot {Jupiters}: {WASP}-76b, {WASP}-82b, and {WASP}-90b⋆},
	volume = {585},
	issn = {0004-6361, 1432-0746},
	shorttitle = {Three irradiated and bloated hot {Jupiters}},
	url = {http://www.aanda.org/10.1051/0004-6361/201527276},
	doi = {10.1051/0004-6361/201527276},
	urldate = {2025-12-05},
	journal = {Astronomy \& Astrophysics},
	author = {West, R. G. and Hellier, C. and Almenara, J.-M. and Anderson, D. R. and Barros, S. C. C. and Bouchy, F. and Brown, D. J. A. and Collier Cameron, A. and Deleuil, M. and Delrez, L. and Doyle, A. P. and Faedi, F. and Fumel, A. and Gillon, M. and Gómez Maqueo Chew, Y. and Hébrard, G. and Jehin, E. and Lendl, M. and Maxted, P. F. L. and Pepe, F. and Pollacco, D. and Queloz, D. and Ségransan, D. and Smalley, B. and Smith, A. M. S. and Southworth, J. and Triaud, A. H. M. J. and Udry, S.},
	month = jan,
	year = {2016},
	pages = {A126},
}

@article{eastman_kelt-4ab_2016,
	title = {{KELT}-{4Ab}: {AN} {INFLATED} {HOT} {JUPITER} {TRANSITING} {THE} {BRIGHT} ( \textit{{V}} ∼ 10) {COMPONENT} {OF} {A} {HIERARCHICAL} {TRIPLE}},
	volume = {151},
	issn = {1538-3881},
	shorttitle = {{KELT}-{4Ab}},
	url = {https://iopscience.iop.org/article/10.3847/0004-6256/151/2/45},
	doi = {10.3847/0004-6256/151/2/45},
	number = {2},
	urldate = {2025-12-05},
	journal = {The Astronomical Journal},
	author = {Eastman, Jason D. and Beatty, Thomas G. and Siverd, Robert J. and Antognini, Joseph M. O. and Penny, Matthew T. and Gonzales, Erica J. and Crepp, Justin R. and Howard, Andrew W. and Avril, Ryan L. and Bieryla, Allyson and Collins, Karen and Fulton, Benjamin J. and Ge, Jian and Gregorio, Joao and Ma, Bo and Mellon, Samuel N. and Oberst, Thomas E. and Wang, Ji and Gaudi, B. Scott and Pepper, Joshua and Stassun, Keivan G. and Buchhave, Lars A. and Jensen, Eric L. N. and Latham, David W. and Berlind, Perry and Calkins, Michael L. and Cargile, Phillip A. and Colón, Knicole D. and Dhital, Saurav and Esquerdo, Gilbert A. and Johnson, John Asher and Kielkopf, John F. and Manner, Mark and Mao, Qingqing and McLeod, Kim K. and Penev, Kaloyan and Stefanik, Robert P. and Street, Rachel and Zambelli, Roberto and DePoy, D. L. and Gould, Andrew and Marshall, Jennifer L. and Pogge, Richard W. and Trueblood, Mark and Trueblood, Patricia},
	month = feb,
	year = {2016},
	pages = {45},
}

@article{borsa_gaps_2019,
	title = {The {GAPS} {Programme} with {HARPS}-{N} at {TNG} - {XIX}. {Atmospheric} {Rossiter}-{McLaughlin} effect and improved parameters of {KELT}-9b},
	volume = {631},
	copyright = {© ESO 2019},
	issn = {0004-6361, 1432-0746},
	url = {https://www.aanda.org/articles/aa/abs/2019/11/aa35718-19/aa35718-19.html},
	doi = {10.1051/0004-6361/201935718},
	language = {en},
	urldate = {2025-11-27},
	journal = {Astronomy \& Astrophysics},
	author = {Borsa, F. and Rainer, M. and Bonomo, A. S. and Barbato, D. and Fossati, L. and Malavolta, L. and Nascimbeni, V. and Lanza, A. F. and Esposito, M. and Affer, L. and Andreuzzi, G. and Benatti, S. and Biazzo, K. and Bignamini, A. and Brogi, M. and Carleo, I. and Claudi, R. and Cosentino, R. and Covino, E. and Damasso, M. and Desidera, S. and Rubio, A. Garrido and Giacobbe, P. and González-Álvarez, E. and Harutyunyan, A. and Knapic, C. and Leto, G. and Ligi, R. and Maggio, A. and Maldonado, J. and Mancini, L. and Fiorenzano, A. F. M. and Masiero, S. and Micela, G. and Molinari, E. and Pagano, I. and Pedani, M. and Piotto, G. and Pino, L. and Poretti, E. and Scandariato, G. and Smareglia, R. and Sozzetti, A.},
	month = nov,
	year = {2019},
	pages = {A34},
}

@article{basinger_pepsi_2025,
	title = {{PEPSI} investigation, retrieval, and atlas of numerous giant atmospheres ({PIRANGA}) – {III}. {Composition} and winds in the atmosphere of {TOI}-1518 b},
	volume = {543},
	copyright = {https://creativecommons.org/licenses/by/4.0/},
	issn = {0035-8711, 1365-2966},
	url = {https://academic.oup.com/mnras/article/543/4/4136/8266785},
	doi = {10.1093/mnras/staf1648},
	language = {en},
	number = {4},
	urldate = {2025-11-27},
	journal = {Monthly Notices of the Royal Astronomical Society},
	author = {Basinger, Connor and Johnson, Marshall C and Wang, Ji and Duck, Alison and Pai Asnodkar, Anusha and Petz, Sydney and Lenhart, Calder and Ilyin, Ilya and Strassmeier, Klaus},
	month = oct,
	year = {2025},
	pages = {4136--4143},
}

@misc{lenhart_pepsi_2025,
	title = {{PEPSI} {Investigation}, {Retrieval}, and {Atlas} of {Numerous} {Giant} {Atmospheres} ({PIRANGA}). {II}. {Phase}-{Resolved} {Cross}-{Correlation} {Transmission} {Spectroscopy} of {KELT}-20b},
	copyright = {Creative Commons Attribution 4.0 International},
	url = {https://arxiv.org/abs/2503.07719},
	doi = {10.48550/ARXIV.2503.07719},
	urldate = {2025-11-27},
	publisher = {arXiv},
	author = {Lenhart, Calder and Johnson, Marshall C. and Wang, Ji and Asnodkar, Anusha Pai and Petz, Sydney and Duck, Alison and Strassmeier, Klaus G. and Ilyin, Ilya},
	year = {2025},
}

@article{zaghoo_size_2018,
	title = {Size and {Strength} of {Self}-excited {Dynamos} in {Jupiter}-like {Extrasolar} {Planets}},
	volume = {862},
	issn = {0004-637X, 1538-4357},
	url = {https://iopscience.iop.org/article/10.3847/1538-4357/aac6e8},
	doi = {10.3847/1538-4357/aac6e8},
	number = {1},
	urldate = {2025-11-27},
	journal = {The Astrophysical Journal},
	author = {Zaghoo, Mohamed and Collins, G. W.},
	month = jul,
	year = {2018},
	pages = {19},
}

@article{dietrich_magnetic_2022,
	title = {Magnetic induction processes in hot {Jupiters}, application to {KELT}-9b},
	volume = {517},
	copyright = {https://creativecommons.org/licenses/by/4.0/},
	issn = {0035-8711, 1365-2966},
	url = {https://academic.oup.com/mnras/article/517/3/3113/6753224},
	doi = {10.1093/mnras/stac2849},
	language = {en},
	number = {3},
	urldate = {2025-11-27},
	journal = {Monthly Notices of the Royal Astronomical Society},
	author = {Dietrich, Wieland and Kumar, Sandeep and Poser, Anna Julia and French, Martin and Nettelmann, Nadine and Redmer, Ronald and Wicht, Johannes},
	month = oct,
	year = {2022},
	pages = {3113--3125},
}

@article{vigano_inflated_2025,
	title = {Inflated hot {Jupiters}: {Inferring} average atmospheric velocity via {Ohmic} models coupled with internal dynamo evolution},
	volume = {701},
	copyright = {https://creativecommons.org/licenses/by/4.0},
	issn = {0004-6361, 1432-0746},
	shorttitle = {Inflated hot {Jupiters}},
	url = {https://www.aanda.org/10.1051/0004-6361/202555219},
	doi = {10.1051/0004-6361/202555219},
	urldate = {2025-11-27},
	journal = {Astronomy \& Astrophysics},
	author = {Viganò, Daniele and Sengupta, Soumya and Soriano-Guerrero, Clàudia and Perna, Rosalba and Elias-López, Albert and Kumar, Sandeep and Akgün, Taner},
	month = sep,
	year = {2025},
	pages = {A8},
}

@article{elias-lopez_rossby_2025,
	title = {Rossby {Number} {Regime}, {Convection} {Suppression}, and {Dynamo}-generated {Magnetism} in {Inflated} {Hot} {Jupiters}},
	volume = {990},
	issn = {0004-637X, 1538-4357},
	url = {https://iopscience.iop.org/article/10.3847/1538-4357/adf057},
	doi = {10.3847/1538-4357/adf057},
	number = {1},
	urldate = {2025-11-27},
	journal = {The Astrophysical Journal},
	author = {Elias-López, Albert and Cantiello, Matteo and Viganò, Daniele and Del Sordo, Fabio and Kaur, Simranpreet and Soriano-Guerrero, Clàudia},
	month = sep,
	year = {2025},
	pages = {38},
}

@article{kilmetis_magnetic_2024,
	title = {Magnetic field evolution of hot exoplanets},
	volume = {535},
	copyright = {https://creativecommons.org/licenses/by/4.0/},
	issn = {0035-8711, 1365-2966},
	url = {https://academic.oup.com/mnras/article/535/4/3646/7881581},
	doi = {10.1093/mnras/stae2505},
	language = {en},
	number = {4},
	urldate = {2025-11-27},
	journal = {Monthly Notices of the Royal Astronomical Society},
	author = {Kilmetis, K and Vidotto, A A and Allan, A and Kubyshkina, D},
	month = nov,
	year = {2024},
	pages = {3646--3655},
}

@article{yadav_estimating_2017,
	title = {Estimating the {Magnetic} {Field} {Strength} in {Hot} {Jupiters}},
	volume = {849},
	issn = {2041-8205, 2041-8213},
	url = {https://iopscience.iop.org/article/10.3847/2041-8213/aa93fd},
	doi = {10.3847/2041-8213/aa93fd},
	number = {1},
	urldate = {2025-11-27},
	journal = {The Astrophysical Journal Letters},
	author = {Yadav, Rakesh K. and Thorngren, Daniel P.},
	month = nov,
	year = {2017},
	pages = {L12},
}

@article{visscher_atmospheric_2010,
	title = {{ATMOSPHERIC} {CHEMISTRY} {IN} {GIANT} {PLANETS}, {BROWN} {DWARFS}, {AND} {LOW}-{MASS} {DWARF} {STARS}. {III}. {IRON}, {MAGNESIUM}, {AND} {SILICON}},
	volume = {716},
	issn = {0004-637X, 1538-4357},
	url = {https://iopscience.iop.org/article/10.1088/0004-637X/716/2/1060},
	doi = {10.1088/0004-637X/716/2/1060},
	number = {2},
	urldate = {2025-11-26},
	journal = {The Astrophysical Journal},
	author = {Visscher, Channon and Lodders, Katharina and Fegley, Bruce},
	month = jun,
	year = {2010},
	pages = {1060--1075},
}

@article{koll_temperature_2016,
	title = {{TEMPERATURE} {STRUCTURE} {AND} {ATMOSPHERIC} {CIRCULATION} {OF} {DRY} {TIDALLY} {LOCKED} {ROCKY} {EXOPLANETS}},
	volume = {825},
	issn = {0004-637X, 1538-4357},
	url = {https://iopscience.iop.org/article/10.3847/0004-637X/825/2/99},
	doi = {10.3847/0004-637X/825/2/99},
	number = {2},
	urldate = {2025-09-16},
	journal = {The Astrophysical Journal},
	author = {Koll, Daniel D. B. and Abbot, Dorian S.},
	month = jul,
	year = {2016},
	pages = {99},
}

@article{kaspi_jupiters_2018,
	title = {Jupiter’s atmospheric jet streams extend thousands of kilometres deep},
	volume = {555},
	issn = {0028-0836, 1476-4687},
	url = {https://www.nature.com/articles/nature25793},
	doi = {10.1038/nature25793},
	language = {en},
	number = {7695},
	urldate = {2025-09-14},
	journal = {Nature},
	author = {Kaspi, Y. and Galanti, E. and Hubbard, W. B. and Stevenson, D. J. and Bolton, S. J. and Iess, L. and Guillot, T. and Bloxham, J. and Connerney, J. E. P. and Cao, H. and Durante, D. and Folkner, W. M. and Helled, R. and Ingersoll, A. P. and Levin, S. M. and Lunine, J. I. and Miguel, Y. and Militzer, B. and Parisi, M. and Wahl, S. M.},
	month = mar,
	year = {2018},
	pages = {223--226},
}

@article{liu_mechanisms_2010,
	title = {Mechanisms of {Jet} {Formation} on the {Giant} {Planets}},
	volume = {67},
	issn = {1520-0469, 0022-4928},
	url = {https://journals.ametsoc.org/doi/10.1175/2010JAS3492.1},
	doi = {10.1175/2010JAS3492.1},
	language = {en},
	number = {11},
	urldate = {2025-09-14},
	journal = {Journal of the Atmospheric Sciences},
	author = {Liu, Junjun and Schneider, Tapio},
	month = nov,
	year = {2010},
	pages = {3652--3672},
}

@article{kesseli_up_2024,
	title = {Up, {Up}, and {Away}: {Winds} and {Dynamical} {Structure} as a {Function} of {Altitude} in the {Ultrahot} {Jupiter} {WASP}-76b},
	volume = {975},
	issn = {0004-637X},
	shorttitle = {Up, {Up}, and {Away}},
	url = {https://ui.adsabs.harvard.edu/abs/2024ApJ...975....9K/abstract},
	doi = {10.3847/1538-4357/ad772f},
	language = {en},
	number = {1},
	urldate = {2025-09-04},
	journal = {The Astrophysical Journal},
	author = {Kesseli, Aurora Y. and Beltz, Hayley and Rauscher, Emily and Snellen, I. a. G.},
	month = nov,
	year = {2024},
	pages = {9},
}

@article{showman_doppler_2013,
	title = {Doppler {Signatures} of the {Atmospheric} {Circulation} on {Hot} {Jupiters}},
	volume = {762},
	issn = {0004-637X},
	url = {https://ui.adsabs.harvard.edu/abs/2013ApJ...762...24S/abstract},
	doi = {10.1088/0004-637X/762/1/24},
	language = {en},
	number = {1},
	urldate = {2025-08-31},
	journal = {The Astrophysical Journal},
	author = {Showman, Adam P. and Fortney, Jonathan J. and Lewis, Nikole K. and Shabram, Megan},
	month = jan,
	year = {2013},
	pages = {24},
}

@article{watkins_gravity_2010,
	title = {Gravity {Waves} on {Hot} {Extrasolar} {Planets}. {I}. {Propagation} and {Interaction} with the {Background}},
	volume = {714},
	issn = {0004-637X},
	url = {https://ui.adsabs.harvard.edu/abs/2010ApJ...714..904W/abstract},
	doi = {10.1088/0004-637X/714/1/904},
	language = {en},
	number = {1},
	urldate = {2025-08-31},
	journal = {The Astrophysical Journal, Volume 714, Issue 1, pp. 904-914 (2010).},
	author = {Watkins, Chris and Cho, J. Y.-K.},
	month = may,
	year = {2010},
	pages = {904},
}

@article{collet_new_2024,
	title = {A {New} {Type} of {Jovian} {Hectometric} {Radiation} {Powered} by {Monoenergetic} {Electron} {Beams}},
	volume = {129},
	issn = {0148-0227},
	url = {https://ui.adsabs.harvard.edu/abs/2024JGRA..12932422C/abstract},
	doi = {10.1029/2024JA032422},
	language = {en},
	number = {5},
	urldate = {2025-08-28},
	journal = {Journal of Geophysical Research: Space Physics, Volume 129, Issue 5, article id. e2024JA032422},
	author = {Collet, B. and Lamy, L. and Louis, C. K. and Zarka, P. and Prangé, R. and Louarn, P. and Kurth, W. S. and Allegrini, F.},
	month = may,
	year = {2024},
	pages = {e2024JA032422},
}

@article{thorngren_bayesian_2018,
	title = {Bayesian {Analysis} of {Hot}-{Jupiter} {Radius} {Anomalies}: {Evidence} for {Ohmic} {Dissipation}?},
	volume = {155},
	issn = {0004-6256},
	shorttitle = {Bayesian {Analysis} of {Hot}-{Jupiter} {Radius} {Anomalies}},
	url = {https://ui.adsabs.harvard.edu/abs/2018AJ....155..214T/abstract},
	doi = {10.3847/1538-3881/aaba13},
	language = {en},
	number = {5},
	urldate = {2025-08-09},
	journal = {The Astronomical Journal, Volume 155, Issue 5, article id. 214, {\textless}NUMPAGES{\textgreater}10{\textless}/NUMPAGES{\textgreater} pp. (2018).},
	author = {Thorngren, Daniel P. and Fortney, Jonathan J.},
	month = may,
	year = {2018},
	pages = {214},
}

@article{stangret_high-resolution_2022,
	title = {High-resolution transmission spectroscopy study of ultra-hot {Jupiters} {HAT}-{P}-57b, {KELT}-17b, {KELT}-21b, {KELT}-7b, {MASCARA}-1b, and {WASP}-189b},
	volume = {662},
	issn = {0004-6361},
	url = {https://ui.adsabs.harvard.edu/abs/2022A&A...662A.101S},
	doi = {10.1051/0004-6361/202141799},
	urldate = {2024-10-09},
	journal = {Astronomy and Astrophysics},
	author = {Stangret, M. and Casasayas-Barris, N. and Pallé, E. and Orell-Miquel, J. and Morello, G. and Luque, R. and Nowak, G. and Yan, F.},
	month = jun,
	year = {2022},
	note = {ADS Bibcode: 2022A\&A...662A.101S},
	pages = {A101},
}

@misc{smith_roasting_2024,
	title = {The {Roasting} {Marshmallows} {Program} with {IGRINS} on {Gemini} {South} {II} -- {WASP}-121 b has super-stellar {C}/{O} and refractory-to-volatile ratios},
	url = {https://ui.adsabs.harvard.edu/abs/2024arXiv241019017S},
	doi = {10.48550/arXiv.2410.19017},
	urldate = {2024-11-23},
	author = {Smith, Peter C. B. and Sanchez, Jorge A. and Line, Michael R. and Rauscher, Emily and Weiner Mansfield, Megan and Kempton, Eliza M. -R. and Savel, Arjun and Wardenier, Joost P. and Pino, Lorenzo and Bean, Jacob L. and Beltz, Hayley and Panwar, Vatsal and Brogi, Matteo and Malsky, Isaac and Fortney, Jonathan and Desert, Jean-Michel and Pelletier, Stefan and Parmentier, Vivien and Kanumalla, Krishna and Welbanks, Luis and Meyer, Michael and Monnier, John},
	month = oct,
	year = {2024},
	note = {Publication Title: arXiv e-prints
ADS Bibcode: 2024arXiv241019017S},
}

@misc{simonnin_time_2025,
	title = {Time {Resolved} {Absorption} of {Six} {Chemical} {Species} {With} {MAROON}-{X} {Points} to {Strong} {Drag} in the {Ultra} {Hot} {Jupiter} {TOI}-1518 b},
	url = {http://arxiv.org/abs/2412.01472},
	doi = {10.48550/arXiv.2412.01472},
	urldate = {2025-06-06},
	publisher = {arXiv},
	author = {Simonnin, A. and Parmentier, V. and Wardenier, J. P. and Chauvin, G. and Chiavassa, A. and N'Diaye, M. and Tan, X. and Heidari, N. and Bean, B. Prinoth J. and Hébrard, G. and Line, M. and Kitzmann, D. and Kasper, D. and Pelletier, S. and Seidel, J. V. and Seifhart, A. and Benneke, B. and Bonfils, X. and Brogi, M. and Désert, J.-M. and Gandhi, S. and Hammond, M. and Lee, E. K. H. and Moutou, C. and Palma-Bifani, P. and Pino, L. and Rauscher, E. and Mansfield, M. Weiner and Bell, J. Serrano and Smith, P.},
	month = apr,
	year = {2025},
	note = {arXiv:2412.01472 [astro-ph]},
}

@article{roth_hot_2024,
	title = {Hot {Jupiter} diversity and the onset of {TiO}/{VO} revealed by a large grid of non-grey global circulation models},
	volume = {531},
	issn = {0035-8711},
	url = {https://ui.adsabs.harvard.edu/abs/2024MNRAS.531.1056R/abstract},
	doi = {10.1093/mnras/stae984},
	language = {en},
	number = {1},
	urldate = {2025-08-05},
	journal = {Monthly Notices of the Royal Astronomical Society, Volume 531, Issue 1, pp.1056-1083},
	author = {Roth, Alexander and Parmentier, Vivien and Hammond, Mark},
	month = jun,
	year = {2024},
	pages = {1056},
}

@article{llama_shocking_2011,
	title = {The shocking transit of {WASP}-12b: modelling the observed early ingress in the near-ultraviolet},
	volume = {416},
	issn = {0035-8711},
	shorttitle = {The shocking transit of {WASP}-12b},
	url = {https://ui.adsabs.harvard.edu/abs/2011MNRAS.416L..41L/abstract},
	doi = {10.1111/j.1745-3933.2011.01093.x},
	language = {en},
	number = {1},
	urldate = {2025-08-09},
	journal = {Monthly Notices of the Royal Astronomical Society: Letters, Volume 416, Issue 1, pp. L41-L44.},
	author = {Llama, J. and Wood, K. and Jardine, M. and Vidotto, A. A. and Helling, Ch and Fossati, L. and Haswell, C. A.},
	month = sep,
	year = {2011},
	pages = {L41},
}

@article{koll_atmospheric_2018,
	title = {Atmospheric {Circulations} of {Hot} {Jupiters} as {Planetary} {Heat} {Engines}},
	volume = {853},
	issn = {0004-637X},
	url = {https://ui.adsabs.harvard.edu/abs/2018ApJ...853..133K},
	doi = {10.3847/1538-4357/aaa3de},
	urldate = {2025-02-26},
	journal = {The Astrophysical Journal},
	publisher = {IOP},
	author = {Koll, Daniel D. B. and Komacek, Thaddeus D.},
	month = feb,
	year = {2018},
	note = {ADS Bibcode: 2018ApJ...853..133K},
	pages = {133},
}

@article{hellier_wasp-south_2019,
	title = {{WASP}-{South} hot {Jupiters}: {WASP}-178b, {WASP}-184b, {WASP}-185b, and {WASP}-192b},
	volume = {490},
	issn = {0035-8711},
	shorttitle = {{WASP}-{South} hot {Jupiters}},
	url = {https://ui.adsabs.harvard.edu/abs/2019MNRAS.490.1479H/abstract},
	doi = {10.1093/mnras/stz2713},
	language = {en},
	number = {1},
	urldate = {2025-07-25},
	journal = {Monthly Notices of the Royal Astronomical Society},
	author = {Hellier, Coel and Anderson, D. R. and Barkaoui, K. and Benkhaldoun, Z. and Bouchy, F. and Burdanov, A. and Collier Cameron, A. and Delrez, L. and Gillon, M. and Jehin, E. and Nielsen, L. D. and Maxted, P. F. L. and Pepe, F. and Pollacco, D. and Pozuelos, F. J. and Queloz, D. and Ségransan, D. and Smalley, B. and Triaud, A. H. M. J. and Turner, O. D. and Udry, S. and West, R. G.},
	month = nov,
	year = {2019},
	pages = {1479--1487},
}

@article{balbus_solar_2000,
	title = {Solar {Nebula} {Magnetohydrodynamics}},
	volume = {92},
	issn = {0038-6308},
	url = {https://ui.adsabs.harvard.edu/abs/2000SSRv...92...39B/abstract},
	doi = {10.1023/A:1005293132737},
	language = {en},
	urldate = {2025-07-28},
	journal = {Space Science Reviews, v. 92, Issue 1/2, p. 39-54 (2000).},
	author = {Balbus, Steven A. and Hawley, John F.},
	month = apr,
	year = {2000},
	pages = {39},
}

@article{osborn_investigating_2020,
	title = {Investigating the {Planet}-{Metallicity} {Correlation} for {Hot} {Jupiters}},
	volume = {491},
	issn = {0035-8711, 1365-2966},
	url = {http://arxiv.org/abs/1911.05830},
	doi = {10.1093/mnras/stz3207},
	number = {3},
	urldate = {2025-08-21},
	journal = {Monthly Notices of the Royal Astronomical Society},
	author = {Osborn, Ares and Bayliss, Daniel},
	month = jan,
	year = {2020},
	note = {arXiv:1911.05830 [astro-ph]},
	pages = {4481--4487},
}

@article{menou_magnetic_2012,
	title = {Magnetic {Scaling} {Laws} for the {Atmospheres} of {Hot} {Giant} {Exoplanets}},
	volume = {745},
	issn = {0004-637X},
	url = {https://ui.adsabs.harvard.edu/abs/2012ApJ...745..138M/abstract},
	doi = {10.1088/0004-637X/745/2/138},
	language = {en},
	number = {2},
	urldate = {2025-08-13},
	journal = {The Astrophysical Journal, Volume 745, Issue 2, article id. 138, {\textless}NUMPAGES{\textgreater}8{\textless}/NUMPAGES{\textgreater} pp. (2012).},
	author = {Menou, Kristen},
	month = feb,
	year = {2012},
	pages = {138},
}

@article{rogers_magnetohydrodynamic_2014,
	title = {Magnetohydrodynamic {Simulations} of the {Atmosphere} of {HD} 209458b},
	volume = {782},
	issn = {0004-637X},
	url = {https://ui.adsabs.harvard.edu/abs/2014ApJ...782L...4R/abstract},
	doi = {10.1088/2041-8205/782/1/L4},
	language = {en},
	number = {1},
	urldate = {2025-08-10},
	journal = {The Astrophysical Journal Letters, Volume 782, Issue 1, article id. L4, {\textless}NUMPAGES{\textgreater}6{\textless}/NUMPAGES{\textgreater} pp. (2014).},
	author = {Rogers, T. M. and Showman, A. P.},
	month = feb,
	year = {2014},
	pages = {L4},
}

@article{marley_thermal_1999,
	title = {Thermal {Structure} of {Uranus}' {Atmosphere}},
	volume = {138},
	issn = {0019-1035},
	url = {https://ui.adsabs.harvard.edu/abs/1999Icar..138..268M/abstract},
	doi = {10.1006/icar.1998.6071},
	language = {en},
	number = {2},
	urldate = {2025-08-10},
	journal = {Icarus, Volume 138, Issue 2, pp. 268-286.},
	author = {Marley, Mark S. and McKay, Christopher P.},
	month = apr,
	year = {1999},
	pages = {268},
}

@article{showman_atmospheric_2009,
	title = {Atmospheric {Circulation} of {Hot} {Jupiters}: {Coupled} {Radiative}-{Dynamical} {General} {Circulation} {Model} {Simulations} of {HD} 189733b and {HD} 209458b},
	volume = {699},
	issn = {0004-637X},
	shorttitle = {Atmospheric {Circulation} of {Hot} {Jupiters}},
	url = {https://ui.adsabs.harvard.edu/abs/2009ApJ...699..564S/abstract},
	doi = {10.1088/0004-637X/699/1/564},
	language = {en},
	number = {1},
	urldate = {2025-08-10},
	journal = {The Astrophysical Journal, Volume 699, Issue 1, pp. 564-584 (2009).},
	author = {Showman, Adam P. and Fortney, Jonathan J. and Lian, Yuan and Marley, Mark S. and Freedman, Richard S. and Knutson, Heather A. and Charbonneau, David},
	month = jul,
	year = {2009},
	pages = {564},
}

@article{adcroft_implementation_2004,
	title = {Implementation of an {Atmosphere} {Ocean} {General} {Circulation} {Model} on the {Expanded} {Spherical} {Cube}},
	volume = {132},
	issn = {0027-0644},
	url = {https://ui.adsabs.harvard.edu/abs/2004MWRv..132.2845A/abstract},
	doi = {10.1175/MWR2823.1},
	language = {en},
	number = {12},
	urldate = {2025-08-10},
	journal = {Monthly Weather Review, vol. 132, issue 12, p. 2845},
	author = {Adcroft, Alistair and Campin, Jean-Michel and Hill, Chris and Marshall, John},
	year = {2004},
	pages = {2845},
}

@article{komacek_atmospheric_2017,
	title = {Atmospheric {Circulation} of {Hot} {Jupiters}: {Dayside}-{Nightside} {Temperature} {Differences}. {II}. {Comparison} with {Observations}},
	volume = {835},
	issn = {0004-637X},
	shorttitle = {Atmospheric {Circulation} of {Hot} {Jupiters}},
	url = {https://ui.adsabs.harvard.edu/abs/2017ApJ...835..198K/abstract},
	doi = {10.3847/1538-4357/835/2/198},
	language = {en},
	number = {2},
	urldate = {2025-08-10},
	journal = {The Astrophysical Journal, Volume 835, Issue 2, article id. 198, {\textless}NUMPAGES{\textgreater}14{\textless}/NUMPAGES{\textgreater} pp. (2017).},
	author = {Komacek, Thaddeus D. and Showman, Adam P. and Tan, Xianyu},
	month = feb,
	year = {2017},
	pages = {198},
}

@article{emanuel_air-sea_1986,
	title = {An {Air}-{Sea} {Interaction} {Theory} for {Tropical} {Cyclones}. {Part} {I}: {Steady}-{State} {Maintenance}.},
	volume = {43},
	issn = {0022-4928},
	shorttitle = {An {Air}-{Sea} {Interaction} {Theory} for {Tropical} {Cyclones}. {Part} {I}},
	url = {https://ui.adsabs.harvard.edu/abs/1986JAtS...43..585E/abstract},
	doi = {10.1175/1520-0469(1986)043<0585:AASITF>2.0.CO;2},
	language = {en},
	number = {6},
	urldate = {2025-08-10},
	journal = {Journal of Atmospheric Sciences, vol. 43, Issue 6, pp.585-605},
	author = {Emanuel, Kerry A.},
	month = mar,
	year = {1986},
	pages = {585},
}

@article{zhou_hats-70b_2019,
	title = {{HATS}-70b: {A} 13 {MJ} {Brown} {Dwarf} {Transiting} an {A} {Star}},
	volume = {157},
	issn = {0004-6256},
	shorttitle = {{HATS}-70b},
	url = {https://ui.adsabs.harvard.edu/abs/2019AJ....157...31Z/abstract},
	doi = {10.3847/1538-3881/aaf1bb},
	language = {en},
	number = {1},
	urldate = {2025-08-10},
	journal = {The Astronomical Journal, Volume 157, Issue 1, article id. 31, {\textless}NUMPAGES{\textgreater}14{\textless}/NUMPAGES{\textgreater} pp. (2019).},
	author = {Zhou, G. and Bakos, G. A. and Bayliss, D. and Bento, J. and Bhatti, W. and Brahm, R. and Csubry, Z. and Espinoza, N. and Hartman, J. D. and Henning, T. and Jordán, A. and Mancini, L. and Penev, K. and Rabus, M. and Sarkis, P. and Suc, V. and de Val-Borro, M. and Rodriguez, J. E. and Osip, D. and Kedziora-Chudczer, L. and Bailey, J. and Tinney, C. G. and Durkan, S. and Lázár, J. and Papp, I. and Sári, P.},
	month = jan,
	year = {2019},
	pages = {31},
}

@article{saha_precise_2023,
	title = {Precise {Transit} {Photometry} {Using} {TESS}: {Updated} {Physical} {Properties} for 28 {Exoplanets} around {Bright} {Stars}},
	volume = {268},
	issn = {0067-0049},
	shorttitle = {Precise {Transit} {Photometry} {Using} {TESS}},
	url = {https://ui.adsabs.harvard.edu/abs/2023ApJS..268....2S/abstract},
	doi = {10.3847/1538-4365/acdb6b},
	language = {en},
	number = {1},
	urldate = {2025-08-10},
	journal = {The Astrophysical Journal Supplement Series, Volume 268, Issue 1, id.2, {\textless}NUMPAGES{\textgreater}17{\textless}/NUMPAGES{\textgreater} pp.},
	author = {Saha, Suman},
	month = sep,
	year = {2023},
	pages = {2},
}

@article{patel_empirical_2022,
	title = {Empirical {Limb}-darkening {Coefficients} and {Transit} {Parameters} of {Known} {Exoplanets} from {TESS}},
	volume = {163},
	issn = {0004-6256},
	url = {https://ui.adsabs.harvard.edu/abs/2022AJ....163..228P/abstract},
	doi = {10.3847/1538-3881/ac5f55},
	language = {en},
	number = {5},
	urldate = {2025-08-10},
	journal = {The Astronomical Journal, Volume 163, Issue 5, id.228, {\textless}NUMPAGES{\textgreater}24{\textless}/NUMPAGES{\textgreater} pp.},
	author = {Patel, Jayshil A. and Espinoza, Néstor},
	month = may,
	year = {2022},
	pages = {228},
}

@article{snellen_orbital_2010,
	title = {The orbital motion, absolute mass and high-altitude winds of exoplanet {HD209458b}},
	volume = {465},
	issn = {0028-0836},
	url = {https://ui.adsabs.harvard.edu/abs/2010Natur.465.1049S/abstract},
	doi = {10.1038/nature09111},
	language = {en},
	number = {7301},
	urldate = {2025-08-10},
	journal = {Nature, Volume 465, Issue 7301, pp. 1049-1051 (2010).},
	author = {Snellen, Ignas A. G. and de Kok, Remco J. and de Mooij, Ernst J. W. and Albrecht, Simon},
	month = jun,
	year = {2010},
	pages = {1049},
}

@article{seifahrt_-sky_2020,
	title = {On-sky commissioning of {MAROON}-{X}: a new precision radial velocity spectrograph for {Gemini} {North}},
	volume = {11447},
	shorttitle = {On-sky commissioning of {MAROON}-{X}},
	url = {https://ui.adsabs.harvard.edu/abs/2020SPIE11447E..1FS/abstract},
	doi = {10.1117/12.2561564},
	language = {en},
	urldate = {2025-08-10},
	journal = {Proceedings of the SPIE, Volume 11447, id. 114471F {\textless}NUMPAGES{\textgreater}21{\textless}/NUMPAGES{\textgreater} pp. (2020).},
	author = {Seifahrt, Andreas and Bean, Jacob L. and Stürmer, Julian and Kasper, David and Gers, Luke and Schwab, Christian and Zechmeister, Mathias and Stefánsson, Gudmundur and Montet, Ben and Dos Santos, Leonardo A. and Peck, Alison and White, John and Tapia, Eduardo},
	month = dec,
	year = {2020},
	pages = {114471F},
}

@article{tan_atmospheric_2019,
	title = {The {Atmospheric} {Circulation} of {Ultra}-hot {Jupiters}},
	volume = {886},
	issn = {0004-637X},
	url = {https://ui.adsabs.harvard.edu/abs/2019ApJ...886...26T/abstract},
	doi = {10.3847/1538-4357/ab4a76},
	language = {en},
	number = {1},
	urldate = {2025-08-09},
	journal = {The Astrophysical Journal, Volume 886, Issue 1, article id. 26, {\textless}NUMPAGES{\textgreater}20{\textless}/NUMPAGES{\textgreater} pp. (2019).},
	author = {Tan, Xianyu and Komacek, Thaddeus D.},
	month = nov,
	year = {2019},
	pages = {26},
}

@article{bell_increased_2018,
	title = {Increased {Heat} {Transport} in {Ultra}-hot {Jupiter} {Atmospheres} through {H}{\textless}{SUB}{\textgreater}2{\textless}/{SUB}{\textgreater} {Dissociation} and {Recombination}},
	volume = {857},
	issn = {0004-637X},
	url = {https://ui.adsabs.harvard.edu/abs/2018ApJ...857L..20B/abstract},
	doi = {10.3847/2041-8213/aabcc8},
	language = {en},
	number = {2},
	urldate = {2025-08-09},
	journal = {The Astrophysical Journal Letters, Volume 857, Issue 2, article id. L20, {\textless}NUMPAGES{\textgreater}6{\textless}/NUMPAGES{\textgreater} pp. (2018).},
	author = {Bell, Taylor J. and Cowan, Nicolas B.},
	month = apr,
	year = {2018},
	pages = {L20},
}

@article{saffe_kelt-17_2020,
	title = {{KELT}-17: a chemically peculiar {Am} star and a hot-{Jupiter} planet},
	volume = {641},
	issn = {0004-6361},
	shorttitle = {{KELT}-17},
	url = {https://ui.adsabs.harvard.edu/abs/2020A%26A...641A.145S/abstract},
	doi = {10.1051/0004-6361/202038843},
	language = {en},
	urldate = {2025-08-09},
	journal = {Astronomy \&amp; Astrophysics, Volume 641, id.A145, {\textless}NUMPAGES{\textgreater}5{\textless}/NUMPAGES{\textgreater} pp.},
	author = {Saffe, C. and Miquelarena, P. and Alacoria, J. and González, J. F. and Flores, M. and Jaque Arancibia, M. and Calvo, D. and Jofré, E. and Collado, A.},
	month = sep,
	year = {2020},
	pages = {A145},
}

@article{heng_atmospheric_2011,
	title = {Atmospheric circulation of tidally locked exoplanets: a suite of benchmark tests for dynamical solvers},
	volume = {413},
	issn = {0035-8711},
	shorttitle = {Atmospheric circulation of tidally locked exoplanets},
	url = {https://ui.adsabs.harvard.edu/abs/2011MNRAS.413.2380H/abstract},
	doi = {10.1111/j.1365-2966.2011.18315.x},
	language = {en},
	number = {4},
	urldate = {2025-08-09},
	journal = {Monthly Notices of the Royal Astronomical Society, Volume 413, Issue 4, pp. 2380-2402.},
	author = {Heng, Kevin and Menou, Kristen and Phillipps, Peter J.},
	month = jun,
	year = {2011},
	pages = {2380},
}

@article{rogers_magnetic_2014,
	title = {Magnetic {Effects} in {Hot} {Jupiter} {Atmospheres}},
	volume = {794},
	issn = {0004-637X},
	url = {https://ui.adsabs.harvard.edu/abs/2014ApJ...794..132R/abstract},
	doi = {10.1088/0004-637X/794/2/132},
	language = {en},
	number = {2},
	urldate = {2025-08-09},
	journal = {The Astrophysical Journal, Volume 794, Issue 2, article id. 132, {\textless}NUMPAGES{\textgreater}12{\textless}/NUMPAGES{\textgreater} pp. (2014).},
	author = {Rogers, T. M. and Komacek, T. D.},
	month = oct,
	year = {2014},
	pages = {132},
}

@article{christensen_energy_2009,
	title = {Energy flux determines magnetic field strength of planets and stars},
	volume = {457},
	issn = {0028-0836},
	url = {https://ui.adsabs.harvard.edu/abs/2009Natur.457..167C/abstract},
	doi = {10.1038/nature07626},
	language = {en},
	number = {7226},
	urldate = {2025-08-09},
	journal = {Nature, Volume 457, Issue 7226, pp. 167-169 (2009).},
	author = {Christensen, Ulrich R. and Holzwarth, Volkmar and Reiners, Ansgar},
	month = jan,
	year = {2009},
	pages = {167},
}

@article{vidotto_prospects_2011,
	title = {Prospects for detection of exoplanet magnetic fields through bow-shock observations during transits},
	volume = {411},
	issn = {0035-8711},
	url = {https://ui.adsabs.harvard.edu/abs/2011MNRAS.411L..46V/abstract},
	doi = {10.1111/j.1745-3933.2010.00991.x},
	language = {en},
	number = {1},
	urldate = {2025-08-09},
	journal = {Monthly Notices of the Royal Astronomical Society: Letters, Volume 411, Issue 1, pp. L46-L50.},
	author = {Vidotto, A. A. and Jardine, M. and Helling, Ch},
	month = feb,
	year = {2011},
	pages = {L46},
}

@article{vidotto_early_2010,
	title = {Early {UV} {Ingress} in {WASP}-12b: {Measuring} {Planetary} {Magnetic} {Fields}},
	volume = {722},
	issn = {0004-637X},
	shorttitle = {Early {UV} {Ingress} in {WASP}-12b},
	url = {https://ui.adsabs.harvard.edu/abs/2010ApJ...722L.168V/abstract},
	doi = {10.1088/2041-8205/722/2/L168},
	language = {en},
	number = {2},
	urldate = {2025-08-09},
	journal = {The Astrophysical Journal Letters, Volume 722, Issue 2, pp. L168-L172 (2010).},
	author = {Vidotto, A. A. and Jardine, M. and Helling, Ch},
	month = oct,
	year = {2010},
	pages = {L168},
}

@article{shkolnik_hot_2005,
	title = {Hot {Jupiters} and {Hot} {Spots}: {The} {Short}- and {Long}-{Term} {Chromospheric} {Activity} on {Stars} with {Giant} {Planets}},
	volume = {622},
	issn = {0004-637X},
	shorttitle = {Hot {Jupiters} and {Hot} {Spots}},
	url = {https://ui.adsabs.harvard.edu/abs/2005ApJ...622.1075S/abstract},
	doi = {10.1086/428037},
	language = {en},
	number = {2},
	urldate = {2025-08-09},
	journal = {The Astrophysical Journal, Volume 622, Issue 2, pp. 1075-1090.},
	author = {Shkolnik, E. and Walker, G. a. H. and Bohlender, D. A. and Gu, P.-G. and Kürster, M.},
	month = apr,
	year = {2005},
	pages = {1075},
}

@article{shkolnik_evidence_2003,
	title = {Evidence for {Planet}-induced {Chromospheric} {Activity} on {HD} 179949},
	volume = {597},
	issn = {0004-637X},
	url = {https://ui.adsabs.harvard.edu/abs/2003ApJ...597.1092S/abstract},
	doi = {10.1086/378583},
	language = {en},
	number = {2},
	urldate = {2025-08-09},
	journal = {The Astrophysical Journal, Volume 597, Issue 2, pp. 1092-1096.},
	author = {Shkolnik, E. and Walker, G. a. H. and Bohlender, D. A.},
	month = nov,
	year = {2003},
	pages = {1092},
}

@article{cuntz_stellar_2000,
	title = {On {Stellar} {Activity} {Enhancement} {Due} to {Interactions} with {Extrasolar} {Giant} {Planets}},
	volume = {533},
	issn = {0004-637X},
	url = {https://ui.adsabs.harvard.edu/abs/2000ApJ...533L.151C/abstract},
	doi = {10.1086/312609},
	language = {en},
	number = {2},
	urldate = {2025-08-09},
	journal = {The Astrophysical Journal, Volume 533, Issue 2, pp. L151-L154.},
	author = {Cuntz, Manfred and Saar, Steven H. and Musielak, Zdzislaw E.},
	month = apr,
	year = {2000},
	pages = {L151},
}

@article{turner_follow-up_2024,
	title = {Follow-up {LOFAR} observations of the τ {Boötis} exoplanetary system},
	volume = {688},
	issn = {0004-6361},
	url = {https://ui.adsabs.harvard.edu/abs/2024A%26A...688A..66T/abstract},
	doi = {10.1051/0004-6361/202450095},
	language = {en},
	urldate = {2025-08-08},
	journal = {Astronomy \&amp; Astrophysics, Volume 688, id.A66, {\textless}NUMPAGES{\textgreater}8{\textless}/NUMPAGES{\textgreater} pp.},
	author = {Turner, Jake D. and Grießmeier, Jean-Mathias and Zarka, Philippe and Zhang, Xiang and Mauduit, Emilie},
	month = aug,
	year = {2024},
	pages = {A66},
}

@article{zarka_plasma_2007,
	title = {Plasma interactions of exoplanets with their parent star and associated radio emissions},
	volume = {55},
	issn = {0032-0633},
	url = {https://ui.adsabs.harvard.edu/abs/2007P%26SS...55..598Z/abstract},
	doi = {10.1016/j.pss.2006.05.045},
	language = {en},
	number = {5},
	urldate = {2025-08-08},
	journal = {Planetary and Space Science, Volume 55, Issue 5, p. 598-617.},
	author = {Zarka, Philippe},
	month = apr,
	year = {2007},
	pages = {598},
}

@article{zarka_ground-based_2008,
	title = {Ground-{Based} and {Space}-{Based} {Radio} {Observations} of {Planetary} {Lightning}},
	volume = {137},
	issn = {0038-6308},
	url = {https://ui.adsabs.harvard.edu/abs/2008SSRv..137..257Z/abstract},
	doi = {10.1007/s11214-008-9366-8},
	language = {en},
	number = {1-4},
	urldate = {2025-08-08},
	journal = {Space Science Reviews, Volume 137, Issue 1-4, pp. 257-269},
	author = {Zarka, P. and Farrell, W. and Fischer, G. and Konovalenko, A.},
	month = jun,
	year = {2008},
	pages = {257},
}

@article{griesmeier_predicting_2007,
	title = {Predicting low-frequency radio fluxes of known extrasolar planets},
	volume = {475},
	issn = {0004-6361},
	url = {https://ui.adsabs.harvard.edu/abs/2007A%26A...475..359G/abstract},
	doi = {10.1051/0004-6361:20077397},
	language = {en},
	number = {1},
	urldate = {2025-08-08},
	journal = {Astronomy and Astrophysics, Volume 475, Issue 1, November III 2007, pp.359-368},
	author = {Grießmeier, J.-M. and Zarka, P. and Spreeuw, H.},
	month = nov,
	year = {2007},
	pages = {359},
}

@article{cowley_solar-wind-magnetosphere-ionosphere_2003,
	title = {Solar-wind-magnetosphere-ionosphere interactions in the {Earth}'s plasma environment},
	volume = {361},
	issn = {1364-503X,0080-4614,0962-8436},
	url = {https://ui.adsabs.harvard.edu/abs/2003RSPTA.361..113C/abstract},
	doi = {10.1098/rsta.2002.1112},
	language = {en},
	number = {1802},
	urldate = {2025-08-08},
	journal = {Science and applications of the space environment: new results and interdisciplinary connections. Papers of a Theme compiled and edited by A. J. Coates, J. L. Culhane and J. C. R. Hunt. Roy Soc of London Phil Tr A, vol. 361, Issue 1802, p.113},
	author = {Cowley, S. W. H. and Davies, J. A. and Grocott, A. and Khan, H. and Lester, M. and McWilliams, K. A. and Milan, S. E. and Provan, G. and Sandholt, P. E. and Wild, J. A. and Yeoman, T. K.},
	month = jan,
	year = {2003},
	pages = {113},
}

@article{khodachenko_impact_2021,
	title = {The impact of intrinsic magnetic field on the absorption signatures of elements probing the upper atmosphere of {HD209458b}},
	volume = {507},
	issn = {0035-8711},
	url = {https://ui.adsabs.harvard.edu/abs/2021MNRAS.507.3626K/abstract},
	doi = {10.1093/mnras/stab2366},
	language = {en},
	number = {3},
	urldate = {2025-08-08},
	journal = {Monthly Notices of the Royal Astronomical Society},
	author = {Khodachenko, M. L. and Shaikhislamov, I. F. and Lammer, H. and Miroshnichenko, I. B. and Rumenskikh, M. S. and Berezutsky, A. G. and Fossati, L.},
	month = nov,
	year = {2021},
	pages = {3626--3637},
}

@article{schreyer_using_2024,
	title = {Using helium 10 830 Å transits to constrain planetary magnetic fields},
	volume = {527},
	issn = {0035-8711},
	url = {https://ui.adsabs.harvard.edu/abs/2024MNRAS.527.5117S/abstract},
	doi = {10.1093/mnras/stad3528},
	language = {en},
	number = {3},
	urldate = {2025-08-08},
	journal = {Monthly Notices of the Royal Astronomical Society},
	author = {Schreyer, Ethan and Owen, James E. and Spake, Jessica J. and Bahroloom, Zahra and Di Giampasquale, Simone},
	month = jan,
	year = {2024},
	pages = {5117--5130},
}

@article{turner_search_2021,
	title = {The search for radio emission from the exoplanetary systems 55 {Cancri}, υ {Andromedae}, and τ {Boötis} using {LOFAR} beam-formed observations},
	volume = {645},
	issn = {0004-6361},
	url = {https://ui.adsabs.harvard.edu/abs/2021A%26A...645A..59T/abstract},
	doi = {10.1051/0004-6361/201937201},
	language = {en},
	urldate = {2025-08-07},
	journal = {Astronomy \&amp; Astrophysics, Volume 645, id.A59, {\textless}NUMPAGES{\textgreater}28{\textless}/NUMPAGES{\textgreater} pp.},
	author = {Turner, Jake D. and Zarka, Philippe and Grießmeier, Jean-Mathias and Lazio, Joseph and Cecconi, Baptiste and Emilio Enriquez, J. and Girard, Julien N. and Jayawardhana, Ray and Lamy, Laurent and Nichols, Jonathan D. and de Pater, Imke},
	month = jan,
	year = {2021},
	pages = {A59},
}

@article{narang_ugmrt_2024,
	title = {A {uGMRT} search for radio emission from planets around evolved stars},
	volume = {529},
	issn = {0035-8711},
	url = {https://ui.adsabs.harvard.edu/abs/2024MNRAS.529.1161N/abstract},
	doi = {10.1093/mnras/stae536},
	language = {en},
	number = {2},
	urldate = {2025-08-07},
	journal = {Monthly Notices of the Royal Astronomical Society, Volume 529, Issue 2, pp.1161-1168},
	author = {Narang, Mayank and Manoj, P. and Chandra, C. H. Ishwara and Banerjee, Bihan and Tyagi, Himanshu and Tamura, Motohide and Henning, Thomas and Mathew, Blesson and Lazio, Joseph and Surya, Arun and Nayak, Prasanta K.},
	month = apr,
	year = {2024},
	pages = {1161},
}

@article{savel_new_2024,
	title = {A {New} {Lever} on {Exoplanetary} {B} {Fields}: {Measuring} {Heavy} {Ion} {Velocities}},
	volume = {969},
	issn = {0004-637X},
	shorttitle = {A {New} {Lever} on {Exoplanetary} {B} {Fields}},
	url = {https://ui.adsabs.harvard.edu/abs/2024ApJ...969L..27S/abstract},
	doi = {10.3847/2041-8213/ad5a0a},
	language = {en},
	number = {2},
	urldate = {2025-08-01},
	journal = {The Astrophysical Journal Letters, Volume 969, Issue 2, id.L27, 8 pp.},
	author = {Savel, Arjun B. and Beltz, Hayley and Komacek, Thaddeus D. and Tsai, Shang-Min and Kempton, Eliza M.-R.},
	month = jul,
	year = {2024},
	pages = {L27},
}

@article{zarka_auroral_1998,
	title = {Auroral radio emissions at the outer planets: {Observations} and theories},
	volume = {103},
	issn = {0148-0227},
	shorttitle = {Auroral radio emissions at the outer planets},
	url = {https://ui.adsabs.harvard.edu/abs/1998JGR...10320159Z/abstract},
	doi = {10.1029/98JE01323},
	language = {en},
	number = {E9},
	urldate = {2025-07-28},
	journal = {Journal of Geophysical Research},
	author = {Zarka, Philippe},
	month = sep,
	year = {1998},
	pages = {20159--20194},
}

@article{knierim_shallowness_2022,
	title = {Shallowness of circulation in hot {Jupiters}. {Advancing} the {Ohmic} dissipation model},
	volume = {658},
	issn = {0004-6361},
	url = {https://ui.adsabs.harvard.edu/abs/2022A%26A...658L...7K/abstract},
	doi = {10.1051/0004-6361/202142588},
	language = {en},
	urldate = {2025-07-28},
	journal = {Astronomy \&amp; Astrophysics, Volume 658, id.L7, {\textless}NUMPAGES{\textgreater}7{\textless}/NUMPAGES{\textgreater} pp.},
	author = {Knierim, H. and Batygin, K. and Bitsch, B.},
	month = feb,
	year = {2022},
	pages = {L7},
}

@article{batygin_inflating_2010,
	title = {Inflating {Hot} {Jupiters} with {Ohmic} {Dissipation}},
	volume = {714},
	issn = {0004-637X},
	url = {https://ui.adsabs.harvard.edu/abs/2010ApJ...714L.238B/abstract},
	doi = {10.1088/2041-8205/714/2/L238},
	language = {en},
	number = {2},
	urldate = {2025-07-28},
	journal = {The Astrophysical Journal Letters, Volume 714, Issue 2, pp. L238-L243 (2010).},
	author = {Batygin, Konstantin and Stevenson, David J.},
	month = may,
	year = {2010},
	pages = {L238},
}

@misc{lupu_correlated_2021,
	title = {Correlated k coefficients for {H2}-{He} atmospheres; 11 spectral windows and 1460 pressure-temperature points},
	url = {https://zenodo.org/records/7542048},
	doi = {10.5281/zenodo.7542048},
	urldate = {2025-07-28},
	publisher = {Zenodo},
	author = {Lupu, Roxana and Freedman, Richard and Gharib-Nezhad, Ehsan and Visscher, Channon and Molliere, Paul},
	month = oct,
	year = {2021},
}

@article{batygin_magnetically_2013,
	title = {Magnetically {Controlled} {Circulation} on {Hot} {Extrasolar} {Planets}},
	volume = {776},
	issn = {0004-637X},
	url = {https://ui.adsabs.harvard.edu/abs/2013ApJ...776...53B/abstract},
	doi = {10.1088/0004-637X/776/1/53},
	language = {en},
	number = {1},
	urldate = {2025-07-27},
	journal = {The Astrophysical Journal, Volume 776, Issue 1, article id. 53, {\textless}NUMPAGES{\textgreater}14{\textless}/NUMPAGES{\textgreater} pp. (2013).},
	author = {Batygin, Konstantin and Stanley, Sabine and Stevenson, David J.},
	month = oct,
	year = {2013},
	pages = {53},
}

@article{rauscher_three-dimensional_2013,
	title = {Three-dimensional {Atmospheric} {Circulation} {Models} of {HD} 189733b and {HD} 209458b with {Consistent} {Magnetic} {Drag} and {Ohmic} {Dissipation}},
	volume = {764},
	issn = {0004-637X},
	url = {https://ui.adsabs.harvard.edu/abs/2013ApJ...764..103R/abstract},
	doi = {10.1088/0004-637X/764/1/103},
	language = {en},
	number = {1},
	urldate = {2025-07-27},
	journal = {The Astrophysical Journal, Volume 764, Issue 1, article id. 103, {\textless}NUMPAGES{\textgreater}18{\textless}/NUMPAGES{\textgreater} pp. (2013).},
	author = {Rauscher, Emily and Menou, Kristen},
	month = feb,
	year = {2013},
	pages = {103},
}

@article{perna_ohmic_2010,
	title = {Ohmic {Dissipation} in the {Atmospheres} of {Hot} {Jupiters}},
	volume = {724},
	issn = {0004-637X},
	url = {https://ui.adsabs.harvard.edu/abs/2010ApJ...724..313P/abstract},
	doi = {10.1088/0004-637X/724/1/313},
	language = {en},
	number = {1},
	urldate = {2025-07-27},
	journal = {The Astrophysical Journal, Volume 724, Issue 1, pp. 313-317 (2010).},
	author = {Perna, Rosalba and Menou, Kristen and Rauscher, Emily},
	month = nov,
	year = {2010},
	pages = {313},
}

@article{cabot_toi-1518b_2021,
	title = {{TOI}-1518b: {A} {Misaligned} {Ultra}-hot {Jupiter} with {Iron} in {Its} {Atmosphere}},
	volume = {162},
	issn = {0004-6256},
	shorttitle = {{TOI}-1518b},
	url = {https://ui.adsabs.harvard.edu/abs/2021AJ....162..218C/abstract},
	doi = {10.3847/1538-3881/ac1ba3},
	language = {en},
	number = {5},
	urldate = {2025-07-25},
	journal = {The Astronomical Journal},
	author = {Cabot, Samuel H. C. and Bello-Arufe, Aaron and Mendonça, João M. and Tronsgaard, René and Wong, Ian and Zhou, George and Buchhave, Lars A. and Fischer, Debra A. and Stassun, Keivan G. and Antoci, Victoria and Baker, David and Belinski, Alexander A. and Benneke, Björn and Bouma, Luke G. and Christiansen, Jessie L. and Collins, Karen A. and Goliguzova, Maria V. and Hagey, Simone and Jenkins, Jon M. and Jensen, Eric L. N. and Kidwell, Richard C. and Laloum, Didier and Massey, Bob and McLeod, Kim K. and Latham, David W. and Morgan, Edward H. and Ricker, George and Safonov, Boris S. and Schlieder, Joshua E. and Seager, Sara and Shporer, Avi and Smith, Jeffrey C. and Srdoc, Gregor and Strakhov, Ivan A. and Torres, Guillermo and Twicken, Joseph D. and Vanderspek, Roland and Vezie, Michael and Winn, Joshua N.},
	month = nov,
	year = {2021},
	pages = {218},
}

@article{talens_mascara-2_2018,
	title = {{MASCARA}-2 b. {A} hot {Jupiter} transiting the m{\textless}{SUB}{\textgreater}{V}{\textless}/{SUB}{\textgreater} = 7.6 {A}-star {HD} 185603},
	volume = {612},
	issn = {0004-6361},
	url = {https://ui.adsabs.harvard.edu/abs/2018A&A...612A..57T/abstract},
	doi = {10.1051/0004-6361/201731512},
	language = {en},
	urldate = {2025-07-25},
	journal = {Astronomy and Astrophysics},
	author = {Talens, G. J. J. and Justesen, A. B. and Albrecht, S. and McCormac, J. and Van Eylen, V. and Otten, G. P. P. L. and Murgas, F. and Palle, E. and Pollacco, D. and Stuik, R. and Spronck, J. F. P. and Lesage, A.-L. and Grundahl, F. and Fredslund Andersen, M. and Antoci, V. and Snellen, I. a. G.},
	month = apr,
	year = {2018},
	pages = {A57},
}

@article{lund_kelt-20b_2017,
	title = {{KELT}-20b: {A} {Giant} {Planet} with a {Period} of {P} ∼ 3.5 days {Transiting} the {V} ∼ 7.6 {Early} {A} {Star} {HD} 185603},
	volume = {154},
	issn = {0004-6256},
	shorttitle = {{KELT}-20b},
	url = {https://ui.adsabs.harvard.edu/abs/2017AJ....154..194L/abstract},
	doi = {10.3847/1538-3881/aa8f95},
	language = {en},
	number = {5},
	urldate = {2025-07-25},
	journal = {The Astronomical Journal},
	author = {Lund, Michael B. and Rodriguez, Joseph E. and Zhou, George and Gaudi, B. Scott and Stassun, Keivan G. and Johnson, Marshall C. and Bieryla, Allyson and Oelkers, Ryan J. and Stevens, Daniel J. and Collins, Karen A. and Penev, Kaloyan and Quinn, Samuel N. and Latham, David W. and Villanueva, Steven and Eastman, Jason D. and Kielkopf, John F. and Oberst, Thomas E. and Jensen, Eric L. N. and Cohen, David H. and Joner, Michael D. and Stephens, Denise C. and Relles, Howard and Corfini, Giorgio and Gregorio, Joao and Zambelli, Roberto and Esquerdo, Gilbert A. and Calkins, Michael L. and Berlind, Perry and Ciardi, David R. and Dressing, Courtney and Patel, Rahul and Gagnon, Patrick and Gonzales, Erica and Beatty, Thomas G. and Siverd, Robert J. and Labadie-Bartz, Jonathan and Kuhn, Rudolf B. and Colón, Knicole D. and James, David and Pepper, Joshua and Fulton, Benjamin J. and McLeod, Kim K. and Stockdale, Christopher and Calchi Novati, Sebastiano and DePoy, D. L. and Gould, Andrew and Marshall, Jennifer L. and Trueblood, Mark and Trueblood, Patricia and Johnson, John A. and Wright, Jason and McCrady, Nate and Wittenmyer, Robert A. and Johnson, Samson A. and Sergi, Anthony and Wilson, Maurice and Sliski, David H.},
	month = nov,
	year = {2017},
	pages = {194},
}

@misc{soriano-guerrero_non-ideal_2025,
	title = {Non-ideal {MHD} simulations of hot {Jupiter} atmospheres},
	url = {http://arxiv.org/abs/2505.14342},
	doi = {10.48550/arXiv.2505.14342},
	urldate = {2025-07-18},
	publisher = {arXiv},
	author = {Soriano-Guerrero, Clàudia and Viganò, Daniele and Perna, Rosalba and Elias-López, Albert and Beltz, Hayley},
	month = may,
	year = {2025},
	note = {arXiv:2505.14342 [astro-ph]},
}

@misc{christie_geometric_2025,
	title = {Geometric {Considerations} in {Hot} {Jupiter} {Magnetic} {Drag} {Models}},
	url = {http://arxiv.org/abs/2507.08511},
	doi = {10.48550/arXiv.2507.08511},
	urldate = {2025-07-18},
	publisher = {arXiv},
	author = {Christie, Duncan A. and Evans-Soma, Tom M. and Mayne, Nathan J. and Kohary, Krisztian},
	month = jul,
	year = {2025},
	note = {arXiv:2507.08511 [astro-ph]},
}

@article{prinoth_titanium_2025,
	title = {Titanium chemistry of {WASP}-121 b with {ESPRESSO} in 4-{UT} mode},
	volume = {694},
	issn = {0004-6361},
	url = {https://ui.adsabs.harvard.edu/abs/2025A%26A...694A.284P/abstract},
	doi = {10.1051/0004-6361/202452405},
	language = {en},
	urldate = {2025-07-10},
	journal = {Astronomy \&amp; Astrophysics, Volume 694, id.A284, 14 pp.},
	author = {Prinoth, B. and Seidel, J. V. and Hoeijmakers, H. J. and Morris, B. M. and Baratella, M. and Borsato, N. W. and Damasceno, Y. C. and Parmentier, V. and Kitzmann, D. and Sedaghati, E. and Pino, L. and Borsa, F. and Allart, R. and Santos, N. and Steiner, M. and Suárez Mascareño, A. and Tabernero, H. and Zapatero Osorio, M. R.},
	month = feb,
	year = {2025},
	pages = {A284},
}

@article{parmentier_thermal_2018,
	title = {From thermal dissociation to condensation in the atmospheres of ultra hot {Jupiters}: {WASP}-121b in context},
	volume = {617},
	issn = {0004-6361},
	shorttitle = {From thermal dissociation to condensation in the atmospheres of ultra hot {Jupiters}},
	url = {https://ui.adsabs.harvard.edu/abs/2018A%26A...617A.110P/abstract},
	doi = {10.1051/0004-6361/201833059},
	language = {en},
	urldate = {2025-06-18},
	journal = {Astronomy \&amp; Astrophysics, Volume 617, id.A110, {\textless}NUMPAGES{\textgreater}17{\textless}/NUMPAGES{\textgreater} pp.},
	author = {Parmentier, Vivien and Line, Mike R. and Bean, Jacob L. and Mansfield, Megan and Kreidberg, Laura and Lupu, Roxana and Visscher, Channon and Désert, Jean-Michel and Fortney, Jonathan J. and Deleuil, Magalie and Arcangeli, Jacob and Showman, Adam P. and Marley, Mark S.},
	month = sep,
	year = {2018},
	pages = {A110},
}

@article{batygin_evolution_2011,
	title = {Evolution of {Ohmically} {Heated} {Hot} {Jupiters}},
	volume = {738},
	issn = {0004-637X},
	url = {https://ui.adsabs.harvard.edu/abs/2011ApJ...738....1B/abstract},
	doi = {10.1088/0004-637X/738/1/1},
	language = {en},
	number = {1},
	urldate = {2025-06-17},
	journal = {The Astrophysical Journal, Volume 738, Issue 1, article id. 1, {\textless}NUMPAGES{\textgreater}10{\textless}/NUMPAGES{\textgreater} pp. (2011).},
	author = {Batygin, Konstantin and Stevenson, David J. and Bodenheimer, Peter H.},
	month = sep,
	year = {2011},
	pages = {1},
}

@misc{snellen_exoplanet_2025,
	title = {Exoplanet atmospheres at high spectral resolution},
	url = {http://arxiv.org/abs/2505.08926},
	doi = {10.48550/arXiv.2505.08926},
	urldate = {2025-05-15},
	publisher = {arXiv},
	author = {Snellen, Ignas},
	month = may,
	year = {2025},
	note = {arXiv:2505.08926 [astro-ph]},
}

@article{seidel_vertical_2025,
	title = {Vertical structure of an exoplanet's atmospheric jet stream},
	volume = {639},
	issn = {0028-0836},
	url = {https://ui.adsabs.harvard.edu/abs/2025Natur.639..902S/abstract},
	doi = {10.1038/s41586-025-08664-1},
	language = {en},
	number = {8056},
	urldate = {2025-05-10},
	journal = {Nature, Volume 639, Issue 8056, pp. 902-908},
	author = {Seidel, Julia V. and Prinoth, Bibiana and Pino, Lorenzo and dos Santos, Leonardo A. and Chakraborty, Hritam and Parmentier, Vivien and Sedaghati, Elyar and Wardenier, Joost P. and Farret Jentink, Casper and Zapatero Osorio, Maria Rosa and Allart, Romain and Ehrenreich, David and Lendl, Monika and Roccetti, Giulia and Damasceno, Yuri and Bourrier, Vincent and Lillo-Box, Jorge and Hoeijmakers, H. Jens and Pallé, Enric and Santos, Nuno and Suárez Mascareño, Alejandro and Sousa, Sergio G. and Tabernero, Hugo M. and Pepe, Francesco A.},
	month = mar,
	year = {2025},
	pages = {902},
}

@article{beltz_effects_2025,
	title = {The {Effects} of {Kinematic} {Magnetohydrodynamics} on the {Atmospheric} {Circulation} of {Eccentric} {Hot} {Jupiters}},
	volume = {984},
	issn = {0004-637X},
	url = {https://ui.adsabs.harvard.edu/abs/2025ApJ...984...90B/abstract},
	doi = {10.3847/1538-4357/adc56c},
	language = {en},
	number = {1},
	urldate = {2025-05-05},
	journal = {The Astrophysical Journal, Volume 984, Issue 1, id.90, 18 pp.},
	author = {Beltz, Hayley and Houck, Willow and Mayorga, L. C. and Komacek, Thaddeus D. and Livesey, Joseph R. and Becker, Juliette},
	month = may,
	year = {2025},
	pages = {90},
}

@misc{pelletier_crires_2024,
	title = {{CRIRES}+ and {ESPRESSO} reveal an atmosphere enriched in volatiles relative to refractories on the ultra-hot {Jupiter} {WASP}-121b},
	url = {http://arxiv.org/abs/2410.18183},
	doi = {10.48550/arXiv.2410.18183},
	urldate = {2024-11-05},
	publisher = {arXiv},
	author = {Pelletier, Stefan and Benneke, Björn and Chachan, Yayaati and Bazinet, Luc and Allart, Romain and Hoeijmakers, H. Jens and Lavail, Alexis and Prinoth, Bibiana and Coulombe, Louis-Philippe and Lothringer, Joshua D. and Parmentier, Vivien and Smith, Peter and Borsato, Nicholas and Thorsbro, Brian},
	month = oct,
	year = {2024},
	note = {arXiv:2410.18183},
}

@article{sedaghati_spectral_2021,
	title = {A spectral survey of {WASP}-19b with {ESPRESSO}},
	volume = {505},
	issn = {0035-8711},
	url = {https://ui.adsabs.harvard.edu/abs/2021MNRAS.505..435S},
	doi = {10.1093/mnras/stab1164},
	urldate = {2024-10-09},
	journal = {Monthly Notices of the Royal Astronomical Society},
	publisher = {OUP},
	author = {Sedaghati, Elyar and MacDonald, Ryan J. and Casasayas-Barris, Núria and Hoeijmakers, H. Jens and Boffin, Henri M. J. and Rodler, Florian and Brahm, Rafael and Jones, Matías and Sánchez-López, Alejandro and Carleo, Ilaria and Figueira, Pedro and Mehner, Andrea and López-Puertas, Manuel},
	month = jul,
	year = {2021},
	note = {ADS Bibcode: 2021MNRAS.505..435S},
	pages = {435--458},
}

@misc{darpa_gaps_2024,
	title = {The {GAPS} programme at {TNG} {LX} {Atmospheric} characterisation of {KELT}-9 b via single-line analysis: {Detection} of six {H} {I} {Balmer} lines, {Na} {I}, {Ca} {I}, {Ca} {II}, {Fe} {I}, {Fe} {II}, {Mg} {I}, {Ti} {II}, {Sc} {II}, and {Cr} {II}},
	shorttitle = {The {GAPS} programme at {TNG} {LX} {Atmospheric} characterisation of {KELT}-9 b via single-line analysis},
	url = {http://arxiv.org/abs/2409.01779},
	doi = {10.48550/arXiv.2409.01779},
	urldate = {2024-09-11},
	publisher = {arXiv},
	author = {D'Arpa, M. C. and Saba, A. and Borsa, F. and Fossati, L. and Micela, G. and Di Maio, C. and Stangret, M. and Tripodo, G. and Affer, L. and Bonomo, A. S. and Benatti, S. and Brogi, M. and Fardella, V. and Lanza, A. F. and Guilluy, G. and Maldonado, J. and Mantovan, G. and Nascimbeni, V. and Pino, L. and Scandariato, G. and Sicilia, D. and Sozzetti, A. and Spinelli, R. and Andreuzzi, G. and Bignamini, A. and Claudi, R. and Desidera, S. and Ghedina, A. and Knapic, C. and Lorenzi, V.},
	month = sep,
	year = {2024},
	note = {arXiv:2409.01779 [astro-ph]},
}

@book{gray_observation_2008,
	title = {The {Observation} and {Analysis} of {Stellar} {Photospheres}},
	url = {https://ui.adsabs.harvard.edu/abs/2008oasp.book.....G},
	urldate = {2024-09-09},
	author = {Gray, David F.},
	month = jun,
	year = {2008},
	note = {Publication Title: The Observation and Analysis of Stellar Photospheres
ADS Bibcode: 2008oasp.book.....G},
}

@article{molliere_petitradtrans_2019,
	title = {{petitRADTRANS}. {A} {Python} radiative transfer package for exoplanet characterization and retrieval},
	volume = {627},
	issn = {0004-6361},
	url = {https://ui.adsabs.harvard.edu/abs/2019A&A...627A..67M},
	doi = {10.1051/0004-6361/201935470},
	urldate = {2024-09-09},
	journal = {Astronomy and Astrophysics},
	author = {Mollière, P. and Wardenier, J. P. and van Boekel, R. and Henning, Th. and Molaverdikhani, K. and Snellen, I. A. G.},
	month = jul,
	year = {2019},
	note = {ADS Bibcode: 2019A\&A...627A..67M},
	pages = {A67},
}

@article{showman_atmospheric_2002,
	title = {Atmospheric {Circulation} and {Tides} of "{51Peg} b-like" {Planets}},
	volume = {385},
	issn = {0004-6361, 1432-0746},
	url = {http://arxiv.org/abs/astro-ph/0202236},
	doi = {10.1051/0004-6361:20020101},
	language = {en},
	number = {1},
	urldate = {2024-08-30},
	journal = {Astronomy \& Astrophysics},
	author = {Showman, Adam P. and Guillot, Tristan},
	month = apr,
	year = {2002},
	note = {arXiv:astro-ph/0202236},
	pages = {166--180},
}

@article{hoeijmakers_mantis_2024,
	title = {The {Mantis} {Network}. {IV}. {A} titanium cold trap on the ultra-hot {Jupiter} {WASP}-121 b},
	volume = {685},
	issn = {0004-6361},
	url = {https://ui.adsabs.harvard.edu/abs/2024A&A...685A.139H},
	doi = {10.1051/0004-6361/202244968},
	urldate = {2024-08-08},
	journal = {Astronomy and Astrophysics},
	author = {Hoeijmakers, H. J. and Kitzmann, D. and Morris, B. M. and Prinoth, B. and Borsato, N. W. and Thorsbro, B. and Pino, L. and Lee, E. K. H. and Akın, C. and Seidel, J. V. and Birkby, J. L. and Allart, R. and Heng, Kevin},
	month = may,
	year = {2024},
	note = {ADS Bibcode: 2024A\&A...685A.139H},
	pages = {A139},
}

@article{kitzmann_mantis_2023,
	title = {The {Mantis} network - {A} standard grid of templates and masks for cross-correlation analyses of ultra-hot {Jupiter} transmission spectra,},
	volume = {669},
	copyright = {© The Authors 2023},
	issn = {0004-6361, 1432-0746},
	url = {https://www.aanda.org/articles/aa/abs/2023/01/aa42969-21/aa42969-21.html},
	doi = {10.1051/0004-6361/202142969},
	language = {en},
	urldate = {2024-08-08},
	journal = {Astronomy \& Astrophysics},
	publisher = {EDP Sciences},
	author = {Kitzmann, D. and Hoeijmakers, H. J. and Grimm, S. L. and Borsato, N. W. and Lueber, A. and Prinoth, B.},
	month = jan,
	year = {2023},
	pages = {A113},
}

@article{seidel_into_2021,
	title = {Into the storm: diving into the winds of the ultra-hot {Jupiter} {WASP}-76 b with {HARPS} and {ESPRESSO}},
	volume = {653},
	issn = {0004-6361},
	shorttitle = {Into the storm},
	url = {https://ui.adsabs.harvard.edu/abs/2021A&A...653A..73S},
	doi = {10.1051/0004-6361/202140569},
	urldate = {2024-08-05},
	journal = {Astronomy and Astrophysics},
	author = {Seidel, J. V. and Ehrenreich, D. and Allart, R. and Hoeijmakers, H. J. and Lovis, C. and Bourrier, V. and Pino, L. and Wyttenbach, A. and Adibekyan, V. and Alibert, Y. and Borsa, F. and Casasayas-Barris, N. and Cristiani, S. and Demangeon, O. D. S. and Di Marcantonio, P. and Figueira, P. and González Hernández, J. I. and Lillo-Box, J. and Martins, C. J. A. P. and Mehner, A. and Molaro, P. and Nunes, N. J. and Palle, E. and Pepe, F. and Santos, N. C. and Sousa, S. G. and Sozzetti, A. and Tabernero, H. M. and Zapatero Osorio, M. R.},
	month = sep,
	year = {2021},
	note = {ADS Bibcode: 2021A\&A...653A..73S},
	pages = {A73},
}

@article{seidel_detection_2023-1,
	title = {Detection of a high-velocity sodium feature on the ultra-hot {Jupiter} {WASP}-121 b},
	volume = {673},
	issn = {0004-6361},
	url = {https://ui.adsabs.harvard.edu/abs/2023A&A...673A.125S},
	doi = {10.1051/0004-6361/202245800},
	urldate = {2024-08-05},
	journal = {Astronomy and Astrophysics},
	author = {Seidel, J. V. and Borsa, F. and Pino, L. and Ehrenreich, D. and Stangret, M. and Zapatero Osorio, M. R. and Palle, E. and Alibert, Y. and Allart, R. and Bourrier, V. and Di Marcantonio, P. and Figueira, P. and González Hernández, J. I. and Lillo-Box, J. and Lovis, C. and Martins, C. J. A. P. and Mehner, A. and Molaro, P. and Nunes, N. J. and Pepe, F. and Santos, N. C. and Sozzetti, A.},
	month = may,
	year = {2023},
	note = {ADS Bibcode: 2023A\&A...673A.125S},
	pages = {A125},
}

@article{buchner_pymultinest_2016,
	title = {{PyMultiNest}: {Python} interface for {MultiNest}},
	shorttitle = {{PyMultiNest}},
	url = {https://ui.adsabs.harvard.edu/abs/2016ascl.soft06005B},
	urldate = {2024-08-05},
	journal = {Astrophysics Source Code Library},
	author = {Buchner, Johannes},
	month = jun,
	year = {2016},
	note = {ADS Bibcode: 2016ascl.soft06005B},
	pages = {ascl:1606.005},
}

@article{kesseli_confirmation_2021,
	title = {Confirmation of {Asymmetric} {Iron} {Absorption} in {WASP}-76b with {HARPS}},
	volume = {908},
	issn = {0004-637X},
	url = {https://ui.adsabs.harvard.edu/abs/2021ApJ...908L..17K},
	doi = {10.3847/2041-8213/abe047},
	urldate = {2024-08-05},
	journal = {The Astrophysical Journal},
	publisher = {IOP},
	author = {Kesseli, Aurora Y. and Snellen, I. A. G.},
	month = feb,
	year = {2021},
	note = {ADS Bibcode: 2021ApJ...908L..17K},
	pages = {L17},
}

@article{smette_molecfit_2015,
	title = {Molecfit: {A} general tool for telluric absorption correction. {I}. {Method} and application to {ESO} instruments},
	volume = {576},
	issn = {0004-6361},
	shorttitle = {Molecfit},
	url = {https://ui.adsabs.harvard.edu/abs/2015A&A...576A..77S},
	doi = {10.1051/0004-6361/201423932},
	urldate = {2024-07-14},
	journal = {Astronomy and Astrophysics},
	author = {Smette, A. and Sana, H. and Noll, S. and Horst, H. and Kausch, W. and Kimeswenger, S. and Barden, M. and Szyszka, C. and Jones, A. M. and Gallenne, A. and Vinther, J. and Ballester, P. and Taylor, J.},
	month = apr,
	year = {2015},
	note = {ADS Bibcode: 2015A\&A...576A..77S},
	pages = {A77},
}

@article{husser_new_2013,
	title = {A new extensive library of {PHOENIX} stellar atmospheres and synthetic spectra},
	volume = {553},
	issn = {0004-6361, 1432-0746},
	url = {http://arxiv.org/abs/1303.5632},
	doi = {10.1051/0004-6361/201219058},
	urldate = {2024-07-14},
	journal = {Astronomy \& Astrophysics},
	author = {Husser, Tim-Oliver and von Berg, Sebastian Wende- and Dreizler, Stefan and Homeier, Derek and Reiners, Ansgar and Barman, Travis and Hauschildt, Peter H.},
	month = may,
	year = {2013},
	note = {arXiv:1303.5632 [astro-ph]},
	pages = {A6},
}

@article{cauley_time-resolved_2021,
	title = {Time-resolved rotational velocities in the upper atmosphere of {WASP}-33 b},
	volume = {161},
	issn = {0004-6256, 1538-3881},
	url = {http://arxiv.org/abs/2010.02118},
	doi = {10.3847/1538-3881/abde43},
	number = {3},
	urldate = {2024-07-14},
	journal = {The Astronomical Journal},
	author = {Cauley, P. Wilson and Wang, Ji and Shkolnik, Evgenya L. and Ilyin, Ilya and Strassmeier, Klaus G. and Redfield, Seth and Jensen, Adam},
	month = mar,
	year = {2021},
	note = {arXiv:2010.02118 [astro-ph]},
	pages = {152},
}

@article{kausch_molecfit_2015,
	title = {Molecfit: {A} general tool for telluric absorption correction {II}. {Quantitative} evaluation on {ESO}-{VLT} {X}-{Shooter} spectra},
	volume = {576},
	issn = {0004-6361, 1432-0746},
	shorttitle = {Molecfit},
	url = {http://arxiv.org/abs/1501.07265},
	doi = {10.1051/0004-6361/201423909},
	urldate = {2024-07-14},
	journal = {Astronomy \& Astrophysics},
	author = {Kausch, W. and Smette, S. Noll A. and Kimeswenger, S. and Barden, M. and Szyszka, C. and Jones, A. M. and Sana, H. and Horst, H. and Kerber, F.},
	month = apr,
	year = {2015},
	note = {arXiv:1501.07265 [astro-ph]},
	pages = {A78},
}

@article{pelletier_vanadium_2023,
	title = {Vanadium oxide and a sharp onset of cold-trapping on a giant exoplanet},
	issn = {0028-0836, 1476-4687},
	url = {http://arxiv.org/abs/2306.08739},
	doi = {10.1038/s41586-023-06134-0},
	urldate = {2023-06-16},
	journal = {Nature},
	author = {Pelletier, Stefan and Benneke, Björn and Ali-Dib, Mohamad and Prinoth, Bibiana and Kasper, David and Seifahrt, Andreas and Bean, Jacob L. and Debras, Florian and Klein, Baptiste and Bazinet, Luc and Hoeijmakers, H. Jens and Kesseli, Aurora Y. and Lim, Olivia and Carmona, Andres and Pino, Lorenzo and Casasayas-Barris, Núria and Hood, Thea and Stürmer, Julian},
	month = jun,
	year = {2023},
	note = {arXiv:2306.08739 [astro-ph]},
}

@article{gandhi_retrieval_2023,
	title = {Retrieval survey of metals in six ultra-hot {Jupiters}: {Trends} in chemistry, rain-out, ionisation and atmospheric dynamics},
	volume = {165},
	issn = {0004-6256, 1538-3881},
	shorttitle = {Retrieval survey of metals in six ultra-hot {Jupiters}},
	url = {http://arxiv.org/abs/2305.17228},
	doi = {10.3847/1538-3881/accd65},
	number = {6},
	urldate = {2023-06-01},
	journal = {The Astronomical Journal},
	author = {Gandhi, Siddharth and Kesseli, Aurora and Zhang, Yapeng and Louca, Amy and Snellen, Ignas and Brogi, Matteo and Miguel, Yamila and Casasayas-Barris, Núria and Pelletier, Stefan and Landman, Rico and Maguire, Cathal and Gibson, Neale P.},
	month = jun,
	year = {2023},
	note = {arXiv:2305.17228 [astro-ph]},
	pages = {242},
}

@article{ehrenreich_nightside_2020,
	title = {Nightside condensation of iron in an ultrahot giant exoplanet},
	volume = {580},
	issn = {0028-0836},
	url = {https://ui.adsabs.harvard.edu/abs/2020Natur.580..597E},
	doi = {10.1038/s41586-020-2107-1},
	urldate = {2023-03-29},
	journal = {Nature},
	author = {Ehrenreich, David and Lovis, Christophe and Allart, Romain and Zapatero Osorio, María Rosa and Pepe, Francesco and Cristiani, Stefano and Rebolo, Rafael and Santos, Nuno C. and Borsa, Francesco and Demangeon, Olivier and Dumusque, Xavier and González Hernández, Jonay I. and Casasayas-Barris, Núria and Ségransan, Damien and Sousa, Sérgio and Abreu, Manuel and Adibekyan, Vardan and Affolter, Michael and Allende Prieto, Carlos and Alibert, Yann and Aliverti, Matteo and Alves, David and Amate, Manuel and Avila, Gerardo and Baldini, Veronica and Bandy, Timothy and Benz, Willy and Bianco, Andrea and Bolmont, Émeline and Bouchy, François and Bourrier, Vincent and Broeg, Christopher and Cabral, Alexandre and Calderone, Giorgio and Pallé, Enric and Cegla, H. M. and Cirami, Roberto and Coelho, João M. P. and Conconi, Paolo and Coretti, Igor and Cumani, Claudio and Cupani, Guido and Dekker, Hans and Delabre, Bernard and Deiries, Sebastian and D'Odorico, Valentina and Di Marcantonio, Paolo and Figueira, Pedro and Fragoso, Ana and Genolet, Ludovic and Genoni, Matteo and Génova Santos, Ricardo and Hara, Nathan and Hughes, Ian and Iwert, Olaf and Kerber, Florian and Knudstrup, Jens and Landoni, Marco and Lavie, Baptiste and Lizon, Jean-Louis and Lendl, Monika and Lo Curto, Gaspare and Maire, Charles and Manescau, Antonio and Martins, C. J. A. P. and Mégevand, Denis and Mehner, Andrea and Micela, Giusi and Modigliani, Andrea and Molaro, Paolo and Monteiro, Manuel and Monteiro, Mario and Moschetti, Manuele and Müller, Eric and Nunes, Nelson and Oggioni, Luca and Oliveira, António and Pariani, Giorgio and Pasquini, Luca and Poretti, Ennio and Rasilla, José Luis and Redaelli, Edoardo and Riva, Marco and Santana Tschudi, Samuel and Santin, Paolo and Santos, Pedro and Segovia Milla, Alex and Seidel, Julia V. and Sosnowska, Danuta and Sozzetti, Alessandro and Spanò, Paolo and Suárez Mascareño, Alejandro and Tabernero, Hugo and Tenegi, Fabio and Udry, Stéphane and Zanutta, Alessio and Zerbi, Filippo},
	month = apr,
	year = {2020},
	note = {ADS Bibcode: 2020Natur.580..597E},
	pages = {597--601},
}

@article{pepe_espresso_2021,
	title = {{ESPRESSO} at {VLT}. {On}-sky performance and first results},
	volume = {645},
	issn = {0004-6361},
	url = {https://ui.adsabs.harvard.edu/abs/2021A&A...645A..96P/abstract},
	doi = {10.1051/0004-6361/202038306},
	language = {en},
	urldate = {2023-03-30},
	journal = {Astronomy and Astrophysics},
	author = {Pepe, F. and Cristiani, S. and Rebolo, R. and Santos, N. C. and Dekker, H. and Cabral, A. and Di Marcantonio, P. and Figueira, P. and Lo Curto, G. and Lovis, C. and Mayor, M. and Mégevand, D. and Molaro, P. and Riva, M. and Zapatero Osorio, M. R. and Amate, M. and Manescau, A. and Pasquini, L. and Zerbi, F. M. and Adibekyan, V. and Abreu, M. and Affolter, M. and Alibert, Y. and Aliverti, M. and Allart, R. and Allende Prieto, C. and Álvarez, D. and Alves, D. and Avila, G. and Baldini, V. and Bandy, T. and Barros, S. C. C. and Benz, W. and Bianco, A. and Borsa, F. and Bourrier, V. and Bouchy, F. and Broeg, C. and Calderone, G. and Cirami, R. and Coelho, J. and Conconi, P. and Coretti, I. and Cumani, C. and Cupani, G. and D'Odorico, V. and Damasso, M. and Deiries, S. and Delabre, B. and Demangeon, O. D. S. and Dumusque, X. and Ehrenreich, D. and Faria, J. P. and Fragoso, A. and Genolet, L. and Genoni, M. and Génova Santos, R. and González Hernández, J. I. and Hughes, I. and Iwert, O. and Kerber, F. and Knudstrup, J. and Landoni, M. and Lavie, B. and Lillo-Box, J. and Lizon, J.-L. and Maire, C. and Martins, C. J. a. P. and Mehner, A. and Micela, G. and Modigliani, A. and Monteiro, M. A. and Monteiro, M. J. P. F. G. and Moschetti, M. and Murphy, M. T. and Nunes, N. and Oggioni, L. and Oliveira, A. and Oshagh, M. and Pallé, E. and Pariani, G. and Poretti, E. and Rasilla, J. L. and Rebordão, J. and Redaelli, E. M. and Santana Tschudi, S. and Santin, P. and Santos, P. and Ségransan, D. and Schmidt, T. M. and Segovia, A. and Sosnowska, D. and Sozzetti, A. and Sousa, S. G. and Spanò, P. and Suárez Mascareño, A. and Tabernero, H. and Tenegi, F. and Udry, S. and Zanutta, A.},
	month = jan,
	year = {2021},
	pages = {A96},
}

@article{cegla_rossiter-mclaughlin_2016,
	title = {The {Rossiter}-{McLaughlin} effect reloaded: {Probing} the {3D} spin-orbit geometry, differential stellar rotation, and the spatially-resolved stellar spectrum of star-planet systems},
	volume = {588},
	issn = {0004-6361},
	shorttitle = {The {Rossiter}-{McLaughlin} effect reloaded},
	url = {https://ui.adsabs.harvard.edu/abs/2016A&A...588A.127C},
	doi = {10.1051/0004-6361/201527794},
	urldate = {2024-01-27},
	journal = {Astronomy and Astrophysics},
	author = {Cegla, H. M. and Lovis, C. and Bourrier, V. and Beeck, B. and Watson, C. A. and Pepe, F.},
	month = apr,
	year = {2016},
	note = {ADS Bibcode: 2016A\&A...588A.127C},
	pages = {A127},
}

@misc{damasceno_atmospheric_2024,
	title = {The atmospheric composition of the ultra-hot {Jupiter} {WASP}-178 b observed with {ESPRESSO}},
	url = {http://arxiv.org/abs/2406.08348},
	urldate = {2024-06-13},
	publisher = {arXiv},
	author = {Damasceno, Y. C. and Seidel, J. V. and Prinoth, B. and Psaridi, A. and Esparza-Borges, E. and Stangret, M. and Santos, N. C. and Zapatero-Osorio, M. R. and Alibert, Y. and Allart, R. and Silva, T. Azevedo and Cointepas, M. and Silva, A. R. Costa and Cristo, E. and Di Marcantonio, P. and Ehrenreich, D. and Hernández, J. I. González and Herrero-Cisneros, E. and Lendl, M. and Lillo-Box, J. and Martins, C. J. A. P. and Micela, G. and Pallé, E. and Sousa, S. G. and Steiner, M. and Vaulato, V. and Zhao, Y. and Pepe, F.},
	month = jun,
	year = {2024},
	note = {arXiv:2406.08348 [astro-ph]},
}

@article{delrez_wasp-121_2016,
	title = {{WASP}-121 b: a hot {Jupiter} close to tidal disruption transiting an active {F} star},
	volume = {458},
	issn = {0035-8711},
	shorttitle = {{WASP}-121 b},
	url = {https://ui.adsabs.harvard.edu/abs/2016MNRAS.458.4025D},
	doi = {10.1093/mnras/stw522},
	urldate = {2024-05-30},
	journal = {Monthly Notices of the Royal Astronomical Society},
	publisher = {OUP},
	author = {Delrez, L. and Santerne, A. and Almenara, J. -M. and Anderson, D. R. and Collier-Cameron, A. and Díaz, R. F. and Gillon, M. and Hellier, C. and Jehin, E. and Lendl, M. and Maxted, P. F. L. and Neveu-VanMalle, M. and Pepe, F. and Pollacco, D. and Queloz, D. and Ségransan, D. and Smalley, B. and Smith, A. M. S. and Triaud, A. H. M. J. and Udry, S. and Van Grootel, V. and West, R. G.},
	month = jun,
	year = {2016},
	note = {ADS Bibcode: 2016MNRAS.458.4025D},
	pages = {4025--4043},
}

@misc{maguire_high-resolution_2024,
	title = {High-resolution atmospheric retrievals of {WASP}-76b transmission spectroscopy with {ESPRESSO}: {Monitoring} limb asymmetries across multiple transits},
	shorttitle = {High-resolution atmospheric retrievals of {WASP}-76b transmission spectroscopy with {ESPRESSO}},
	url = {http://arxiv.org/abs/2404.10463},
	doi = {10.48550/arXiv.2404.10463},
	urldate = {2024-04-23},
	publisher = {arXiv},
	author = {Maguire, Cathal and Gibson, Neale P. and Nugroho, Stevanus K. and Fortune, Mark and Ramkumar, Swaetha and Gandhi, Siddharth and de Mooij, Ernst},
	month = apr,
	year = {2024},
	note = {arXiv:2404.10463 [astro-ph]},
}

@misc{nortmann_crires_2024,
	title = {{CRIRES}\${\textasciicircum}+\$ transmission spectroscopy of {WASP}-127b. {Detection} of the resolved signatures of a supersonic equatorial jet and cool poles in a hot planet},
	url = {http://arxiv.org/abs/2404.12363},
	doi = {10.48550/arXiv.2404.12363},
	urldate = {2024-04-23},
	publisher = {arXiv},
	author = {Nortmann, L. and Lesjak, F. and Yan, F. and Cont, D. and Czesla, S. and Lavail, A. and Rains, A. D. and Nagel, E. and Boldt-Christmas, L. and Hatzes, A. and Reiners, A. and Piskunov, N. and Kochukhov, O. and Heiter, U. and Shulyak, D. and Rengel, M. and Seemann, U.},
	month = apr,
	year = {2024},
	note = {arXiv:2404.12363 [astro-ph]},
}

@article{prinoth_atlas_2024,
	title = {An atlas of resolved spectral features in the transmission spectrum of {WASP}-189 b with {MAROON}-{X}},
	issn = {0004-6361, 1432-0746},
	url = {http://arxiv.org/abs/2403.08863},
	doi = {10.1051/0004-6361/202349125},
	urldate = {2024-03-17},
	journal = {Astronomy \& Astrophysics},
	author = {Prinoth, B. and Hoeijmakers, H. J. and Morris, B. M. and Lam, M. and Kitzmann, D. and Sedaghati, E. and Seidel, J. V. and Lee, E. K. H. and Thorsbro, B. and Borsato, N. W. and Damasceno, Y. C. and Pelletier, S. and Seifahrt, A.},
	month = feb,
	year = {2024},
	note = {arXiv:2403.08863 [astro-ph]},
}

@article{maguire_high-resolution_2022,
	title = {High-resolution atmospheric retrievals of {WASP}-121b transmission spectroscopy with {ESPRESSO}: {Consistent} relative abundance constraints across multiple epochs and instruments},
	volume = {519},
	issn = {0035-8711, 1365-2966},
	shorttitle = {High-resolution atmospheric retrievals of {WASP}-121b transmission spectroscopy with {ESPRESSO}},
	url = {http://arxiv.org/abs/2211.09621},
	doi = {10.1093/mnras/stac3388},
	number = {1},
	urldate = {2024-01-22},
	journal = {Monthly Notices of the Royal Astronomical Society},
	author = {Maguire, Cathal and Gibson, Neale P. and Nugroho, Stevanus K. and Ramkumar, Swaetha and Fortune, Mark and Merritt, Stephanie R. and de Mooij, Ernst},
	month = dec,
	year = {2022},
	note = {arXiv:2211.09621 [astro-ph]},
	pages = {1030--1048},
}

@article{heng_influence_2012,
	title = {The {Influence} of {Atmospheric} {Scattering} and {Absorption} on {Ohmic} {Dissipation} in {Hot} {Jupiters}},
	volume = {748},
	issn = {0004-637X},
	url = {https://ui.adsabs.harvard.edu/abs/2012ApJ...748L..17H},
	doi = {10.1088/2041-8205/748/1/L17},
	urldate = {2024-01-20},
	journal = {The Astrophysical Journal},
	author = {Heng, Kevin},
	month = mar,
	year = {2012},
	note = {ADS Bibcode: 2012ApJ...748L..17H},
	pages = {L17},
}

@article{fromang_shear-driven_2016,
	title = {Shear-driven instabilities and shocks in the atmospheres of hot {Jupiters}},
	volume = {591},
	issn = {0004-6361},
	url = {https://ui.adsabs.harvard.edu/abs/2016A&A...591A.144F},
	doi = {10.1051/0004-6361/201527600},
	urldate = {2024-01-20},
	journal = {Astronomy and Astrophysics},
	author = {Fromang, Sébastien and Leconte, Jeremy and Heng, Kevin},
	month = jul,
	year = {2016},
	note = {ADS Bibcode: 2016A\&A...591A.144F},
	pages = {A144},
}

@article{borsa_atmospheric_2021,
	title = {Atmospheric {Rossiter}-{McLaughlin} effect and transmission spectroscopy of {WASP}-121b with {ESPRESSO}},
	volume = {645},
	issn = {0004-6361, 1432-0746},
	url = {http://arxiv.org/abs/2011.01245},
	doi = {10.1051/0004-6361/202039344},
	urldate = {2023-12-20},
	journal = {Astronomy \& Astrophysics},
	author = {Borsa, F. and Allart, R. and Casasayas-Barris, N. and Tabernero, H. and Osorio, M. R. Zapatero and Cristiani, S. and Pepe, F. and Rebolo, R. and Santos, N. C. and Adibekyan, V. and Bourrier, V. and Demangeon, O. D. S. and Ehrenreich, D. and Pallé, E. and Sousa, S. and Lillo-Box, J. and Lovis, C. and Micela, G. and Oshagh, M. and Poretti, E. and Sozzetti, A. and Prieto, C. Allende and Alibert, Y. and Amate, M. and Benz, W. and Bouchy, F. and Cabral, A. and Dekker, H. and D'Odorico, V. and Di Marcantonio, P. and Figueira, P. and Santos, R. Genova and Hernández, J. I. González and Curto, G. Lo and Manescau, A. and Martins, C. J. A. P. and Mégevand, D. and Mehner, A. and Molaro, P. and Nunes, N. and Riva, M. and Mascareño, A. Suárez and Udry, S. and Zerbi, F.},
	month = jan,
	year = {2021},
	note = {arXiv:2011.01245 [astro-ph]},
	pages = {A24},
}

@misc{yang_high-resolution_2023,
	title = {High-resolution transmission spectroscopy of ultra-hot {Jupiter} {WASP}-33b with {NEID}},
	url = {http://arxiv.org/abs/2312.02413},
	doi = {10.48550/arXiv.2312.02413},
	urldate = {2023-12-07},
	publisher = {arXiv},
	author = {Yang, Yuanheng and Chen, Guo and Wang, Songhu and Yan, Fei},
	month = dec,
	year = {2023},
	note = {arXiv:2312.02413 [astro-ph]},
}

@misc{pagano_constraining_2023,
	title = {Constraining the reflective properties of {WASP}-178b using {Cheops} photometry},
	url = {http://arxiv.org/abs/2309.09037},
	doi = {10.48550/arXiv.2309.09037},
	urldate = {2023-10-05},
	publisher = {arXiv},
	author = {Pagano, I. and Scandariato, G. and Singh, V. and Lendl, M. and Queloz, D. and Simon, A. E. and Sousa, S. G. and Brandeker, A. and Cameron, A. Collier and Sulis, S. and Van Grootel, V. and Wilson, T. G. and Alibert, Y. and Alonso, R. and Anglada, G. and Bárczy, T. and Navascues, D. Barrado and Barros, S. C. C. and Baumjohann, W. and Beck, M. and Beck, T. and Benz, W. and Billot, N. and Bonfils, X. and Borsato, L. and Broeg, C. and Bruno, G. and Carone, L. and Charnoz, S. and van Damme, C. Corral and Csizmadia, Sz and Cubillos, P. E. and Davies, M. B. and Deleuil, M. and Deline, A. and Delrez, L. and Demangeon, O. D. S. and Demory, B.-O. and Ehrenreich, D. and Erikson, A. and Fortier, A. and Fossati, L. and Fridlund, M. and Gandolfi, D. and Gillon, M. and Güdel, M. and Günther, M. N. and Helling, Ch and Hoyer, S. and Isaak, K. G. and Kiss, L. L. and Kopp, E. and Lam, K. W. F. and Laskar, J. and Etangs, A. Lecavelier des and Magrin, D. and Maxted, P. F. L. and Mordasini, C. and Munari, M. and Nascimbeni, V. and Olofsson, G. and Ottensamer, R. and Pallé, E. and Peter, G. and Piotto, G. and Pollacco, D. and Ragazzoni, R. and Rando, N. and Rauer, H. and Reimers, C. and Ribas, I. and Rieder, M. and Santos, N. C. and Ségransan, D. and Smith, A. M. S. and Stalport, M. and Steller, M. and Szabó, Gy M. and Thomas, N. and Udry, S. and Venturini, J. and Walton, N. A.},
	month = sep,
	year = {2023},
	note = {arXiv:2309.09037 [astro-ph]},
}

\appendix

\section{Data preparation}
\label{app:data_prep}

In Figure \ref{fig:data_overview} we show all considered targets as a function of their equilibrium temperature which we have re-calculated for homogenity assuming zero albedo instead of relying on reported values that include dayside measurements. The overview of the data can be found in Table \ref{tab:observation_log}. All masses stem from the respective discovery papers and their RV analysis, except KELT-20~b and HAT-P-70~b which stem from the transit observation scaling in \citep{gandhi_retrieval_2023}. We found no trend in planetary mass and radius, highlighted by the similar value for the gravity in our sample (see Figure \ref{fig:data_overview}). Nonetheless, we report the values for the planet with an iron signature in Table \ref{tab:mass}. The rejected targets (mostly due to a lack of iron signature) show that the window of observable iron signatures lies in the equilibrium temperature window identified by \cite{koll_atmospheric_2018} where the dissipation of kinetic energy in the atmosphere is likely dominated by Ohmic dissipation ($T_{\text{eq}}>1400~K$). We specify in the following which targets were deemed suitable for the survey and the individual rejection reasons per targets and dataset.

\begin{figure}
   \centering
    \includegraphics[width=\linewidth]{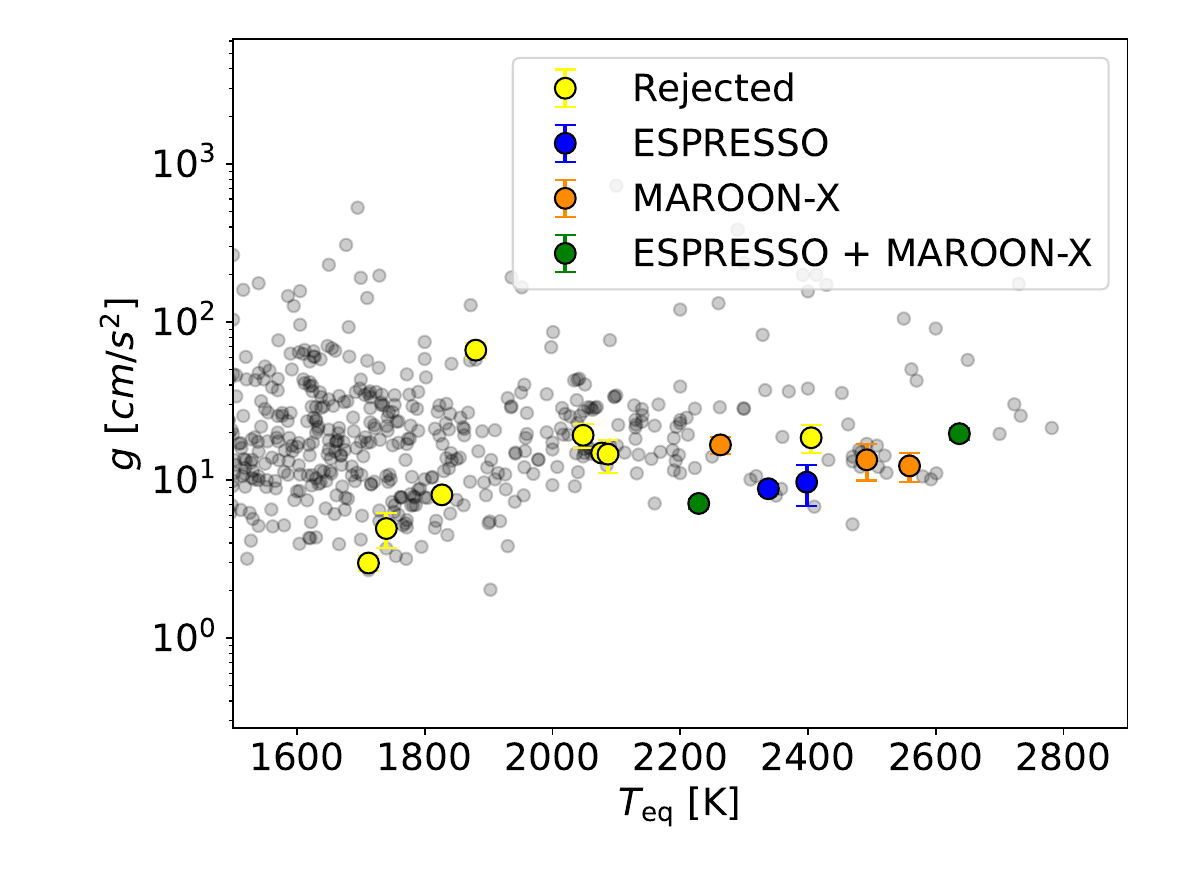}
   \caption{Exoplanet population overview with the studied targets highlighted in the temperature vs gravity space. All selected targets are Jupiter-like targets with resolved iron detections from 8m-class telescopes. The background targets in transparent black are all available entries from the NASA exoplanet archive with constrained masses and radii (accessed 27-May-2025). The errorbars represent the propagated uncertainties of measured mass and radius.}
    \label{fig:data_overview}
\end{figure}

\begin{table*}[h!]
	\small
	\begin{center}
		\begin{tabular}{llllllll}
			\toprule
			{Planet }& {Date}  & {Instrument}    & {PI, \#}     & {S/N$^a$} & {T$_{eq}$ [K]} & {Atmospheric signature published in$^b$} \\
             \midrule
			KELT-11  b*       & 2024-03-31	& MAROON-X     & Parmentier, GN-2024A-Q-130         & 22.2-302.5  & 1712 & \\	
                             & 2024-04-19	& MAROON-X     & Parmentier, GN-2024A-Q-130         & 22.2-297.6  & \\
			\midrule
			\\[-0.95em] 
			WASP-172 b*		 & 2022-06-01 & ESPRESSO & Albrecht, 109.22Z4.006 &  14.3-63.3 & 1740 & \cite{seidel_detection_2023-1} \\       
             \midrule
			KELT-4 A b* & 2023-12-07	& MAROON-X  & Parmentier, GN-2023B-Q-127    &  28.0-82.7  & 1827 & \\	
            & 2023-12-10	& MAROON-X     & Parmentier, GN-2023B-Q-127         &  26.4-77.6   &  & \\
            & 2023-12-16	& MAROON-X     & Parmentier, GN-2023B-Q-127         &  24.7-76.3    &  & \\
            & 2023-12-31	& MAROON-X     & Parmentier, GN-2023B-Q-127         &   -   &  & \\
		    \midrule
			WASP-173A b* &  2022-07-24 & ESPRESSO & Albrecht, 109.22Z4.003  & 12.4-54.9  & 1880 & \\
	
             \midrule
			KELT-7 b* & 2023-12-28	& MAROON-X     & Parmentier, GN-2023B-Q-127   &  22.2-158.1 & 2048 & \\
			\midrule
			WASP-19 b* 			& 2019-01-14 & ESPRESSO & Sedaghati, 0102.C-0311 &  13.2-60.7 & 2077 & \cite{sedaghati_spectral_2021}\\
		      & 2019-03-03 & ESPRESSO & Sedaghati, 0102.C-0311 &  11.8-56.1 && \cite{sedaghati_spectral_2021}\\
		      & 2019-03-22 & ESPRESSO & Sedaghati, 0102.C-0311 &  13.0-66.3 && \cite{sedaghati_spectral_2021}\\
			& 2020-01-11 & ESPRESSO & Sedaghati, 0102.C-0311 &   6.8-48.9 && \cite{sedaghati_spectral_2021}\\
			\midrule
			KELT-17 b*	& 2020-12-25	& ESPRESSO & Seidel, 0106.C-0126   & 22.7.107.2 & 2087  & \\
							& 2021-01-31	& ESPRESSO & Seidel, 0106.C-0126 &  32.0-134.9 && \\
								& 2021-03-06	& ESPRESSO & Seidel, 0106.C-0126  & 25.5-111.4 & & \\
			\midrule
			WASP-76 b&2018-09-02   & ESPRESSO & Pepe, 1102.C-0744       & 13.1-83.7 &2229& \cite{ehrenreich_nightside_2020}\\ 
			&2018-10-30   & ESPRESSO & Pepe, 1102.C-0744                 & 10.5-67.5&  & \cite{ehrenreich_nightside_2020}\\
            &2019-10-18   & ESPRESSO & Gibson, 0104.C-0642              & 29-55 && \cite{maguire_high-resolution_2024}\\
            &2020-09-03   & MAROON-X & Pelletier, GN-2020B-Q-122     &  20.0-68.4 && \cite{pelletier_vanadium_2023}\\
            &2020-09-12   & MAROON-X & Pelletier, GN-2020B-Q-122     &  17.6-63.1 && \cite{pelletier_vanadium_2023}\\
            &2021-10-28   & MAROON-X & Debras, GN-2021B-Q-138        &  29.4-99.0 && \cite{pelletier_vanadium_2023}\\
            \midrule
            KELT-20 b  & 2023-07-07 & MAROON-X &  Parmentier, GN-2023A-Q-224 & 71.3-190.6 & 2263 & de Lia et al. in prep.\\
			\midrule
			WASP-121 b & 2021-01-26	 & ESPRESSO & Gibson,  0106.C-0516  &  27-37 & 2338 & \cite{maguire_high-resolution_2022}\\
			& 2021-03-04 & ESPRESSO & Gibson,  0106.C-0516 & 21-36  && \cite{maguire_high-resolution_2022}\\
           & 2019-01-06 & ESPRESSO & Pepe, 1102.C-0744 &  21-36  && \cite{borsa_atmospheric_2021} \\
           & 2018-11-30 & ESPR-4UT & 60.A-9128 (Comm.) &   23.1-163.3  & 2358 & \cite{borsa_atmospheric_2021} \\
           & 2023-09-23 & ESPR-4UT & Seidel, 111.24J8 &   26.6-174.5 && \cite{seidel_vertical_2025}\\
			\midrule
			MASCARA-4 b*	  & 2020-02-12 & ESPRESSO & Wyttenbach, 0104.C-0605 300 & 44.9-201.7 & 2405& \cite{zhang_transmission_2022}\\
			& 2020-02-29 & ESPRESSO & Wyttenbach, 0104.C-0605  &  46.3-207.6 && \cite{zhang_transmission_2022}\\
            \midrule
            WASP-178 b & 2021-05-03 & ESPRESSO & Pepe, 1104.C-0350 &   11.3-48.6  & 2398 & \cite{damasceno_atmospheric_2024}\\
            & 2021-07-09 & ESPRESSO & Pepe, 1104.C-0350 & 9.3-37.7 &  & \cite{damasceno_atmospheric_2024}\\
             \midrule
			TOI-1518 b & 2022-08-13	& MAROON-X     & Parmentier, GN-2022B-Q-128     &23.4-64.9 &  2491 & \cite{simonnin_time_2025}\\
             & 2023-10-19	                & MAROON-X     & Parmentier, GN-2023B-Q-127     &  27.4-80.2 &  &   \cite{simonnin_time_2025}   \\
              & 2024-06-26	                & MAROON-X     & Parmentier, GN-2024A-Q-130   &  24.0-65.1 &  &  \cite{simonnin_time_2025}    \\ 
			\midrule
           HAT-P-70 b  &  2023-12-13 & 	MAROON-X & Parmentier, GN-2022B-Q-128 &  22.2-64.8 & 2559  & \\
                     & & & Pelletier, GN-2022B-Q-127 & &  & \\
           &  2023-12-24 & 	MAROON-X & Parmentier, GN-2022B-Q-128 &  22.2-72.5 & \\
                     & & & Pelletier, GN-2022B-Q-127 & & & \\
			\midrule
			WASP-189 b &2021-06-04   & ESPRESSO  & Prinoth, 107.22QF     &  107.7-432.6 & 2638 & \cite{prinoth_time-resolved_2023}\\
             &2022-04-03  & MAROON-X  & Pelletier, GN-2022A-FT-208      & 84.1-231.6 && \cite{prinoth_atlas_2024}\\ 
             &2022-06-02  & MAROON-X  & Pelletier, GN-2022A-FT-208     & 84.7-233.5 && \cite{prinoth_atlas_2024}\\ 
          
			\bottomrule
		\end{tabular}
	\end{center}
	\vspace{-1.2em}
	\captionof{table}{Overview of the observations. Stars mark rejected datasets. $^{(a)}$ Minimum and maximum SNR value. $^{(b)}$ First publication of the exoplanet's atmosphere from the specific dataset studied here, if existing.}
	\label{tab:observation_log}
\end{table*}

\begin{table}[ht]
\centering
\begin{tabular}{llcc}
\toprule
Planet         & Mass [$M_\mathrm{Jup}$]  & Radius [$R_\mathrm{Jup}$] & Ref. \\
\midrule
\hline
WASP-76b & $0.92 \pm 0.03$ & $1.83 \pm 0.06$ & (1)\\                    
\hline
KELT-20b & $2.16 \pm 0.21$ & $1.83 \pm 0.07$ & (2)\\
\hline
WASP-121b & $1.18 \pm 0.07$ & $1.87 \pm 0.04$ & (3)\\
\hline
WASP-178b & $1.41 \pm 0.40$ & $1.94 \pm 0.06$ & (4)\\
\hline
TOI-1518b & $1.83 \pm 0.47$ & $1.88 \pm 0.04$ & (5)\\
\hline
HAT-P-70b & $1.66 \pm 0.22$ & $1.87 \pm 0.15$ & (2)\\
\hline
WASP-189b & $1.99 \pm 0.16$ & $1.62 \pm 0.02$ & (6)\\            
\bottomrule
\end{tabular}

	\captionof{table}{Masses and radii of the planets in our survey. \footnotesize (1) \cite{west_three_2016} (2) \cite{gandhi_retrieval_2023} (3) \cite{delrez_wasp-121_2016} (4) \cite{martinez_kelt-25b_2019} (5) \cite{simonnin_time_2025} (6) \cite{lendl_hot_2020}}
\label{tab:mass}
\end{table}

\subsection{Rejected planets from the survey}

KELT-4~b, KELT-7~b, and MASCARA-4~b are orbiting pulsating host stars \citep{eastman_kelt-4ab_2016, stangret_high-resolution_2022, zhang_transmission_2022}. The proper extraction of atmospheric signatures for systems around pulsating host stars is an ongoing investigation within the field and not suitable as of yet for a robust population study. Three other planets in our sample did not show a detectable iron signature. WASP-173~b is an extremely heavy target leading to a severely reduced signal which can explain the non-detection, while KELT-11~b and WASP-172~b did not show a significant enough iron feature to be included in the survey. KELT-17 is a chemically peculiar Am star, making the recovery of any planetary signal extremely challenging and - if achieved - prone to extreme stellar contamination \citep{saffe_kelt-17_2020}. We excluded the observations of WASP-19~b entirely due to low S/N across all four nights and encourage additional observations of this object. Similarly, we discarded the night of 18 Oct 2019 for WASP-76~b for the same reason. For WASP-121~b, we only used the data taken in 4-UT mode, as its superior photon-collecting power dominates the observed signal (see S/N for WASP-121~b in Table~\ref{tab:observation_log}). The constraint of the resolved iron detections naturally downsizes to a sample to ultra-hot Jupiters only, where iron is present in the atmosphere in atomic form \citep{tan_atmospheric_2019, bell_increased_2018}.

Two planets which are not included in our survey but technically classified as ultra-hot Jupiters are WASP-33~b and KELT-9~b. For WASP-33~b, PEPSI data exist which report a wind speed measurement \citep{cauley_time-resolved_2021}, however, WASP-33 is a pulsating host star as evidenced by the clear pulsation pattern of the host star in \citep{yang_high-resolution_2023}. In the mentioned paper the impact of the pulsating host star is reflected in the excessively large uncertainty, making WASP-33~b unsuitable for a robust population survey. 
KELT-9~b, as a northern target, unfortunately only has transit observations available with HARPS-N \citep{darpa_gaps_2024} as the large majority of KELT-9~b observations are in emission, unsuitable for our survey. Additionally, with an equilibrium temperature of more than 1000~K above even the hottest ultra-hot Jupiter in our survey \citep{borsa_gaps_2019}, KELT-9~b represents a class of its own and should be treated as such. Higher order effects mentioned in the main paper likely impact KELT-9~b which make any wind speeds measured for this planet non representative of the derived trend. 

\subsection{Data cleaning steps}

The ESPRESSO observations were reduced using the dedicated ESPRESSO pipeline (ESPR v3.2.0) via \texttt{esoreflex} (v2.11.5), provided by ESO and the ESPRESSO consortium. For our analysis, we used the non-blaze-corrected, two-dimensional order-by-order spectra from fibre A (S2D) for cross-correlation, and the one-dimensional flux-calibrated spectra from fibre A (S1D) for telluric correction.

The MAROON-X observations were reduced using the standard pipeline described in \cite{seifahrt_-sky_2020}, which extracts wavelength-calibrated, two-dimensional spectra order-by-order from each exposure in the time series, separately for the blue and red arms. Due to differences in exposure times, we treated the two arms independently. As the pipeline does not output stitched one-dimensional spectra (analogous to ESPRESSO’s S1D), we manually combined the orders to create a single spectrum for telluric correction. 

We corrected for telluric contamination in all datasets using the standalone version of \texttt{molecfit} (v1.5.9; \citealt{smette_molecfit_2015,kausch_molecfit_2015}). We fitted the telluric model to each exposure individually, using spectral regions with strong telluric \ch{H2O} and \ch{O2} absorption near the sodium doublet at 590, and around 630, and 650~\si{\nm}. The resulting models were interpolated onto the corresponding wavelength grids and divided out.

\subsection{Cross-correlation}
\label{sec:crosscorr}
We searched for atomic absorption by neutral iron using the cross-correlation technique \citep{ snellen_orbital_2010} as implemented in \texttt{tayph} \citep[see e.g.][]{hoeijmakers_hot_2020,prinoth_titanium_2022,borsato_mantis_2023,hoeijmakers_mantis_2024}.

After pipeline reduction and telluric correction, we shifted the spectra to the stellar rest frame, correcting for the barycentric velocity\footnote{Note that this is only needed for MAROON-X observations, as the ESPRESSO pipeline provides barycentre-velocity-corrected spectra} and the stellar reflex motion due to the orbiting planet. Following \cite{hoeijmakers_hot_2020}, we removed outliers via an order-by-order sigma-clipping algorithm, using a running median absolute deviation over 40-pixel sections of the time series and rejecting $5\sigma$ outliers. Additionally, we manually masked residual telluric contamination where the flux dropped by 50\% or more, primarily in the saturated \ch{O2} band where \texttt{molecfit} corrections are inadequate. This affected on average 8.5\% of the pixels for each of the observations, and at most 17.5\%. We rejected individual exposures on each night if they showed significant deviations in S/N compared to the rest, or if the S/N was generally too low (approx S/N $\leq$ 20) -- i.e., entering the red noise regime instead of being limited by photon-noise.

Before cross-correlation, we corrected for the spectrograph response via a colour correction: we divided each order by the mean out-of-transit spectrum to assess wavelength-dependent differences, fit the residual with a third-order polynomial, and divided this fit from the original order. Note that we do not divide out the mean out-of-transit spectrum at this stage, but rather cross-correlate first and then divide the mean out-of-transit cross-correlation function. In the limit of independent stellar and planetary spectra, the order of these operations is interchangeable.

We then cross-correlated the corrected 2D spectra with a neutral iron template from \cite{kitzmann_mantis_2023} at \SI{3000}{\kelvin}, appropriate for the range of limb temperatures in our sample. The templates assume an isothermal atmosphere in hydrostatic and chemical equilibrium and were broadened to the approximate line-spread functions of the instruments: \num{2.14} and \SI{3.53}{\km\per\second} for ESPRESSO (1UT-mode HR) and MAROON-X, respectively. For WASP-121~b, some of the observations were taken in ESPRESSO 4UT MR mode (see Table~\ref{tab:observation_log}), and the templates were instead broadened to \SI{4.28}{\km\per\second}. For the planets of the survey where iron was found we provide the K$_p$-V$_{\rm sys}$ diagrams in Figure \ref{fig:kpvsys}.

\begin{figure*}
    \centering
    \includegraphics[width=0.85\linewidth, trim=0 0 0 0, clip]{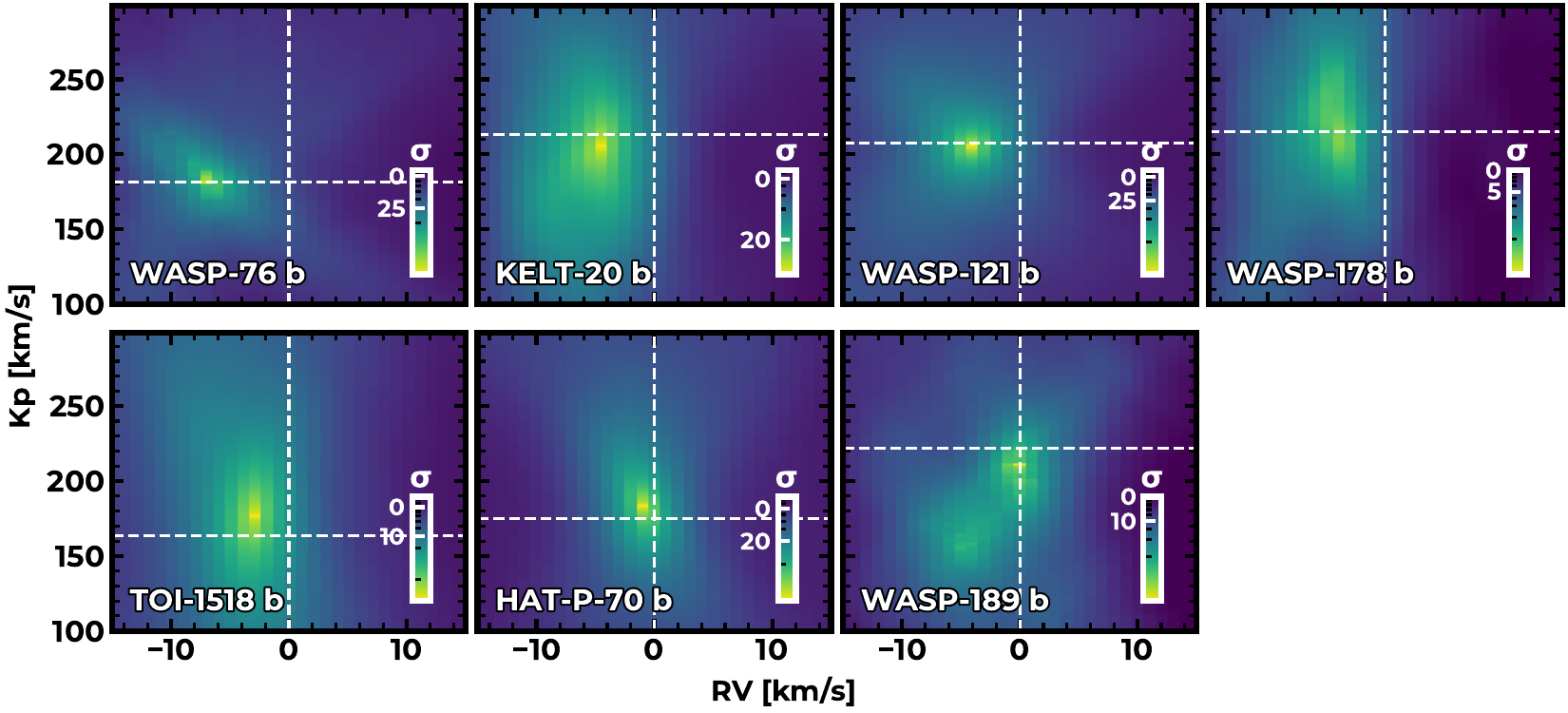}
   \caption{K${\rm p}$–v${\rm sys}$ diagrams showing detected Fe I absorption in the sample. Each panel displays the stacked K${\rm p}$–v${\rm sys}$ diagram after correcting individual observations for their system velocities. The colour bars indicates the detection significance, calculated by dividing the diagram by its standard deviation outside the velocity ranges affected by the star, planet, Rossiter–McLaughlin effect, and tellurics.}
    \label{fig:kpvsys}
\end{figure*}

\subsection{Rossiter-McLaughlin fit}
We corrected for the Rossiter-McLaughlin effect (Doppler shadow) using \texttt{StarRotator} \citep[see][]{prinoth_time-resolved_2023}, following the same approach as described in \cite{prinoth_titanium_2025}.

Briefly, we assumed that the stellar line is well-approximated by a Gaussian profile, implying that the stellar cross-correlation function also adopts a Gaussian shape. We used a grid size of 25 pixels in \texttt{StarRotator}, resulting in a total of 50 $\times$ 50 grid cells. For each cell, we computed a Gaussian cross-correlation function as
\begin{equation}
    \text{CCF}_{i}(v) = 1 - A \exp \left(- \frac{(v_i + v_{\rm sys} - v)^2}{2\sigma^2} \right),
\end{equation}
where $v_i$ is the velocity of cell $i$ due to stellar rotation (assuming no differential rotation in the vertical direction; see \citealt{cegla_rossiter-mclaughlin_2016,prinoth_time-resolved_2023}), $A$ is the Gaussian amplitude (ranging from 0 for no absorption to 1 for saturation), and $\sigma$ is the Gaussian width, determined by the instrumental profile.
We accounted for limb darkening using a quadratic law, adopting coefficients from the large survey in \cite{patel_empirical_2022}. If unavailable, we adopted limb-darkening coefficients from \cite{eastman_kelt-4ab_2016, saha_precise_2023, pagano_constraining_2023}.

\subsection{Measuring the system velocity}
\label{sec:vsys}
The apparent system velocity (the movement of the barycenter of the observed system with respect to the Solar System barycenter) varies slightly between spectrographs due to the lack of absolute wavelength calibration. This results in a constant offset in the wavelength solution, which we correct by independently measuring the system velocity for each observation using the out-of-transit cross-correlation functions. This also accounts for velocity offsets introduced by, for example, instrument interventions as described for ESPRESSO in \cite{ehrenreich_nightside_2020}.

We determined the system velocity for each dataset by fitting a rotationally broadened Voigt profile \citep{gray_observation_2008} to the average out-of-transit cross-correlation functions, obtained by cross-correlating the processed spectra with a suitable PHOENIX model of the host star \citep{husser_new_2013}. We have opted for the more complex profile in comparison to a Gaussian due to the better results for fast rotating stars with blended stellar lines ($vsini > 80~km/s$). Prior to cross-correlation, the spectra were corrected for the stellar reflex motion induced by the orbiting planet and for the barycentric velocity (if not already handled by the reduction pipeline) and cleaned as described in Section \ref{sec:crosscorr}. The fits were performed using \texttt{pymultinest} \citep{buchner_pymultinest_2016}. The results, as well as literature values for reference, are shown in Table \ref{tab:vsys}. The uncertainty of our results in the table are solely the fitting error and do not take into account that the system velocity cannot be estimated better than the $\sim 100$~m/s level due to the photosphere of the star itself.

\begin{table}[ht]
\centering
\begin{tabular}{llcc}
\toprule
Planet         & Date         & $v_{\mathrm{sys}}$ [km/s] & Instrument \\
\midrule
\hline
\multirow{6}{*}{WASP-76b} & 02 Sep 2018 & $-0.81 \pm 0.01$ & ESPR\\
                          & 30 Oct 2018 & $-0.99 \pm 0.01$ & ESPR\\
                          & 03 Sep 2020 & $-0.75 \pm 0.33$ & M-X\\
                          & 12 Sep 2020 & $-0.78 \pm 0.31$ & M-X\\
                          & 28 Oct 2021 & $-0.72 \pm 0.34$ & M-X\\
                          & ref (1) & $-1.11\pm0.50$ & ESPR\\        
\hline
\multirow{3}{*}{KELT-20b} & 07 Jul 2023 & $-24.63 \pm 0.01$ & M-X\\
                           & ref (2) & $-23.3\pm0.3$ & TRES\\
                           & ref (3) & $-21.3\pm0.3$ & SONG\\
\hline
\multirow{3}{*}{WASP-121b} & 30 Nov 2018 & $38.64 \pm 0.01$ & ESPR-4UT\\
                           & 23 Sep 2023 & $38.61 \pm 0.01$ & ESPR-4UT\\
                           & ref (4) & $38.35\pm0.02$ & CORALIE\\
\hline
\multirow{3}{*}{WASP-178b} & 03 May 2021 & $-23.58 \pm 0.01$ & ESPR\\
                           & 09 Jul 2021 & $-23.56 \pm 0.01$ & ESPR\\
                           & ref (5) & $-23.91\pm0.01$ & CORALIE\\
\hline
\multirow{4}{*}{TOI-1518b} & 13 Aug 2022 & $-11.91 \pm 0.02$ & M-X\\
                           & 19 Oct 2023 & $-11.84 \pm 0.01$ & M-X\\
                           & 26 Jun 2024 & $-11.87 \pm 0.02$ & M-X\\
                           & ref (6) & $-11.74 \pm 0.17$ & SOPHIE\\
\hline
\multirow{3}{*}{HAT-P-70b} & 13 Dec 2023 & $21.93 \pm 0.56$ & M-X\\
                           & 24 Dec 2023 & $21.60 \pm 0.55$ & M-X\\
                           & ref (7) & $25.26\pm0.11$ & TRES\\
\hline
\multirow{3}{*}{WASP-189b} & 04 Jun 2021 & $-20.61 \pm 0.01$ & ESPR\\
                           & 03 Apr 2022 & $-21.41 \pm 1.52$ & M-X\\
                           & 02 Jun 2022 & $-21.81 \pm 1.66$ & M-X\\             
\bottomrule
\end{tabular}

	\captionof{table}{Systemic velocity overview of our derived values and the fitting uncertainties for all observing nights as well as references found in the literature. \footnotesize (1) \cite{ehrenreich_nightside_2020} (2) \cite{lund_kelt-20b_2017} (3) \cite{talens_mascara-2_2018} (4) \cite{delrez_wasp-121_2016} (5) \cite{hellier_wasp-south_2019} (6) \cite{simonnin_time_2025} (7) \cite{zhou_hats-70b_2019}}
\label{tab:vsys}
\end{table}

Overall, we find good agreement between values measured with the same instrument even over year-long timescales. Nonetheless, when velocity precision is key, measuring the velocity in each night is necessary, see e.g. the $0.2$~km/s difference for the two ESPRESSO nights of WASP-76~b. More importantly, different instruments will likely show different systemic velocities as they also include instrumental effects such as drifts and shifts in the "zero" radial velocity due to instrument interventions. As a consequence, measuring the systemic velocity becomes crucial when combining spectra from different spectrographs. Overall, the literature values and our derived values between instruments and nights are within $1\sigma$ with notable exceptions:
For TOI-1518~b, we highlight that the reference value from \cite{simonnin_time_2025} is the value as derived from SOPHIE measurements in their Appendix A, in their Table 2 a higher value is given, likely copied from \cite{cabot_toi-1518b_2021} and a typo in the manuscript. 
TOI-1518, KELT-20, HAT-P-70, WASP-189 all fall, to varying degrees of severity, in the category of fast-rotating host stars which can lead to larger uncertainties. All uncertainties were derived from the $1\sigma$ envelope of the posterior distribution of the Bayesian retrieval. Most notably, the only literature value we were able to find for HAT-P-70~b from \cite{zhou_hats-70b_2019} with the TRES spectrograph at a 1m-class telescope is in disagreement with our observed value. Given the overall agreement of our derived values with the order of magnitude of literature values, we assume that this offset is due to the TRES spectrograph or a difference in methodology. In \cite{zhou_hats-70b_2019} they opted for fitting stellar line profiles derived from a least-squares deconvolution to the data which usually gives acceptable values for the RVs of fast-rotating stars. Given the goal of this study, to derive trends from the population of planets, this difference should have no impact on our results, since we correct our dataset for the measured velocity and additionally use the same method for all datasets. This means we would at most introduce an overall bias on the absolute velocity value but not change the population trend. Given the excellent agreement between literature values and our measurements, we can likely rule out such a potential bias. Nonetheless, to assure that the observed trend does not hinge on an over-confidence in the system velocity measurement we do not use the fit uncertainty to propagate uncertainties to our final velocity measurement, but instead use the variance between the literature values and our observed values as the maximum possible uncertainty (excluding the TrES spectrograph due to the mentioned discrepancies).

\subsection{Measured offset in velocity}
The offset relative to the measured system velocity, which encodes the atmospheric movement, is determined from the stacked K$_p$-V$_{\rm sys}$ diagrams. To generate these diagrams, we scan over a range of possible projected orbital velocities $K_p = v_{\rm orb} \sin{i}$, avoiding assumptions about the true orbital velocity or inclination and preventing bias from their uncertainties. Additionally, the map is corrected for the system velocity as derived for each individual night prior to stacking in K$_p$-V$_{\rm sys}$ space based on S/N per night of observation. This ensures that differences in the system velocity per spectrograph or even per night due to interventions have been accounted for. 

\section{Ohmic dissipation model}
\label{sec:assumptions}

\begin{figure}
    \centering
    \includegraphics[width=1.0\linewidth]{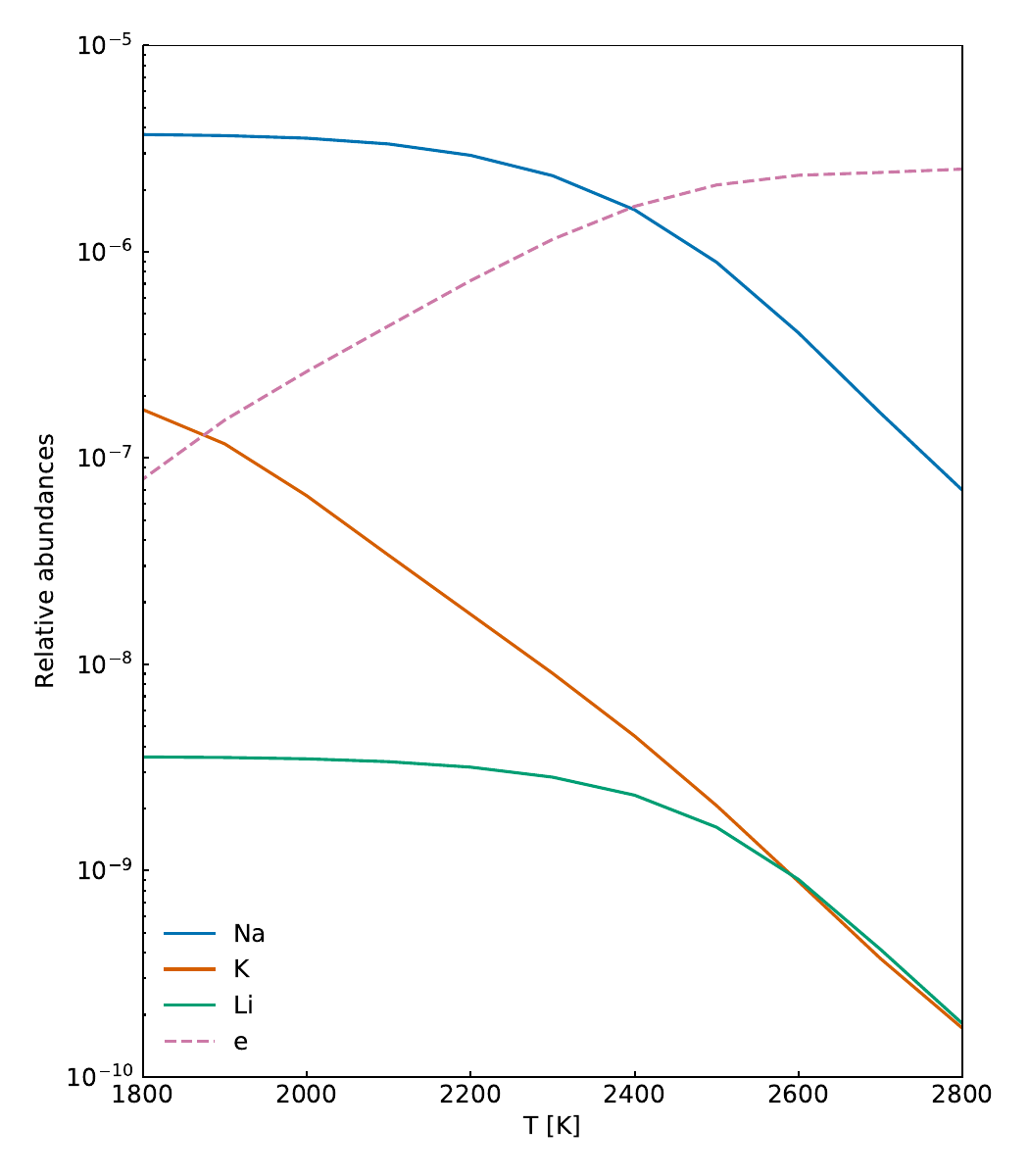}
    \caption{Order of magnitude change in relative electron abundance over the selected temperature range. The relative abundance of the alkali metals as a function of temperature at the mbar pressure level provided by \cite{lupu_correlated_2021} is overplotted with the relative electron abundance (dashed). The underlying atmospheric chemistry is described in \cite{visscher_atmospheric_2010}.}
    \label{fig:alkali_contribution}
\end{figure}

\cite{koll_atmospheric_2018} relies on various assumptions to create their models of which the most important is their treatment of hot Jupiters, compared to the ultra-hot Jupiters in our study. While they state that their work also holds for hotter planets, we will highlight some differences.
Most importantly, the application of the Saha equation in \cite{koll_atmospheric_2018} relies on the assumption that the main contribution to the electron abundance in the atmosphere stems from potassium. However, at the mbar pressure level and for temperatures above $\sim 2000~$K, potassium is fully ionised and the contribution of the partial ionisation of sodium becomes dominant, as well as the first and second valence electron removal from Fe. At these pressure levels, the electric conductivity can, as a consequence, reach values up to $\sim 1 S m^{-1}$ \citep{batygin_evolution_2011, rauscher_three-dimensional_2013, heng_influence_2012}. We have opted to update the literature approach to assume potassium as the main source of free electrons with the simplified Saha equation from \cite{balbus_solar_2000} as applied in e.g. \cite{perna_Ohmic_2010, koll_atmospheric_2018}. Instead, we employ the pre-calculated relative abundance tables from the full Saha equation calculation as provided by \cite{lupu_correlated_2021} for an atmosphere containing all alkali metals, as well as C2H2, C2H4, C2H6, CH4, CO, CO2, CrH, Fe, FeH, H2, H3+, H2O, H2S, HCN, LiCl, LiF, LiH, MgH, N2, NH3, OCS, PH3, SiO, TiO, and VO. Their pre-computed grid provides steps of half orders of magnitude in pressure and 100~K in temperature, sufficient for our purposes. The top panel of Figure \ref{fig:alkali_contribution} shows the overall relative abundance of the alkalis and electrons as a function of temperature at the mbar pressure level, the bottom panel highlights the Na dominant contribution compared to K for ultra-hot Jupiters above 2000~K.

\begin{figure}
    \centering
    \includegraphics[width=1.0\linewidth]{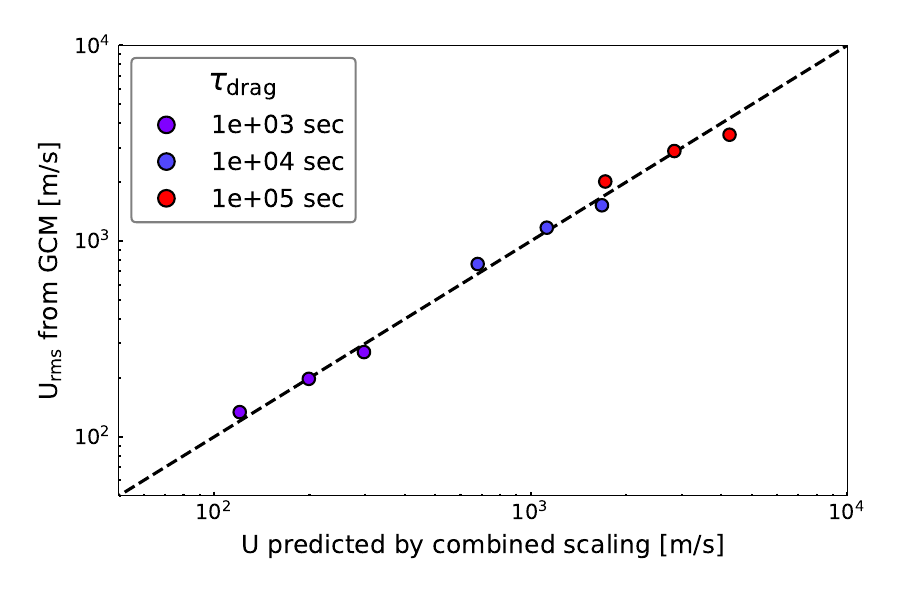}
    \caption{Recreation of the calculation of $k_0$, in line with Figure 4, panel b from \cite{koll_atmospheric_2018}. The y-axis corresponds to the line of sight velocity at the terminator at the mbar pressure level in the GCM, in line with what is probed observationally. The x-axis is the calculated velocity from Equation \ref{eq:rayleigh} with the applied scaling. The black, dashed line indicates 1:1 equality between the GCM simulation and the scaled model. The colours highlight the applied $\tau$ time scales for different temperatures, namely 1400, 1800, and 2200~K.}
    \label{fig:fudgefactor}
\end{figure}

While the application of the Saha equation fundamentally impacts the form of the model via the ionsation fraction as a function of temperature, the scaling factor $k_0$ in Equation \ref{eq:rayleigh} governs the vertical anchoring of the models and thus the absolute magnetic field value since $U \sim k_0/B$. $k_0$ is an important scale factor, as it encodes various model assumptions, most importantly the application of an isothermal temperature profile in the atmosphere, as well as any simplifications regarding the efficiency of the heat engine. $k_0$ was set in \cite{koll_atmospheric_2018} by scaling the values derived from Equation \ref{eq:rayleigh} for different values of $\tau$ to the output of GCM simulations as described in \cite{komacek_atmospheric_2017}. Simply put, they anchor the calculation of the velocity of the atmosphere from the heat engine theory to GCM outputs calculated from first principle \textit{before} introducing magnetic drag as the governing force of the drag timescale. This allows to reset the assumptions going into the heat engine calculation and give absolute values of the magnetic field strength tied to the level of realism of the applied GCM. In consequence, the choice of GCM for the applied scaling is crucial. \cite{koll_atmospheric_2018} opted for the MITgcm \citep{adcroft_implementation_2004} which solves the atmospheric fluid dynamics equations with a double-gray radiative transfer approach as applied in \cite{komacek_atmospheric_2017}. Given the rapid evolution of the field and additional assumptions on the heat engine efficiency from the treatment of ultra-hot Jupiters vs hot Jupiters in \cite{koll_atmospheric_2018}, we have decided to re-calculate $k_0$ by scaling to the non-grey SPARC/MITgcm GCM \citep{showman_atmospheric_2009} as applied in \cite{roth_hot_2024}, where the MITgcm is coupled to the plane-parallel radiative transfer code of \cite{marley_thermal_1999}. \cite{roth_hot_2024} provides their pre-calculated grid of GCMs for a variety of temperatures and time scales, which we have used here for convenience. Most importantly, their GCMs include TiO/VO chemistry, although they show that including this important chemical aspect of ultra-hot atmospheres has a negligible impact on the equatorial, zonal wind speeds. The resulting scaling factor is $k_0 = 0.25$, of the same order of magnitude as the scaling factor in \cite{koll_atmospheric_2018}, highlighting the negligible impact of our additional assumptions. For future applications by the modelling community, we provide here the corresponding magnetic drag timescale from the observed wind speeds in Figure \ref{fig:tau_drag}.

\begin{figure}
    \centering
    \includegraphics[width=1.0\linewidth]{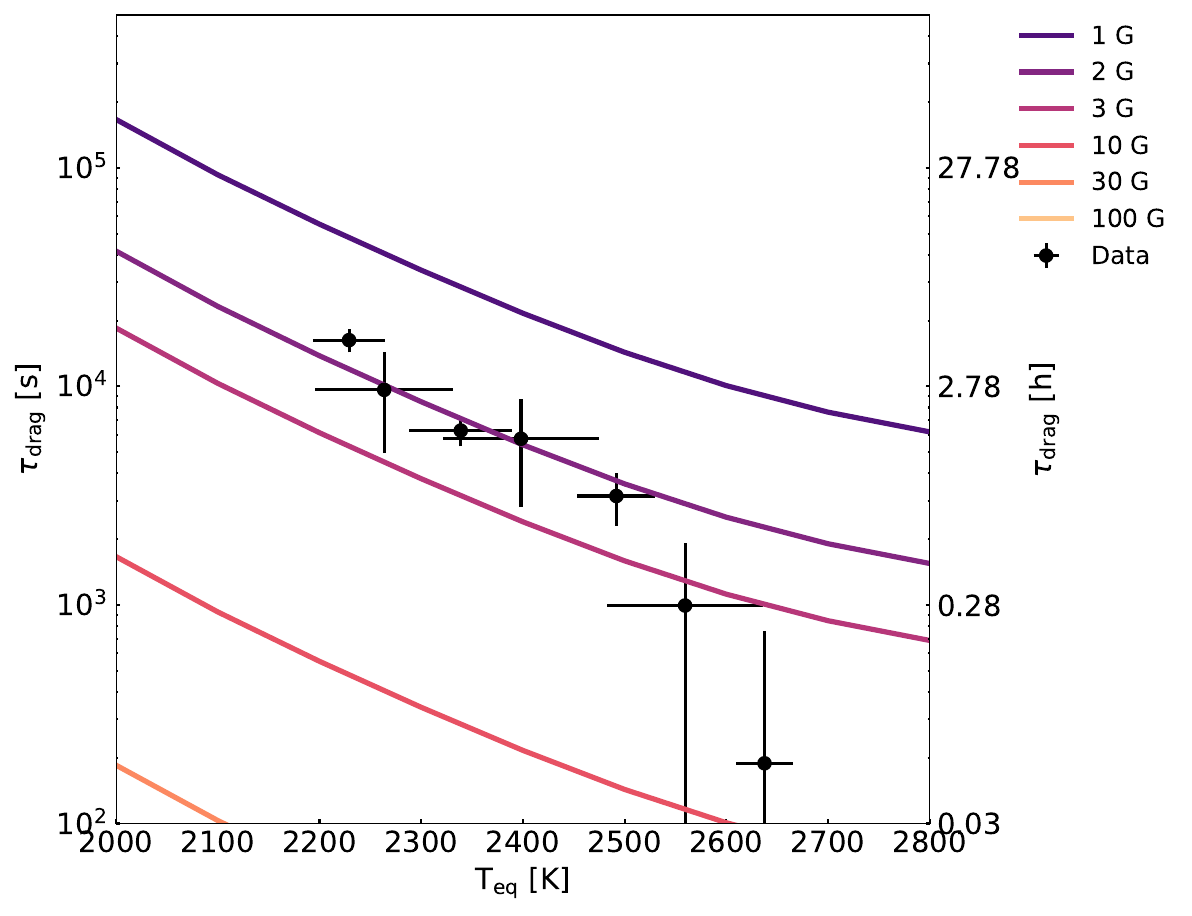}
    \caption{Calculated values of the drag timescale with models assuming Ohmic drag. The uncertainty is the propagated uncertainty of the measured velocity in Figure \ref{fig:master_vsys_model_real}.}
    \label{fig:tau_drag}
\end{figure}

Equally, \cite{roth_hot_2024} discusses the impact of any change in surface gravity on the equatorial zonal wind speed and finds no meaningful difference (see Figure \ref{fig:data_overview} for an overview of our targets in equilibrium temperature vs gravity space). However, they identify a dependence of the equatorial zonal wind speed on the rotation period of the planet. We found no correlation between the measured wind speed and the planetary periods. This is likely due to the short range of periods our data covers, making them an excellent sample to study effects without the additional dependency on the period. Equally, we found no trend with planet mass or radius.

\begin{figure}
    \centering
    \includegraphics[width=1.0\linewidth]{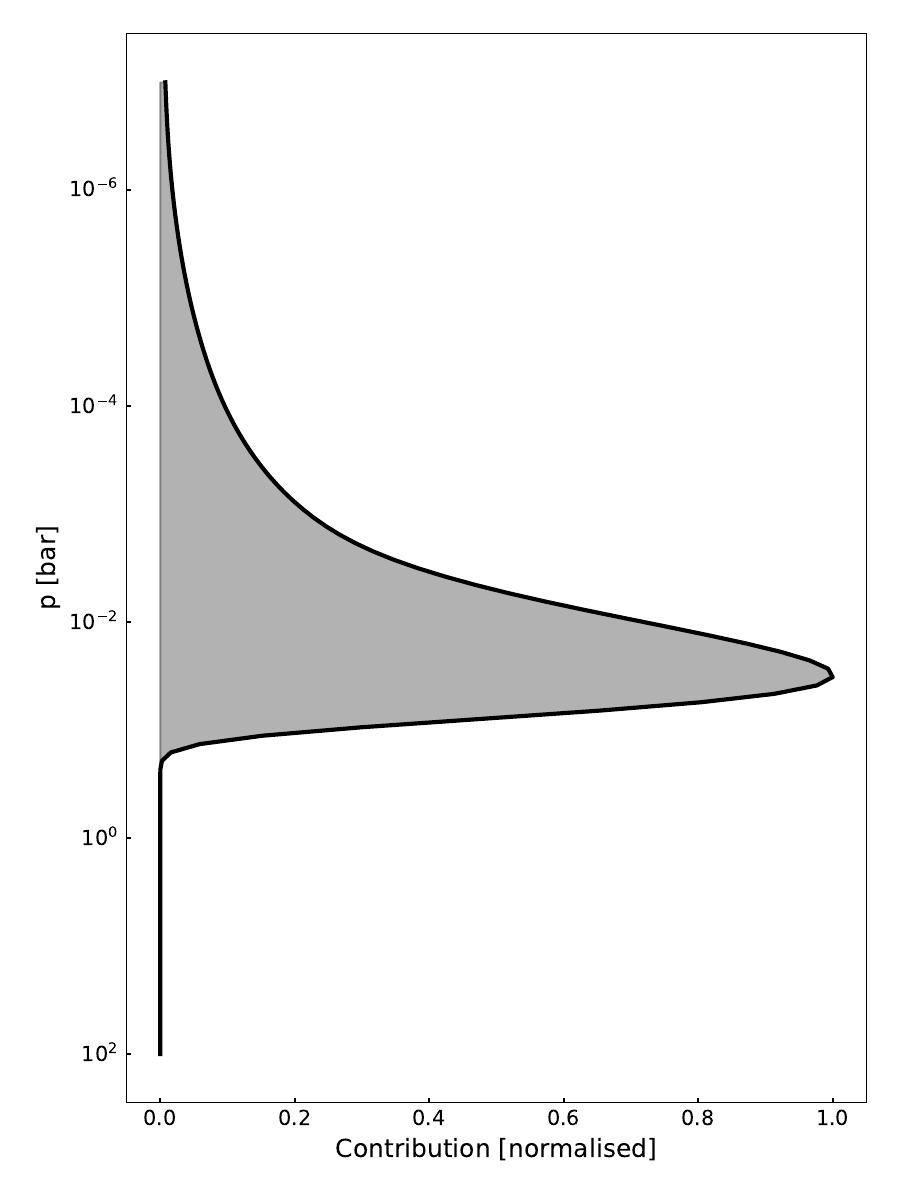}
    \caption{The normalised contribution function as derived from {\tt petitRadTrans} used for weighting of the different pressure layers with contributions above $10\%$.}
    \label{fig:contrib_func}
\end{figure}

\subsection{Assumptions and uncertainty estimate on the absolute atmospheric magnetic field strength}
\label{sec:mag_field_assumptions}

\cite{koll_atmospheric_2018} assesses the impact of the temperature-pressure profile on the absolute value of their models and finds a relatively small effect which they attribute to the inverse effect of pressure and temperature on the magnetic drag timescale. However, the pressure scale has a profound impact on the absolute value of the Ohmic dissipation models due to the pressure dependency of the ionisation in the magnetic diffusivity $H_e$. The dependency of the velocity in Equation \ref{eq:rayleigh} on $1/p$ pressure is offset by the linear dependency of the density on pressure in the ideal gas law.

In order to estimate the range of pressures that are probed by the observations, we developed a new technique to determine the contribution function of the cross-correlation function. Using the radiative transfer code {\tt petitRadTrans} \citep{molliere_petitradtrans_2019}, we calculate a first reference spectra, and then spectra where the opacity in each atmospheric layer is set to zero. Both spectra are then cross-correlated with the same template used to cross-correlate the observations. The contribution function is the normalised ratio of the maxima of the two cross-correlation functions. The resulting contribution function, shown in Figure~\ref{fig:contrib_func} shows that our observations probe from 1~to~100~mbar with a maximum at 30~mbar. In order to obtain the mean ionisation necessary to estimate the magnetic field, we weigh the precomputed ionisation fractions from \cite{lupu_correlated_2021} discussed in the previous section by the normalised contribution of each pressure layer, accounting for all pressure levels that contribute more than $10\%$. 
The current approach to contribution functions for cross-correlation is unfortunately flawed, as a simple setting of the opacity to zero creates a non-physical continuum contribution from aliasing. Likely, the true probing regions of the lines sit higher in the atmosphere \citep[see e.g.][ under the caveat that the line wings are not accounted for here, meaning this is an upper limit]{kesseli_up_2024}. Given that we do not see iron depletion in our observations for any of the targets, we cannot be higher than at most two orders of magnitude in pressure, which is when iron ionisation becomes dominant. In order to provide the most conservative estimates, we re-calculated the magnetic field assuming that the contribution functions is inaccurate by one or two orders of magnitude in pressure. The impact of this difference can be seen in Figure \ref{fig:Bfield_strength}, where a two order of magnitude difference in pressure would bias us to lower magnetic fields by roughly 1~G. 


While this gives us an accurate profile of the ionisation fraction of the probed atmosphere in the line of sight, it hinges on the metallicity, as an order of magnitude difference in metallicity has the same impact as an order of magnitude shift in pressure. \cite{gandhi_retrieval_2023} has shown that ultra-hot Jupiters have stellar metallicity, most importantly for iron, our atmospheric probe. The stellar metallicity of ultra-hot Jupiter host stars is known and diverges from the overall exoplanet host star population, peaking at $0.100\pm0.012$~dex (or roughly 1.25x solar) \citep{osborn_investigating_2020}. Any deviation below the order-of-magnitude level has a negligible impact on the derived magnetic field strength.

More important than the metallicity is the applied temperature as the magnetic diffusivity directly depends on $\sqrt{T}$ as well as indirectly via the ionisation fraction. The work performed by the drag force in the heat engine model is proportional to the integral over $1/\tau$ from the sub- to anti-stellar point (from the day to the night side). While we assume $\tau$ to be constant with temperature, it is unclear to which temperature the averaged $\tau$ corresponds to and if the limb, and the equilibrium temperature, is a good approximation of the work performed to create the drag. To better constrain which temperature is an accurate reflection of reality, we have parametrised the temperature profile of WASP-121~b from sub- to anti-stellar point following the GCM models from \cite{parmentier_thermal_2018}, which to first order follows a sinusoidal in longitude. Calculating $\tau$ as a function of longitude and comparing the normalised integrated value of $\tau$ shows that the main contribution to the work performed by the drag force is at approximately 75 degrees, slightly offset from the limb at 90 degrees. For WASP-121~b this offset leads to an underestimation of the temperature by at most 200~K. This offset is the maximum systematic bias across the dataset and would lead to higher ionisation levels with subsequently lower magnetic field strengths (see Figure \ref{fig:Bfield_strength}). 

Lastly, in transmission spectroscopy when observing iron, we are sensitive to the line of sight velocity component of the day-to-night side wind, which, due to the nature of the horizontal wind direction, is largely aligned with the absolute wind speed but will always be an under-estimation. We assessed the difference between the observed velocity compared to the absolute velocity from the GCM models published in \citep{roth_hot_2024} and find that for drag timescales below $10^5$ s the difference between the line of sight velocity and the real wind speed is at most $30\%$ (see the impact on the atmospheric magnetic field strength in Figure \ref{fig:Bfield_strength}). All of the mentioned effects that could impact our estimation of the strength of the atmospheric magnetic field would reduce its real strength and we have provided the impact of all cumulated effects with the dark crosses under each calculated atmospheric magnetic field strength in Figure \ref{fig:Bfield_strength}. We find that, conservatively opting for the maximum offset from our assumptions, the maximal atmospheric magnetic field strength will be at most reduced to the order of magnitude of half a gauss and thus remains comparable to the Jovian field and overall Solar System planets.

\end{document}